\documentclass[11pt]{article}

\usepackage{amsmath,amsthm,amsfonts,amssymb,srcltx,empheq}
\usepackage[dvipdfmx]{graphicx}
\usepackage[all]{xy}
\usepackage{color}
\usepackage{longtable}
\usepackage{hhline}

\usepackage{caption}
\def\IN{\mathbb {N}}
\def\IZ{\mathbb {Z}}

\def\IR{\mathbb {R}}
\def\IC{\mathbb {C}}
\def\IP{\mathbb {P}}

\def\rank{\operatorname{rank}}
\def\Sym{\operatorname{Sym}}

\theoremstyle{plain}
  \newtheorem{prob}{Problem}[section]

  \newtheorem{thm}[prob]{Theorem}

\theoremstyle{remark}
  \newtheorem{remark}[prob]{\bf Remark}

\newtheorem{exam}[prob]{\bf Example}

\renewcommand{\thefootnote}{\fnsymbol{footnote}}

\makeatletter
\def\Left#1#2\Right{\begingroup%
   \def\ts@r{\nulldelimiterspace=0pt \mathsurround=0pt}%
   \let\@hat=#1%
   \def\sht@im{#2}%
   \def\@t{{\mathchoice{\def\@fen{\displaystyle}\k@fel}%
          {\def\@fen{\textstyle}\k@fel}%
          {\def\@fen{\scriptstyle}\k@fel}%
          {\def\@fen{\scriptscriptstyle}\k@fel}}}%
   \def\g@rin{\ts@r\left\@hat\vphantom{\sht@im}\right.}%
   \def\k@fel{\setbox0=\hbox{$\@fen\g@rin$}\hbox{%
      $\@fen \kern.3875\wd0 \copy0 \kern-.3875\wd0%
      \llap{\copy0}\kern.3875\wd0$}}%
      \def\pt@h{\mathopen\@t}\pt@h\sht@im%
      \Right}%
\def\Right#1{\let\@hat=#1%
   \def\st@m{\mathclose\@t}%
   \st@m\endgroup}
\makeatother

\makeatletter
 \renewcommand{\theequation}{%
       \thesection.\arabic{equation}}
 \@addtoreset{equation}{section}
\makeatother

\makeatletter
\def\eqnarray{%
 \stepcounter{equation}%
 \let\@currentlabel=\theequation
 \global\@eqnswtrue
 \global\@eqcnt\z@
 \tabskip\@centering
 \let\\=\@eqncr
 $$\halign to \displaywidth\bgroup\@eqnsel\hskip\@centering
 $\displaystyle\tabskip\z@{##}$&\global\@eqcnt\@ne
 \hfil$\displaystyle{{}##{}}$\hfil
 &\global\@eqcnt\tw@$\displaystyle\tabskip\z@{##}$\hfil
 \tabskip\@centering&\llap{##}\tabskip\z@\cr}
\makeatother

\begin{document}
\begin{titlepage}

\begin{center}
\vspace*{1cm}
{\Large \bf
Determinantal Calabi-Yau varieties in Grassmannians\\[0.8em]
and the Givental $I$-functions}
\vskip 1.5cm
{\large Yoshinori Honma${}^{a}$\footnote[2]{yhonma@law.meijigakuin.ac.jp} and 
Masahide Manabe${}^b$\footnote[3]{masahidemanabe@gmail.com}}
\vskip 1.0em
{\it 
${}^a$%
Department of Current Legal Studies, \\
Meiji Gakuin University, \\
Yokohama, Kanagawa 244-8539, Japan \\

${}^b$%
School of Mathematics and Statistics, \\
University of Melbourne, \\
Royal Parade, Parkville, VIC 3010, Australia\\}
\end{center}
\vskip2.5cm

\begin{abstract}
We examine a class of Calabi-Yau varieties of the determinantal type 
in Grassmannians 
and clarify what kind of examples can be constructed explicitly. 
We also demonstrate how to compute their genus-0 Gromov-Witten invariants 
from the analysis of the Givental $I$-functions. By constructing $I$-functions 
from the supersymmetric localization formula for the two dimensional
gauged linear sigma models, we describe an algorithm to evaluate the 
genus-0 A-model correlation functions appropriately. 
We also check 
that our results for the Gromov-Witten invariants are consistent with previous results for known examples included in our construction.
\end{abstract}
\end{titlepage}


\renewcommand{\thefootnote}{\arabic{footnote}} \setcounter{footnote}{0}

\section{Introduction}\label{sec:introduction}

String compactifications to lower dimensions preserving 
supersymmetry motivate 
us to study Calabi-Yau varieties as initiated in \cite{Candelas:1985en}. 
To analyze various properties of Calabi-Yau backgrounds, it is useful to consider the two dimensional gauged linear 
sigma model (GLSM) \cite{Witten:1993yc}. This model corresponds to the UV description of the non-linear sigma models 
on Calabi-Yau backgrounds and has provided a powerful technique to compute the correlation functions exactly (see for example \cite{Morrison:1994fr}). 
Utilizing the duality property of the model, a physical understanding about mirror symmetry 
has also been developed \cite{Hori:2000kt} 
(see also \cite{Gu:2018fpm,Chen:2018wep} for several recent developments).

On the other hand, by using the supersymmetric localization techniques \cite{Pestun:2016zxk}, exact formula for the 
GLSM partition functions \cite{Benini:2012ui,Doroud:2012xw} (see also \cite{Jockers:2012dk,Honma:2013hma}) and correlation functions 
\cite{Closset:2015rna,Benini:2015noa} on the 
2-sphere backgrounds has been clarified. 
This means that one can evaluate the genus-0 Gromov-Witten invariants from the 
GLSM calculation in a direct fashion. 
This methodology has also been applied to the 
GLSMs with non-abelian gauge groups and the study of the complete intersection Calabi-Yau varieties in 
Grassmannians has progressed in the last few years \cite{Ueda:2016wfa,Kim16}.
Several notable aspects of the GLSM correlation functions have also been clarified in \cite{Gerhardus:2018zwb,Honma:2018qcr}.

While the toric complete intersection varieties described by abelian GLSMs have been thoroughly investigated in various contexts, 
a comprehensive understanding about the non-complete intersection varieties 
requires further efforts. In \cite{Jockers:2012zr}, as an example of a class of non-complete intersections, 
the determinantal varieties \cite{Harris1995,Fulton1998} and the associated non-abelian GLSMs have been investigated. 
The aim of our work is to make advances in the study of the determinantal varieties and provides a further step toward the comprehensive 
understanding of general Calabi-Yau backgrounds with non-abelian GLSM descriptions.

In this paper we explicitly clarify 
what kind of determinantal Calabi-Yau varieties in Grassmannians 
can be constructed while satisfying several requirements.
We mainly focus on the 3-fold examples and the analysis for the determinantal Calabi-Yau 2-folds and 4-folds
is summarized in Appendix.
We will also study the quantum aspects of the GLSM associated with determinantal varieties. 
Our method is based on the analysis of the so-called Givental $I$-functions 
\cite{Givental:1995,Givental,Coates:2001ewh} which can be extracted from the localization formula for the GLSM on a supersymmetric 2-sphere. 
In particular, we will compute the genus-0 Gromov-Witten invariants of determinantal Calabi-Yau varieties 
by using a conjectural handy formula, 
and check that our results coincide with previous results for known examples included in our classification.

This paper is organized as follows. First we examine 
a class of determinantal Calabi-Yau varieties in Grassmannians 
satisfying several requirements and specify the possible examples in Section \ref{sec:det_cy}. 
In Section \ref{sec:gw_I_fn}, we briefly review the computation of the
genus-0 invariants of 
complete intersection Calabi-Yau varieties in complex projective spaces and Grassmannians utilizing the $I$-functions, 
and provide a conjectural formula for 
Grassmannian Calabi-Yau varieties. 
In Section \ref{sec:det_I_fn},  
we compute the genus-0 invariants of determinantal Calabi-Yau varieties 
with GLSM realizations 
by using the algorithm described in Section \ref{sec:gw_I_fn}. 
Section \ref{sec:conclusion} is devoted to conclusions and discussions. 
In Appendix \ref{app:det_cy24} we summarize our results of the analysis for determinantal Calabi-Yau 2-folds and 4-folds.  
In Appendix \ref{app:hodge}, we take a brief look at the computation of Hodge numbers 
using the Koszul complex and demonstrate several computations explicitly. 
In Appendix \ref{app:gw_det} we summarize the data of the genus-0 invariants of 
several determinantal Calabi-Yau 4-folds.

\section{Determinantal Calabi-Yau varieties in Grassmannians}\label{sec:det_cy}

In this section, we first review a minimal ingredient of determinantal 
varieties, following \cite{Jockers:2012zr} 
(see also \cite{Harris1995,Fulton1998}). 
Afterwards we classify a class of determinantal Calabi-Yau 3-folds in Grassmannians satisfying appropriate conditions.

\subsection{Definitions}

Let $V$ be a compact algebraic variety, and $A: \mathcal{E}_p \to \mathcal{F}_q$ 
be a linear map from a rank $p$ vector bundle $\mathcal{E}_p$ on $V$ to a rank $q$ vector bundle $\mathcal{F}_q$ on $V$. Here we assume that the linear map $A$ is a global holomorphic section of the rank $pq$-bundle
$\mathrm{Hom}(\mathcal{E}_p, \mathcal{F}_q) \cong \mathcal{E}_p^* \otimes \mathcal{F}_q$ 
with maximal rank at a generic point of $V$. 
By representing the linear map $A$ locally 
as a $q \times p$ matrix $A(\phi)$ of the holomorphic sections, a determinantal variety 
is defined as
\begin{align}
Z(A,\ell)=\left\{\ \phi \in V \ |\ \rank\,A(\phi) \le \ell\ \right\}, \ \ 0 \le \ell < \min(p,q),
\end{align}
where $\phi$ denotes the homogeneous coordinates on $V$.
The complex codimension of $Z(A,\ell)$ in $V$ is given by 
\begin{align}
\mathrm{codim}\, Z(A,\ell) =(p-\ell)(q-\ell).
\end{align}
Here $(\ell+1) \times (\ell+1)$ 
minors of $A(\phi)$ generate the ideal $I(Z(A,\ell))$. 
Since $\mathrm{codim}\,Z(A,\ell)=(p-\ell)(q-\ell) < 
\binom{p}{\ell+1}\binom{q}{\ell+1}$ for $\ell \ge 1$, 
the ideal $I(Z(A,\ell))$ has non-trivial relations called syzygies 
and the determinantal variety $Z(A,\ell)$ for $\ell \ge 1$ is not 
a complete intersection. 
As argued in \cite{Jockers:2012zr}, a simple analysis of the Jacobian matrix implies that $Z(A,\ell)$ for $\ell \ge 1$ has singular loci 
along $Z(A,\ell-1)\subset Z(A,\ell)$ only. 
One can resolve these singularities by the so-called 
\textit{incidence correspondence} \cite{Jockers:2012zr,Harris1995},
\begin{align}
X_A^V=\left\{\ (\phi,x) \in V_{\mathcal{E}_p,p-\ell} \ |\ A(\phi)x=0\ \right\}\ 
\longrightarrow\ Z(A,\ell),
\label{det_pax_o}
\end{align}
where $V_{\mathcal{E}_p,p-\ell}$ denotes the fibration
\begin{align}
G(p-\ell,\mathcal{E}_p) \longrightarrow V_{\mathcal{E}_p,p-\ell} \xrightarrow{ \ \pi \ } V,
\end{align} 
with Grassmannian fibers 
$G(p-\ell,\mathcal{E}_p)$ of $(p-\ell)$-planes with respect to 
the $p$-dimensional fibers of $\mathcal{E}_p$. 
It is worth noting that the codimension of the singular loci in $V$ is 
$\mathrm{codim}\, Z(A,\ell-1)=\mathrm{codim}\, Z(A,\ell)+p+q-2\ell+1$, and then the determinantal variety $Z(A,\ell)$ with 
the dimension less than $p+q-2\ell+1$ does not have singular 
loci \cite{Jockers:2012zr}.\footnote{
As noted in \cite{Jockers:2012zr}, one can also use the incidence correspondence \eqref{det_pax_o} to describe the determinantal varieties without singular loci.}

\begin{remark}[\cite{Jockers:2012zr}]\label{rem:sing_det}
Since $\ell < \min(p,q)$, the determinantal varieties with 
dimension less than $2$ do not have singular loci. 
The determinantal 3-folds have singular points only when $p=q=\ell+1$. 
The determinantal 4-folds have singular lines only 
when $p=q=\ell+1$, and have singular points only when $(p,q)=(\ell+1,\ell+2)$ 
or $(p,q)=(\ell+2,\ell+1)$.
\end{remark}

In this paper we only consider the square ($p=q$) determinantal varieties with
\begin{align}
V=G(k,n),\qquad
\mathcal{E}_p= \mathcal{O}_V^{\oplus p},\qquad
\rank\,\mathcal{F}_p = p,
\label{det_grass}
\end{align}
where $G(k,n)$ is the complex Grassmannian defined by the set of $k$-planes 
in ${\IC}^n$, and $\mathcal{O}_V$ is the structure sheaf of $V$. 
Then the variety $V_{\mathcal{E}_p,p-\ell}$ can be described by a product variety $V_{\mathcal{E}_p,p-\ell}\cong G(k,n)\times G(p-\ell,p)$ and the incidence correspondence (\ref{det_pax_o}) becomes 
\begin{align}
X_A:=X_A^{G(k,n)}=\left\{\ (\phi,x) \in G(k,n)\times G(\ell_p^{\vee},p) \ |\ A(\phi)x=0\ \right\},\qquad
\ell_p^{\vee}:= p-\ell.
\label{det_pax}
\end{align}
In addition we require 
$n=\ell_p^{\vee} \mathfrak{c}_1(\mathcal{F}_p)$ derived from Calabi-Yau condition \cite{Jockers:2012zr}. 
Here 
$$
c_1(\mathcal{F}_p)=\mathfrak{c}_1(\mathcal{F}_p)\, \sigma_1
$$
is the first Chern class of $\mathcal{F}_p$ 
and $\sigma_1=c_1(\mathcal{Q})$ 
is the Schubert class of $G(k,n)$. 
$\mathcal{Q}$ is 
the universal quotient bundle on $G(k,n)$. 
The dimension of $X_A$ is 
given by 
$\dim X_A = k(n-k) - (p-\ell)^2=
\ell_p^{\vee} \left(k \mathfrak{c}_1(\mathcal{F}_p)-\ell_p^{\vee}\right)-k^2$. 
By taking the duality $G(k,n)\cong G(n-k,n)$ into consideration, 
here we only consider the case with $2k \le n$. 
Furthermore, the rank condition $0\le \ell < p$ 
can be rephrased as $0 <\ell_p^{\vee} \le p=\rank\,\mathcal{F}_p$. 

In summary, we have seen that the following conditions\footnote{Note that we do not impose \textit{irreducibility} or the conditions 
$H^i(X_A,\mathcal{O}_{X_A})=0$ in our analysis.} 
must be satisfied in order to realize the determinantal varieties appropriately.
\begin{align}
&
{\bf 1.}\quad
\textrm{{\bf Dimensional condition:}}
&&\ell_p^{\vee} \left(k \mathfrak{c}_1(\mathcal{F}_p)-\ell_p^{\vee}\right)=
k^2+\dim X_A.
\label{cy_dim_inc}
\\
&
{\bf 2.}\quad
\textrm{{\bf Calabi-Yau condition:}}
&&n=\ell_p^{\vee} \mathfrak{c}_1(\mathcal{F}_p).
\label{cy_cd_inc}
\\
&
{\bf 3.}\quad
\textrm{{\bf Duality condition:}}
&&2k \le n.
\label{cy_dualc}
\\
&
{\bf 4.}\quad
\textrm{{\bf Rank condition:}}
&&0 <\ell_p^{\vee} \le p=\rank\,\mathcal{F}_p.
\label{cy_det_inc}
\end{align}
In the following, we will classify the determinantal Calabi-Yau 3-folds 
satisfying the above four conditions. 
Although we consider the desingularized determinantal varieties $X_A$, 
the following analysis also gives a classification of $Z(A,\ell)$.\footnote{
See Appendix \ref{app:det_cy24} for 
the analysis of determinantal Calabi-Yau 2-folds and 4-folds.}

\subsection{General dimensions}

Before moving on to the discussion about the determinantal Calabi-Yau 3-folds, let us consider general 
implications of the above requirements. In general dimensions, obviously the following two 
ansatz always satisfy the dimensional condition (\ref{cy_dim_inc}). 
\begin{align}
&\textrm{Ansatz \ (I)}: &
\left(\ell_p^{\vee}, k \mathfrak{c}_1(\mathcal{F}_p)\right) &=
\left(1, \dim X_A+k^2+1\right),
\label{sol_1}
\\
&\textrm{Ansatz (II)}: &
\left(\ell_p^{\vee}, k \mathfrak{c}_1(\mathcal{F}_p)\right) &=
\left(\dim X_A+k^2, \dim X_A+k^2+1\right).
\label{sol_2}
\end{align}
In the following, we will illustrate what kind of setups for $\left(k,n;\ell_p^{\vee}, \mathfrak{c}_1(\mathcal{F}_p)\right)$
satisfy all the above requirements if we start from the Ansatz (I) or (II).\footnote{Of course there exist other solutions
which do not belong to the Ansatz (I) or (II). 
In Section \ref{subsec:cy3} we have also taken into account this kind of solutions 
and checked, by Mathematica and Maple, 
that our result exhausted all the possible solutions up to $k=50$.
}

\subsubsection{Ansatz (I)}

In this case, the Calabi-Yau condition (\ref{cy_cd_inc}) becomes
\begin{align}
n=\frac{1}{k}\left(\dim X_A+k^2+1\right).
\end{align}
Then the duality condition (\ref{cy_dualc}) implies 
\begin{align}
k^2 \le \dim X_A+1,
\label{const1}
\end{align}
which means that examples with $k \ge 3$ provide determinantal varieties with $\dim X_A \ge 8$.

When $k=1$, $V$ is given by $G(1,n)\cong {\IP}^{n-1}$ and one obtains the solutions with
\begin{align}
\left(k,n;\ell_p^{\vee}, \mathfrak{c}_1(\mathcal{F}_p)\right)=
\left(1,\dim X_A+2;1,\dim X_A+2\right).
\label{class_1}
\end{align}
In this case the rank condition (\ref{cy_det_inc}) is trivially satisfied, and 
appropriate $\mathcal{F}_p$ on $V$ are given by the following vector bundles 
associated with the integer partitions of $\dim X_A+2:$
\begin{align}
\mathcal{F}_p=\oplus_{i=1}^r\mathcal{O}_V(p_i),\qquad
p_1\ge p_2 \ge \cdots \ge p_r > 0,\quad \sum_{i=1}^rp_i=\dim X_A+2.
\label{class_1f}
\end{align}

When $k=2$, $V$ becomes $G(2,n)$ and one finds the solutions with
\begin{align}
\left(k,n;\ell_p^{\vee}, \mathfrak{c}_1(\mathcal{F}_p)\right)=
\left(2,(\dim X_A+5)/2;1,(\dim X_A+5)/2\right).
\label{class_11}
\end{align}
Since $n$ has to be an integer, this type of solution 
can exist only when the dimension of $X_A$ is odd. 
Moreover, (\ref{const1}) requires $\dim X_A \ge 3$.  

\subsubsection{Ansatz (II)}

In this case, the Calabi-Yau condition (\ref{cy_cd_inc}) becomes
\begin{align}
n=\frac{1}{k}\left(\dim X_A+k^2\right)\left(\dim X_A+k^2+1\right),
\label{sol_2_nn}
\end{align}
and the duality condition (\ref{cy_dualc}) is trivially satisfied. 
Thus we only need to consider the rank condition 
given by
\begin{align}
\dim X_A+k^2 \le \rank\,\mathcal{F}_p.
\label{sol_2_rank}
\end{align}

For example, when $k=1$, we obtain the following solutions 
\begin{align}
\begin{split}
&
\left(k,n;\ell_p^{\vee}, \mathfrak{c}_1(\mathcal{F}_p)\right)=
\left(1,(\dim X_A+1)(\dim X_A+2);\dim X_A+1,\dim X_A+2\right)
\\
&\ \textrm{with} \ 
\mathcal{F}_p=\mathcal{O}_V(1)^{\oplus \dim X_A} \oplus \mathcal{O}_V(2),\
\mathcal{O}_V(1)^{\oplus (\dim X_A+2)}.
\label{class_2}
\end{split}
\end{align}
Note that the rank condition (\ref{sol_2_rank}) strongly constrain the possible vector bundles.

\subsection{Determinantal Calabi-Yau 3-folds}\label{subsec:cy3}

Here we will focus on the square determinantal Calabi-Yau 3-folds and determine what kind of setups satisfy the above four conditions.
Let us start with the dimensional condition given by
\begin{align}
\ell_p^{\vee} \left(k \mathfrak{c}_1(\mathcal{F}_p)-\ell_p^{\vee}\right)=k^2+3.
\label{cy_dim3_ns}
\end{align}
Then we will find out which type of choices for $\left(k,n;\ell_p^{\vee}, \mathfrak{c}_1(\mathcal{F}_p)\right)$
can be possible while changing the parameter $k$.

\subsubsection{$k=1$}

In this case we have $V=G(1,n)\cong {\IP}^{n-1}$. 
From (\ref{class_1}) and (\ref{class_1f}) one finds that there exists a
``quintic family'' (see for example \cite{Hosono:2011np}) given by
\begin{align}
\boxed{
\left(k,n;\ell_p^{\vee}, \mathfrak{c}_1(\mathcal{F}_p)\right)=
\left(1,5;1,5\right) \ \textrm{with} \ 
\mathcal{F}_p=\oplus_{i=1}^r\mathcal{O}_V(p_i),\ 
p_1\ge p_2 \ge \cdots \ge p_r > 0,\ \sum_{i=1}^rp_i=5.}
\label{class_cy3_1}
\end{align}
The example constructed from $\mathcal{F}_p=\mathcal{O}_V(5)$ 
with $p=1$ (\textit{i.e.} $\ell=0$) is the well-known quintic Calabi-Yau 3-fold, 
which is the zero locus of a holomorphic section of $\mathcal{O}_{{\IP}^4}(5)$.

Since $\ell_p^{\vee}=1$ (\textit{i.e.} $p=\ell+1$), 
according to the Remark \ref{rem:sing_det}, generically the above quintic families have singular points. 
The determinantal Calabi-Yau 3-folds in this class are connected by 
the deformations of complex structures, and 
it is known that the desingularized 3-folds are related by 
the so-called extremal transitions.\footnote{The comparison of topological invariants in Section \ref{subsubapp:quintic} 
makes this point clearly understandable.}

Apart from the above quintic family, 
one can also find the following solutions 
\begin{align}
\boxed{
\left(k,n;\ell_p^{\vee}, \mathfrak{c}_1(\mathcal{F}_p)\right)=
(1,8;2,4) \ \textrm{with} \ 
\mathcal{F}_p=\mathcal{O}_V(1)\oplus \mathcal{O}_V(3),\ 
\mathcal{O}_V(2)^{\oplus 2},\
\mathcal{O}_V(1)^{\oplus 2}\oplus \mathcal{O}_V(2),\
\mathcal{O}_V(1)^{\oplus 4}.}
\label{class_cy3_2}
\end{align}
Here the first two examples with $p=2$ (\textit{i.e.} $\ell=0$) 
in \eqref{class_cy3_2} can be identified with the well-known
complete intersection Calabi-Yau 3-folds as
$$
X_A\ \textrm{with}\ 
\mathcal{O}_V(1)\oplus \mathcal{O}_V(3)\ \longleftrightarrow\
X_{3,3} \subset {\IP}^5,\qquad
X_A\ \textrm{with}\ 
\mathcal{O}_V(2)^{\oplus 2}\ \longleftrightarrow\
X_{2,2,2,2} \subset {\IP}^7,
$$
where $X_{d_1,\ldots,d_r}\subset {\IP}^{n-1}$ denotes the complete intersection variety defined by the zero locus of a holomorphic section of the vector bundle 
$\oplus_{a=1}^r \mathcal{O}_{{\IP}^{n-1}}(d_a)$. 
The last two examples in \eqref{class_cy3_2} are 
Gulliksen-Neg{\aa}rd type 3-folds studied in \cite{Bertin0701}.

Moreover, (\ref{class_2}) provides another type of solutions given by
\begin{align}
\boxed{
\left(k,n;\ell_p^{\vee}, \mathfrak{c}_1(\mathcal{F}_p) \right)=
(1,20;4,5) \ \textrm{with} \ 
\mathcal{F}_p=
\mathcal{O}_V(1)^{\oplus 3}\oplus \mathcal{O}_V(2),\
\mathcal{O}_V(1)^{\oplus 5}.}
\label{class_cy3_3}
\end{align}
The Calabi-Yau 3-fold $X_A$ in \eqref{class_cy3_3} 
constructed from $\mathcal{F}_p=\mathcal{O}_V(1)^{\oplus 3}\oplus \mathcal{O}_V(2)$ 
with $p=4$ (\textit{i.e.} $\ell=0$) has the following isomorphism:
$$
X_A\ \textrm{with}\ 
\mathcal{O}_V(1)^{\oplus 3}\oplus \mathcal{O}_V(2)\ \longleftrightarrow\
X_{2,2,2,2} \subset {\IP}^7.
$$
The other example constructed from $\mathcal{F}_p=\mathcal{O}_V(1)^{\oplus 5}$ has 
been studied in \cite{Jockers:2012zr}.

\subsubsection{$k=2$}

In this case, $V$ becomes the Grassmannians $V=G(2,n)$. 
Compared with the complex projective spaces, there exist 
additional components for the vector bundles on the Grassmannians,
as explained in the followings.

When $k\ge 2$, beside the line bundle $\mathcal{O}_V(d)$ on $V=G(k,n)$, 
one can also consider vector bundles with rank greater than one denoted by
$$
\mathcal{S}^*\ \ \textrm{and}\ \ \mathcal{Q}. 
$$
These are known as the dual of the universal subbundle 
and the universal quotient bundle on $G(k,n)$, respectively. 
Note that they fulfill the relation $\wedge^k\mathcal{S}^*\cong \mathcal{O}_V(1)$ and satisfy
$$
\rank\mathcal{S}^*=k,\quad
\mathfrak{c}_1(\mathcal{S}^*)=1,\quad
\rank\mathcal{Q}=n-k,\quad
\mathfrak{c}_1(\mathcal{Q})=1.
$$
Accordingly, general irreducible vector bundles can be constructed as
\begin{align}
&
\Sym^m \mathcal{S}^*(d):=\Sym^m\mathcal{S}^*\otimes \mathcal{O}_V(d),\quad
\wedge^m \mathcal{S}^*(d):=\wedge^m\mathcal{S}^*\otimes \mathcal{O}_V(d),
\nonumber
\\
&
\Sym^m \mathcal{Q}(d):=\Sym^m\mathcal{Q}^*\otimes \mathcal{O}_V(d),\quad
\wedge^m \mathcal{Q}(d):=\wedge^m\mathcal{Q}^*\otimes \mathcal{O}_V(d),
\nonumber
\end{align}
where
\begin{small}
\begin{align}
&
\rank \Sym^m \mathcal{S}^*(d)=\binom{k+m-1}{m},\qquad 
\mathfrak{c}_1\left(\Sym^m \mathcal{S}^*(d)\right)=\binom{k+m-1}{k}+d \binom{k+m-1}{m},
\nonumber\\
&
\rank \wedge^m \mathcal{S}^*(d)=\binom{k}{m},\qquad
\mathfrak{c}_1\left(\wedge^m \mathcal{S}^*(d)\right)=\binom{k-1}{m-1}+d \binom{k}{m},
\nonumber\\
&
\rank \Sym^m \mathcal{Q}(d)=\binom{n-k+m-1}{m},\qquad 
\mathfrak{c}_1\left(\Sym^m \mathcal{Q}(d)\right)=\binom{n-k+m-1}{n-k}+d \binom{n-k+m-1}{m},
\nonumber\\
&
\rank \wedge^m \mathcal{Q}(d)=\binom{n-k}{m},\qquad
\mathfrak{c}_1\left(\wedge^m \mathcal{Q}(d)\right)=\binom{n-k-1}{m-1}+d \binom{n-k}{m}.
\nonumber
\end{align}
\end{small}

Returning to the main subject of the classification, (\ref{class_11}) with the rank condition (\ref{cy_det_inc}) implies that
the following setups are possible
\begin{empheq}[box=\fbox]{equation}
\begin{split}
&
\left(k,n;\ell_p^{\vee}, \mathfrak{c}_1(\mathcal{F}_p) \right)=
(2,4;1,4) \ \textrm{with} \\
&\ 
\mathcal{F}_p=\mathcal{O}_V(4),\
\mathcal{O}_V(1)\oplus \mathcal{O}_V(3),\
\mathcal{O}_V(2)^{\oplus 2},\
\mathcal{O}_V(1)^{\oplus 2}\oplus \mathcal{O}_V(2),\
\mathcal{S}^*\oplus\mathcal{O}_V(3),
\\
&\qquad\ \
\mathcal{S}^*(1)\oplus\mathcal{O}_V(1),\
\mathcal{O}_V(1)^{\oplus 4},\
\mathcal{S}^*\oplus\mathcal{O}_V(1)\oplus \mathcal{O}_V(2),\
\Sym^2 \mathcal{S}^*\oplus\mathcal{O}_V(1),
\\
&\qquad\ \
\mathcal{S}^*\oplus\mathcal{O}_V(1)^{\oplus 3},\
\left(\mathcal{S}^*\right)^{\oplus 2}\oplus\mathcal{O}_V(2),\
\left(\mathcal{S}^*\right)^{\oplus 2}\oplus\mathcal{O}_V(1)^{\oplus 2},\
\left(\mathcal{S}^*\right)^{\oplus 3}\oplus\mathcal{O}_V(1),\
\left(\mathcal{S}^*\right)^{\oplus 4}.
\label{det3_k2}
\end{split}
\end{empheq}
Note that $\mathcal{S}^*\cong \mathcal{Q}$ on $G(2,4)$ and $\wedge^2\mathcal{S}^*\cong \mathcal{O}_V(1)$ when $k=2$. 
The example constructed from $\mathcal{F}_p=\mathcal{O}_V(4)$ with $p=1$ (\textit{i.e.} $\ell=0$) 
is the complete intersection Grassmannian Calabi-Yau 3-fold in $G(2,4)$ with the vector bundle 
$\mathcal{O}_{G(2,4)}(4)$.
Since $\ell_p^{\vee}=1$ (\textit{i.e.} $p=\ell+1$), 
as discussed in the case for the quintic family \eqref{class_cy3_1}, 
these determinantal Calabi-Yau 3-folds generically have singular points 
and the desingularized 3-folds are connected through the extremal transitions 
(see also Section \ref{subsubapp:det_gr3}).


Another class of solutions can be obtained by the ansatz (II) in \eqref{sol_2} and the result is
\begin{empheq}[box=\fbox]{equation}
\begin{split}
&
\left(k,n;\ell_p^{\vee}, \mathfrak{c}_1(\mathcal{F}_p) \right)=
(2,28;7,4) \ \textrm{with} \\
&\ 
\mathcal{F}_p=
\left(\mathcal{S}^*\right)^{\oplus 3}\oplus\mathcal{O}_V(1),\
\left(\mathcal{S}^*\right)^{\oplus 4},\
\mathcal{Q} \oplus \mathcal{O}_V(3),\
\mathcal{Q} \oplus \mathcal{O}_V(1) \oplus \mathcal{O}_V(2),\
\mathcal{Q} \oplus \mathcal{O}_V(1)^{\oplus 3},
\\
&\qquad\ \
\mathcal{Q}^{\oplus 2} \oplus \mathcal{O}_V(2),\ 
\mathcal{Q}^{\oplus 2} \oplus \mathcal{O}_V(1)^{\oplus 2},\
\mathcal{Q}^{\oplus 3} \oplus \mathcal{O}_V(1),\ 
\mathcal{Q}^{\oplus 4}.
\end{split}
\end{empheq}
Here the Calabi-Yau 3-fold constructed from 
$\mathcal{F}_p=\left(\mathcal{S}^*\right)^{\oplus 3}\oplus\mathcal{O}_V(1)$ 
with $p=7$ (\textit{i.e.} $\ell=0$) can be identified with the complete intersection Grassmannian Calabi-Yau 3-fold 
in $G(2,7)$ with the vector bundle $\mathcal{O}_{G(2,7)}(1)^{\oplus 7}$.

\subsubsection{$k\ge 3$}

In this case, we have $V=G(k,n)$. Interestingly, there exist four
``infinite families'' given by
\begin{empheq}[box=\fbox]{equation}
\begin{split}
&
\left(k,n;\ell_p^{\vee}, \mathfrak{c}_1(\mathcal{F}_p)\right)=
(k_i, 5\ell_i ; \ell_i, 5),\ \ i\in{\IN}
\\
&\
\textrm{with}\ 
k_{1}=4,\ \ell_{1}=19,\
k_{i+1}=\ell_{i},\ \ell_{i+1}=-k_{i}+5\ell_{i}:\
\mathcal{F}_p=\left(\mathcal{S}^*\right)^{\oplus 5},\
\mathcal{Q}^{\oplus 5},
\\
&
\left(k,n;\ell_p^{\vee}, \mathfrak{c}_1(\mathcal{F}_p)\right)=
(k_i, 4\ell_i ; \ell_i, 4),\ \ i\in{\IN}
\\
&\
\textrm{with}\ 
k_{1}=7,\ \ell_{1}=26,\
k_{i+1}=\ell_{i},\ \ell_{i+1}=-k_{i}+4\ell_{i}:\
\mathcal{F}_p=\left(\mathcal{S}^*\right)^{\oplus 4},\
\mathcal{Q}^{\oplus 4}.
\label{inf_det_3}
\end{split}
\end{empheq}
By using mathematical induction, one can check that the duality condition 
\eqref{cy_dualc}, 
the rank condition \eqref{cy_det_inc}, 
and in particular $\ell_p^{\vee} < \rank\,\mathcal{F}_p$, 
are maintained. Since we do not impose the irreducibility condition in our analysis, 
it is still possible that the above infinite families can be reduced to other trivial or non-trivial examples. In 
any case, it is required to thoroughly investigate various topological invariants of these higher rank examples, and 
this would require a considerable effort and we leave this issue as an open problem.


\section{$I$-functions and 
Gromov-Witten invariants}\label{sec:gw_I_fn}

In this section, we 
briefly overview the computation of genus-0 Gromov-Witten invariants 
using the Givental $I$-functions \cite{Givental:1995,Givental,Coates:2001ewh} (see also \cite{Cox:2000vi}). 
We will also provide a handy formula 
for the computations of the Gromov-Witten invariants of Grassmannian 
Calabi-Yau varieties, 
which is also applicable to the determinantal varieties.

\subsection{Building blocks of $I$-functions}\label{subsec:building_I_fn}

When a Fano or a Calabi-Yau variety $X$ has a GLSM realization with gauge group $G$, 
one can easily construct the Givental $I$-function of $X$ by using the supersymmetric localization formula
(see \cite{Bonelli:2013mma,Ueda:2016wfa,Kim16,Gerhardus:2018zwb,Honma:2018qcr}). 
Here we clarify the building blocks of the $I$-function of $X$ 
associated with such a GLSM on the $\Omega$-deformed 2-sphere $S_{\hbar}^2$ 
which has a vector multiplet and chiral matter multiplets with 
$R$-charge 0 or 2 under $U(1)_R$. The deformation parameter $\hbar$ is identified with 
an equivariant parameter.

Let $\mathbf{x}=(x_1,\ldots,x_{\mathrm{rk}(\mathfrak{g})}) \in \mathfrak{h}\otimes_{\IR}{\IC}$ be Coulomb branch parameters 
and $\mathbf{q}=(q_1,\ldots,q_{\mathrm{rk}(\mathfrak{g})}) \in {\IZ}^{\mathrm{rk}(\mathfrak{g})} \subset i \mathfrak{h}$ be magnetic charges 
for Cartan subalgebra $\mathfrak{h}$ of a Lie algebra $\mathfrak{g}$ associated with $G$, where 
$\mathrm{rk}(\mathfrak{g})$ denotes the rank of $\mathfrak{g}$. 
Here the parameters $\mathbf{x}$ are identified with the Chern roots of $X$ which give the total Chern class of $X$ as
\begin{align}
c(X)=\prod_{i=1}^{\mathrm{rk}(\mathfrak{g})}\left(1+x_i\right).
\label{chern_rt}
\end{align}
To construct the $I$-function of $X$, first we need a ``classical block'' 
associated with the subgroup $U(1)^{\mathsf{c}}\subset G$, where 
$\mathsf{c}$ is the number of the central. 
The Fayet-Iliopoulos (FI) parameters $\xi_a$ and theta angles $\theta_a$, $a=1,\ldots,\mathsf{c},$ are associated
with each $U(1)^{\mathsf{c}}$ factor, and the classical block of the $I$-function is given by
\begin{align}
I_{\mathbf{q}}^{\mathrm{c}}(\mathbf{z};\mathbf{x};\hbar)= 
\mathrm{e}^{2\pi \sqrt{-1}\, {\boldsymbol\tau}(\mathbf{x}/\hbar + \mathbf{q})},\qquad
{\boldsymbol\tau}=\{ \tau_a \} := \sqrt{-1}\, \xi_a + \frac{1}{2\pi} \theta_a.
\label{build_classic}
\end{align}
Here the parameters $\mathbf{z} = \{ z_a \} =\mathrm{e}^{2\pi \sqrt{-1}\, \tau_a}$ represent the exponentiated K\"ahler moduli of $X$, and 
the canonical pairing ${\boldsymbol\tau}(*)$ is defined by embedding ${\boldsymbol\tau}$ into $\mathfrak{h}^* \otimes_{\IR}{\IC}$.

Other contributions come from the 1-loop determinants of multiplets of the GLSM. 
The vector multiplet provides a block given by
\begin{align}
I_{\mathbf{q}}^{\mathrm{vec}}(\mathbf{x};\hbar)&=
\prod_{\alpha\in \Delta_+}(-1)^{\alpha(\mathbf{q})}\,
\frac{\alpha(\mathbf{x})+\alpha(\mathbf{q})\hbar}{\alpha(\mathbf{x})},
\label{build_vec}
\end{align}
where $\Delta_+$ is the set of positive roots of $\mathfrak{g}$. 
In general, the GLSM also has chiral matter multiplets $\Phi$ 
with $R$-charge 0 and 
$P$ with $R$-charge 2 in a certain representation $\mathbf{R}$. 
Note that one can turn on a twisted mass parameter $\lambda$ 
while preserving supersymmetry, which is identified 
with an equivariant parameter. 
Their contributions are given as follows:
\begin{align}
I_{\mathbf{q}}^{\Phi}(\mathbf{x},\lambda;\hbar)=
\begin{cases}
\prod_{\rho\in \mathbf{R}}\prod_{p=1}^{\rho(\mathbf{q})}
\left(\rho(\mathbf{x})+\lambda+p\hbar\right)^{-1},\quad
&
\textrm{for}\ \ \rho(\mathbf{q})\ge 1,
\\
1,
&
\textrm{for}\ \ \rho(\mathbf{q})= 0,
\\
\prod_{\rho\in \mathbf{R}}\prod_{p=0}^{-\rho(\mathbf{q})-1}
\left(\rho(\mathbf{x})+\lambda-p\hbar\right),\quad
&
\textrm{for}\ \ \rho(\mathbf{q})\le -1,
\end{cases}
\label{build_mat0}
\end{align}
and
\begin{align}
I_{\mathbf{q}}^{P}(\mathbf{x},\lambda;\hbar)=
\begin{cases}
\prod_{\rho\in \mathbf{R}}\prod_{p=1}^{-\rho(\mathbf{q})}
\left(-\rho(\mathbf{x})-\lambda+p\hbar\right),\quad
&
\textrm{for}\ \ \rho(\mathbf{q})\le -1,
\\
1,
&
\textrm{for}\ \ \rho(\mathbf{q})= 0,
\\
\prod_{\rho\in \mathbf{R}}\prod_{p=0}^{\rho(\mathbf{q})-1}
\left(-\rho(\mathbf{x})-\lambda-p\hbar\right)^{-1},\quad
&
\textrm{for}\ \ \rho(\mathbf{q})\ge 1,
\end{cases}
\label{build_mat2}
\end{align}
where $\rho$ denotes the weight of $\mathbf{R}$.
Note that the products $\alpha(*)$ and $\rho(*)$ are defined by the canonical pairing. 

Combining all the above building blocks \eqref{build_classic}, 
\eqref{build_vec}, \eqref{build_mat0}, and \eqref{build_mat2},  
after taking a sum over the magnetic charges $\mathbf{q}$, 
one can construct the Givental $I$-function as 
\eqref{I_ex1}, \eqref{I_ex2}, and \eqref{i_fun_det}. 
As we will see next, the genus-0 Gromov-Witten invariants can be extracted from this function.

\subsection{Examples}

Here we will demonstrate how to 
compute genus-0 Gromov-Witten invariants via the $I$-functions 
for well-studied examples, and find out 
a useful formula for treating Grassmannian Calabi-Yau varieties.

\subsubsection{Complete intersections in ${\IP}^{n-1}$}\label{subsub:cy_cp}

Let us consider a complete intersection variety 
$X_1=X_{d_1,\ldots,d_r}\subset {\IP}^{n-1}$ defined by the zero locus of 
a holomorphic section of a vector bundle $\mathcal{E}=\oplus_{a=1}^r \mathcal{O}_V(d_a)$ on 
$V={\IP}^{n-1}$ satisfying Fano or Calabi-Yau condition $\sum_{a=1}^r d_a\le n$. 
Note that $\rank\mathcal{E}=r$ 
and $\mathfrak{c}_1(\mathcal{E})=\sum_{a=1}^r d_a$. This variety 
has a complex dimension 
\begin{align}
\dim X_1= n-r-1,
\end{align}
and is described by a $U(1)$ GLSM whose matter content is shown in Table \ref{ex_1}. This model has a superpotential 
$W=\sum_{a=1}^r P_aG_a(\Phi)$ where 
$G_a(\Phi)$ are homogeneous degree $d_a$ polynomials of the chiral matter multiplets $\Phi_i$. 
\begin{table}[t]
\begin{center}
\begin{tabular}{|c|c|c|c|}
\hline
Field & U(1) & twisted mass& $U(1)_R$ \\ \hline
$\Phi_i$ & +1 & $-w_i$ & $0$ \\
$P_a$ & $-d_a$ & $\lambda_a$ & $2$ \\ \hline
\end{tabular}
\caption{Matter content of the $U(1)$ GLSM for the complete intersection variety $X_1$ in ${\IP}^{n-1}$. Here $i=0, \ldots, n-1$ and $a=1, \ldots, r$.}
\label{ex_1}
\end{center}
\end{table}

For each matter multiplet we assign twisted masses and $U(1)$ $R$-charges 
as described in Table \ref{ex_1}. Combining the building blocks \eqref{build_classic}, \eqref{build_mat0} and \eqref{build_mat2}  
with the assignment in Table \ref{ex_1}, the $I$-function 
in the geometric large volume phase with FI parameter $\xi > 0$ 
is constructed as \cite{Givental:1995,Givental,Coates:2001ewh}
\begin{align}
I_{X_1}^{\{w_i\},\{\lambda_a\}}(z;x;\hbar)&=
\sum_{q=0}^{\infty}
I_{q}^{\mathrm{c}}(z;x;\hbar)
\left(\prod_{i=0}^{n-1} I_{q}^{\Phi_i}(x,w_i;\hbar)\right)
\left(\prod_{a=1}^r I_{q}^{P_a}(x,\lambda_a;\hbar)\right)
\nonumber\\
&=
z^{x/\hbar}\sum_{q=0}^{\infty}
\frac{\prod_{a=1}^r \prod_{p=1}^{d_aq}\left(d_ax-\lambda_a+p\hbar\right)}
{\prod_{i=0}^{n-1}\prod_{p=1}^{q}(x-w_i+p\hbar)}\,z^q.
\label{I_ex1}
\end{align}
Geometrically $z=\mathrm{e}^{-2\pi \xi + \sqrt{-1}\theta}$ provides 
the K\"ahler moduli parameter of $X_1$, 
and $x$ is identified with the equivariant second cohomology element of $X_1$ 
satisfying $\prod_{i=0}^{n-1}(x-w_i)=0$, 
where the twisted masses $w_i$ give the equivariant parameters 
acting on ${\IP}^{n-1}$. 
The twisted masses $\lambda_a$ correspond to the equivariant parameters acting on $\mathcal{E}=\oplus_{a=1}^r \mathcal{O}_V(d_a)$.

Then it can be shown that the $I$-function \eqref{I_ex1} obeys the ordinary differential equation
\begin{align}
\left[
\prod_{i=0}^{n-1}\left(\hbar \Theta_z-w_i\right)
-z\prod_{a=1}^{r}\prod_{p=1}^{d_a}\left(\hbar d_a \Theta_z
-\lambda_a+p\hbar\right)\right]I_{X_1}^{\{w_i\},\{\lambda_a\}}(z;x;\hbar)=0,\quad
\Theta_z:=z\frac{d}{dz}.
\label{diff_eq_ex1}
\end{align}
In the Calabi-Yau case $\sum_{a=1}^r d_a=n$ 
with vanishing equivariant parameters $w_i=\lambda_a=0$,
the differential equation \eqref{diff_eq_ex1} yields the Picard-Fuchs equation for the periods of the holomorphic $(n-r-1)$-form on $X_1$ 
given by \cite{Hosono:1993qy,Hosono:1994ax,Greene:1993vm,Klemm:2007in}
\begin{align}
\Theta_z^r \left[\Theta_z^{n-r}
-\left(\prod_{a=1}^rd_a\right) z 
\prod_{a=1}^{r}\prod_{p=1}^{d_a-1}\left(d_a \Theta_z+p\right)\right]
I_{X_1}(z;x;\hbar)=0,
\label{pf_cicy_proj}
\end{align}
where $I_{X_1}(z;x;\hbar) :=I_{X_1}^{\{\mathbf{0}\},\{\mathbf{0}\}}(z;x;\hbar)$. 
If we expand the $I$-function around $\hbar=\infty$ as
\begin{align}
I_{X_1}(z;x;\hbar)=
\sum_{k=0}^{n-r-1} I_k(z) \left(\frac{x}{\hbar}\right)^k,
\end{align}
the coefficients $I_k(z)$ precisely give the solutions to the Picard-Fuchs 
equation. 
One can also obtain the flat coordinate $q$ on the K\"ahler moduli space of $X_1$ through the relation
\begin{align}
\log q = \frac{I_1(z)}{I_0(z)}=\log z + O(z),
\label{mirror_x1}
\end{align}
called the mirror map. 
It has been shown in \cite{PopaZinger_co}
that the genus-0 3-point A-model correlators 
$\left<\mathcal{O}_{h}\mathcal{O}_{h^k}\mathcal{O}_{h^{n-r-k-2}}\right>_{{\IP}^1}$, $k=1,\ldots,n-r-1$, which 
enumerate the number of rational curves are given by 
\begin{align}
\begin{split}
\left<\mathcal{O}_{h}\mathcal{O}_{h^k}\mathcal{O}_{h^{n-r-k-2}}\right>_{{\IP}^1}&=
\kappa\, \frac{\widehat{I}_{k+1}\left(z(q)\right)}{\widehat{I}_{1}\left(z(q)\right)}
\\
&=\kappa+\sum_{d=1}^{\infty} n_d(h^{1},h^{k},h^{n-r-k-2})\, \frac{q^d}{1-q^d},
\label{3pt_ex1}
\end{split}
\end{align}
where $\widehat{I}_k(z)$ are inductively constructed from the $I$-functions as
\begin{align}
\begin{split}
&
\widehat{I}_0(z)=I_0(z),\\
&
\widehat{I}_k(z)=\Theta_z \frac{1}{\widehat{I}_{k-1}(z)}\Theta_z \frac{1}{\widehat{I}_{k-2}(z)}
\cdots \Theta_z \frac{1}{\widehat{I}_{1}(z)}\Theta_z \frac{I_k(z)}{\widehat{I}_{0}(z)},\qquad
k=1,\ldots,n-r-1.
\end{split}
\end{align}
Here the observable $\mathcal{O}_{h^p}$ is associated with
the hyperplane class $h \in H^{1,1}(X_1)$, and
\begin{align}
\kappa=\int_{X_1}h^{n-r-1}=\left(\prod_{a=1}^{r}d_a \right)
\int_{{\IP}^{n-1}}h^{n-1}=\prod_{a=1}^{r}d_a
\label{proj_cy_int}
\end{align}
is the classical intersection number of $X_1$. $\widehat{I}_k(z)$ have relations
\begin{align}
\widehat{I}_k(z)=\widehat{I}_{n-r-k}(z),\qquad k=1,\ldots,n-r-1.
\end{align}
Note that there is a selection rule $\sum_{i=1}^np_i=\dim X_1+n-3$ 
to realize non-trivial genus-0 $n$-point correlators
$\left<\mathcal{O}_{h^{p_1}}\cdots\mathcal{O}_{h^{p_n}}\right>_{{\IP}^1}$
arising from the index theorem. 
The number $n_d(h^{1},h^{k},h^{n-r-k-2})$ in (\ref{3pt_ex1}) 
is an integer and 
enumerates the number of degree $d$ holomorphic maps 
intersecting with the cycles dual to $h$, $h^k$, and $h^{n-r-k-2}$ 
\cite{Greene:1993vm,Klemm:2007in,PandharipandeZinger}
(see also \cite{Gopakumar:1998jq,Cao:2018wmd}).

In a special case with $k=1$, 
the relation $\Theta_z=\widehat{I}_1(z)\Theta_q$ and the so-called divisor equation 
$\left<\mathcal{O}_{h}\cdots\right>_{{\IP}^1}=
\Theta_q \left<\cdots\right>_{{\IP}^1}$ imply that
\begin{align}
\left<\mathcal{O}_{h^{n-r-3}}\right>_{{\IP}^1}=
\kappa\,\frac{I_2\left(z(q)\right)}{I_{0}\left(z(q)\right)}
=\frac{\kappa}{2} \left(\log q\right)^2 + 
\sum_{d=1}^{\infty} n_d(h^{n-r-3})\,\mathrm{Li}_2(q^d),
\label{1pt_ex1}
\end{align}
where $\mathrm{Li}_p(z)=\sum_{k=1}^{\infty}\frac{z^k}{k^p}$. Here $n_d(h^{n-r-3})=n_d(h,h,h^{n-r-3})/d^2$ is an integer which enumerates 
the number of degree $d$ holomorphic maps intersecting with the cycle dual to $h^{n-r-3}$. 
When $n-r=4$ (\textit{i.e.} $\dim X_1=3$), by the divisor equation, 
\eqref{1pt_ex1} yields \cite{Candelas:1990rm,Aspinwall:1991ce}
\begin{align}
\left< * \right>_{{\IP}^1}=
\kappa\, 
\int^q \frac{I_2\left(z(q')\right)}{I_{0}\left(z(q')\right)}\,\frac{dq'}{q'}
=\frac{\kappa}{3!} \left(\log q\right)^3 + 
\sum_{d=1}^{\infty} n_d \,\mathrm{Li}_3(q^d),
\label{0pt_ex1}
\end{align}
where the number $n_d=n_d(h)/d$ is a genus-0 integer invariant.

\subsubsection{Complete intersections in $G(k,n)$}\label{subsub:cy_gr}

Let us consider a complete intersection variety $X_2$ defined by the zero locus of 
a holomorphic section of $\mathcal{E}=\oplus_{a=1}^r \mathcal{O}_V(d_a)$ on Grassmannian $V=G(k,n)$ satisfying 
Fano or Calabi-Yau condition $\sum_{a=1}^r d_a\le n$. 
Note that $\rank\mathcal{E}=r$ and $\mathfrak{c}_1(\mathcal{E})=\sum_{a=1}^r d_a$.
The variety $X_2$ has a complex dimension 
\begin{align}
\dim X_2= k(n-k)-r,
\end{align}
and can be described by a $U(k)$ GLSM whose matter content is given in Table \ref{ex_2}. This model has a superpotential 
$W=\sum_{a=1}^rP_aG_a(B)$ where $G_a(B)$ are homogeneous degree $d_a$ polynomials of the baryonic variables 
$B_{I_1\ldots I_k}=\epsilon_{i_1\ldots i_k}\Phi_{I_1}^{i_1}\cdots \Phi_{I_k}^{i_k}$ called the Pl\"ucker coordinates \cite{Hori:2006dk}. 
\begin{table}[t]
\begin{center}
\begin{tabular}{|c|c|c|c|}
\hline
Field & U(k) & twisted mass& $U(1)_R$ \\ \hline
$\Phi_I$ & $\mathbf{k}$ & $-w_I$ & $0$ \\
$P_a$ & $\det^{-d_a}$ & $\lambda_a$ & $2$ \\ \hline
\end{tabular}
\caption{Matter content of the $U(k)$ GLSM for the complete intersection variety $X_2$ in $G(k,n)$. Here $I=1, \ldots, n$ and $a=1, \ldots, r$.}
\label{ex_2}
\end{center}
\end{table}

For each matter multiplet we assign twisted masses and $U(1)$ $R$-charges 
as described in Table \ref{ex_2}. 
Combining the associated building blocks \eqref{build_classic}, \eqref{build_vec}, \eqref{build_mat0} and \eqref{build_mat2} 
for the $U(k)$ vector multiplet and the chiral multiples in Table \ref{ex_2}, 
we can construct the $I$-function for $X_2$ in 
the geometric phase with large FI parameter $\xi > 0$ 
as \cite{Inoue16}
\begin{align}
I_{X_2}^{\{w_I\},\{\lambda_a\}}(z;\mathbf{x};\hbar)
&=
\sum_{\mathbf{q}\in ({\IZ}_{\ge 0})^k}
I_{\mathbf{q}}^{\mathrm{c}}(z;\mathbf{x};\hbar)\,
I_{\mathbf{q}}^{\mathrm{vec}}(\mathbf{x};\hbar)
\left(\prod_{I=1}^{n} I_{\mathbf{q}}^{\Phi_I}(\mathbf{x},w_I;\hbar)\right)
\left(\prod_{a=1}^r I_{\mathbf{q}}^{P_a}(\mathbf{x},\lambda_a;\hbar)\right)
\nonumber\\
&=z^{\sum_{i=1}^kx_i/\hbar}
\sum_{\mathbf{q}\in ({\IZ}_{\ge 0})^k}
\left((-1)^{k-1}z\right)^{\sum_{i=1}^kq_i}
\prod_{1\le i<j\le k}\frac{{x}_i-{x}_j+(q_i-q_j)\hbar}{{x}_i-{x}_j}
\nonumber\\
&\qquad\qquad\qquad\qquad \times
\frac{\prod_{a=1}^r\prod_{p=1}^{d_a\sum_{i=1}^kq_i}\left(d_a\sum_{i=1}^k{x}_i-\lambda_a+p\hbar\right)}
{\prod_{I=1}^{n}\prod_{i=1}^k\prod_{p=1}^{q_i}({x}_i-w_I+p\hbar)}.
\label{I_ex2}
\end{align}
Geometrically $z=\mathrm{e}^{-2\pi \xi + \sqrt{-1}\theta}$ provides the K\"ahler 
moduli parameter of $X_2$, 
and $x_i$ are identified with the degree 2 elements in 
the equivariant cohomology of $X_2$. 
The twisted masses $w_I$ and $\lambda_a$ correspond to the equivariant parameters 
acting on $G(k,n)$ and $\mathcal{E}=\oplus_{a=1}^r \mathcal{O}_V(d_a)$, 
respectively.

\begin{remark}\label{rem:coh_rel_grass}
For $w_I=0$, the cohomology ring of $G(k,n)$ is given by 
\cite{Martin:1999ng} (see \cite{Mihalcea:2008} for the equivariant quantum cohomology ring):
\begin{align}
H^{*}\big(G(k,n)\big)\cong
{\IC}[x_1,\ldots,x_k]^{S_k}/
\big(h_{n-k+1}(\mathbf{x}),\ldots,h_{n}(\mathbf{x})\big),
\label{gr_coh_rel}
\end{align}
where $S_k$ is the symmetric group on $k$ elements and
$$
h_p(\mathbf{x})=
\sum_{1\le i_1 \le i_2 \le \ldots \le i_p \le k}
x_{i_1}x_{i_2}\cdots x_{i_p}
$$
are the complete symmetric polynomials.
\end{remark}

For the Calabi-Yau case $\sum_{a=1}^r d_a=n$ with
vanishing equivariant parameters $w_I=\lambda_a=0$, 
the $I$-function 
$I_{X_2}(z;\mathbf{x};\hbar)=
I_{X_2}^{\{\mathbf{0}\},\{\mathbf{0}\}}(z;\mathbf{x};\hbar)$ 
can be expanded around $\hbar=\infty$ as
\begin{align}
I_{X_2}(z;\mathbf{x};\hbar)=\sum_{|P|=0}^{\infty} I_P(z)\, \frac{s_P(\mathbf{x})}{\hbar^{|P|}},\qquad
|P|=\sum_{i=1}^kp_i,
\label{gr_cy_I_expnd}
\end{align}
where $s_P(\mathbf{x})=s_P(x_1,\ldots,x_k)$ is the Schur polynomial for 
a partition $P=\{p_1,\ldots,p_k\}$, and note that 
$s_{p,0,\ldots,0}(\mathbf{x})=h_p(\mathbf{x})$ and 
$s_{1,1,\ldots,1}(\mathbf{x})=e_p(\mathbf{x})=
\sum_{1\le i_1 < i_2 < \ldots < i_p \le k}
x_{i_1}x_{i_2}\cdots x_{i_p}$.
The flat coordinate which provides the mirror map is given by
\begin{align}
\log q =\frac{I_1(z)}{I_0(z)}=\log z + O(z).
\end{align}
As a non-abelian generalization of the formula \eqref{1pt_ex1}, 
here we conjecture that 
the genus-0 1-point A-model correlator 
\begin{align}
\left<\mathcal{O}_{H}\right>_{{\IP}^1} = \frac{\kappa_H}{2} \left(\log q\right)^2+
\sum_{d=1}^{\infty} n_d(H)\, \mathrm{Li}_2(q^d),
\label{genuszeroAcorr}
\end{align}
for the Grassmannian Calabi-Yau variety $X_2$ 
is given by
\begin{empheq}[box=\fbox]{equation}
\left<\mathcal{O}_{H}\right>_{{\IP}^1} =
\int_{X_2}H \, \left(
\frac{I_{2}\left(z(q)\right)}{I_{0}\left(z(q)\right)}\,
\sigma_{2}+
\frac{I_{1,1}\left(z(q)\right)}{I_{0}\left(z(q)\right)}\,
\sigma_{1,1}\right).
\label{grass_gw_i}
\end{empheq}
Here 
\begin{align}
\kappa_H=
\int_{X_2}H \, \left(\sigma_{2}+\sigma_{1,1}\right)=
\int_{X_2}H \, \sigma_{1}^2
\label{x2_intersection}
\end{align}
is the classical intersection number associated with the Poincar\'e dual $H$ 
of a codimension $\dim X_2-2$ cycle in $X_2$, 
where $\sigma_P$ denotes 
the Poincar\'e dual of a Schubert cycle of codimension $|P|$ in $G(k,n)$ \cite{Griffiths:1978}. 
The numbers $n_d(H)$ are integer invariants associated with $H$ 
which are related to Gromov-Witten invariants of $X_2$ 
\cite{Batyrev:1998kx,Haghighat:2008ut,Honma:2013hma,Gerhardus:2016iot}.

The Giambelli's formula and the definition of Schur polynomials yield
$$
\sigma_{1,1} = \sigma_{1}^2 - \sigma_{2},\qquad
s_{1,1}(\mathbf{x}) = \sum_{1\le i < j \le k} x_{i}x_{j}
= s_{1}(\mathbf{x})^2 - s_{2}(\mathbf{x}).
$$
Then one can reformulate the expression in \eqref{grass_gw_i} 
in terms of the classes $\sigma_{1}$ and $\sigma_{2}$ as
\begin{align}
\begin{split}
\frac{I_{2}(z)}{I_{0}(z)}\, \sigma_{2}+
\frac{I_{1,1}(z)}{I_{0}(z)}\, \sigma_{1,1}
&=
\frac{I_{1,1}(z)}{I_{0}(z)}\, \sigma_{1}^2 +
\frac{I_{2}(z)-I_{1,1}(z)}{I_{0}(z)}\, \sigma_{2}
\\
&=
\frac{-I_{X_2}[x_1^2] + I_{X_2}[x_1x_2]}{I_{X_2}[1]}\, 
\sigma_{1}^2 +
\frac{2I_{X_2}[x_1^2] - I_{X_2}[x_1x_2]}{I_{X_2}[1]}\, \sigma_{2},
\label{base_I_comp}
\end{split}
\end{align}
where $I_{X_2}[t]$ denotes the coefficient of $t$ at $\hbar=1$ 
in the expansion \eqref{gr_cy_I_expnd}. 
Then, it is obvious that the first term in \eqref{genuszeroAcorr} 
is determined from the classical block. 
One can also see that for $k=1$ the relations $I_{X_2}[x_1x_2]=0$ and $\sigma_{2}=\sigma_{1}^2$ 
imply that the formula \eqref{grass_gw_i} reduces to \eqref{1pt_ex1}.

Now we claim that the conjectural formula \eqref{grass_gw_i} is also 
applicable not only for 
complete intersection Grassmannian Calabi-Yau varieties 
but also for the determinantal Calabi-Yau varieties, 
as we will see in the next Section.

\begin{remark}
Instead of Pieri's formula for Schubert cycles, 
the intersection numbers of Grassmannian $G(k,n)$ can also be 
computed by Martin's formula \cite{Martin:1999ng}:
\begin{align}
\int_{G(k,n)}\prod_{|R|=k(n-k)}\sigma_R=
\frac{(-1)^{\frac12 k(k-1)}}{k!}
\bigg(\prod_{i=1}^{k}\oint_{x_i=0}\frac{dx_i}{2\pi \sqrt{-1}}\bigg)
\frac{\prod_{1\le i<j\le k}(x_i-x_j)^2}{\prod_{i=1}^kx_i^n}
\prod_{|R|=k(n-k)}s_R(\mathbf{x}).
\label{martin_f}
\end{align}
Utilizing this formula, one can then compute the intersection numbers of complete intersection 
Grassmannian Calabi-Yau varieties by 
considering the top Chern class of 
their normal bundles in $G(k,n)$. 
For example, \eqref{x2_intersection} for $X_2$ is computed as
$$
\kappa_H=
\int_{X_2}H \, \sigma_1^2=
\int_{G(k,n)}H \, \sigma_1^2 \, 
\prod_{a=1}^r d_a \sigma_1.
$$
Generic case can also be treated with a slight modification. 
Suppose that a Grassmannian Calabi-Yau variety $X$, 
defined by the zero locus of a holomorphic section of a vector bundle on $G(k,n)$, 
has a GLSM realization with a massless matter multiplet $P$ in 
a representation $\overline{\mathbf{R}}$ of $U(k)$ with 
$R$-charge 2. 
Then one obtains
\begin{align}
\int_{X}\prod_{|R|=\dim X}\sigma_R=
\frac{(-1)^{\frac12 k(k-1)}}{k!}
\bigg(\prod_{i=1}^{k}\oint_{x_i=0}\frac{dx_i}{2\pi \sqrt{-1}}\bigg)
\frac{\prod_{1\le i<j\le k}(x_i-x_j)^2}{\prod_{i=1}^kx_i^n}\,
I^{P}(\mathbf{x})
\prod_{|R|=\dim X}s_R(\mathbf{x}),
\label{martin_norm_bdl}
\end{align}
where
$$
I^{P}(\mathbf{x})=
\prod_{\rho \in \mathbf{R}}\, \rho(\mathbf{x}).
$$
\end{remark}

\vspace{1em}

Let us consider the dual of the universal subbundle $\mathcal{E}=\mathcal{S}^*$ on $G(k,n)$.
A Grassmannian Calabi-Yau variety defined by the zero locus of a holomorphic section of $\mathcal{E}=\mathcal{S}^*$ is 
described by a $U(k)$ GLSM with a superpotential $W=P_iG(\Phi)^i$. Here $G(\Phi)$ is a homogeneous degree 1 
polynomial of $\Phi_I \ (I=1,\ldots,n)$ in $\mathbf{k}$ 
of the $U(k)$ gauge group with $R$-charge 0, 
and $P$ in $\overline{\mathbf{k}}$ 
with $R$-charge 2. For the matter multiplet $P$ with twisted mass $\lambda$, 
\eqref{build_mat2} becomes
\begin{align}
I_{\mathbf{q}}^{P}(\mathbf{x},\lambda;\hbar)=
\prod_{i=1}^k\prod_{p=1}^{q_i}\left(x_i-\lambda+p\hbar\right),\ \
\textrm{for}\ P\ \textrm{in}\ \overline{\mathbf{k}}\ \textrm{with}\ \textrm{$R$-charge 2}.
\end{align}
Similarly, for instance, for vector bundles 
$\mathcal{E}=\Sym^m\mathcal{S}^*(d)$ and $\mathcal{E}=\wedge^m\mathcal{S}^*(d)$ we get
\begin{align}
\begin{split}
I_{\mathbf{q}}^{P}(\mathbf{x},\lambda;\hbar)&=
\prod_{1\le i_1\le\cdots\le i_m\le k}
\prod_{p=1}^{\sum_{j=1}^mq_{i_j}+d \sum_{i=1}^kq_i}
\left(\sum_{j=1}^mx_{i_j}+d \sum_{i=1}^kx_i-\lambda+p\hbar\right),\\
&\qquad\qquad\qquad\qquad
\label{sym_p_gra}
\end{split}
\end{align}
with $P\ \textrm{in}\ \Sym^m \overline{\mathbf{k}}\otimes {\det}^{-d}$ and
\begin{align}
\begin{split}
I_{\mathbf{q}}^{P}(\mathbf{x},\lambda;\hbar)&=
\prod_{1\le i_1<\cdots< i_m\le k}
\prod_{p=1}^{\sum_{j=1}^mq_{i_j}+d \sum_{i=1}^kq_i}
\left(\sum_{j=1}^mx_{i_j}+d \sum_{i=1}^kx_i-\lambda+p\hbar\right),\\
&\qquad\qquad\qquad\qquad
\label{a_sym_p_gra}
\end{split}
\end{align}
with $P\ \textrm{in}\ \wedge^m \overline{\mathbf{k}}\otimes {\det}^{-d}$, respectively. 
Using these building blocks with the help of
our formula \eqref{grass_gw_i}, 
one can obtain the genus-0 Gromov-Witten invariants of 
Grassmannian Calabi-Yau varieties 
computed \textit{e.g.} in \cite{Ueda:2016wfa}. 

To consider a Grassmannian Calabi-Yau variety associated with 
the universal quotient bundle $\mathcal{E}=\mathcal{Q}$ on $G(k,n)$, 
a little ingenuity is needed. 
By tensoring $\mathcal{O}_{G(k,n)}(d)$ with the short exact sequence 
\begin{align}
0 \longrightarrow \mathcal{S} \longrightarrow
\mathcal{O}_{G(k,n)}^{\oplus n} \longrightarrow 
\mathcal{Q} \longrightarrow 0,
\label{ex_univ_grass}
\end{align}
as
\begin{align}
0 \longrightarrow \mathcal{S}(d) \overset{f}{\longrightarrow}
\mathcal{O}_{G(k,n)}(d)^{\oplus n} \overset{g}{\longrightarrow} 
\mathcal{Q}(d) \longrightarrow 0,
\label{ex_univ_grass_t}
\end{align}
one can realize a corresponding GLSM for the vector bundle 
$\mathcal{E}=\mathcal{Q}(d)$ 
from the viewpoint of a quotient $\mathcal{O}_{G(k,n)}(d)^{\oplus n}/\mathrm{im}(f)$. 
The resulting model consists of 
$n$ matter multiplets $P_i$ in ${\det}^{-d}$, $i=1,\ldots,n$ of $U(k)$ 
with $R$-charge 2, 
and a single matter multiplet $Y$ in $\overline{\mathbf{k}}\otimes {\det}^{d}$ 
of $U(k)$ with $R$-charge 0 \cite{Jia:2014ffa}. The associated building block of the $I$-function without twisted mass is then given by
\begin{align}
I_{\mathbf{q}}^{\{P_i\}/Y}(\mathbf{x};\hbar)=
\frac{\prod_{p=1}^{d \sum_{i=1}^kq_i}
\left(d \sum_{i=1}^k x_i + p\hbar\right)^n}
{\prod_{i=1}^k\prod_{p=1}^{-q_{i} + d \sum_{j=1}^kq_j}
\left(-x_{i} + d \sum_{j=1}^kx_j + p\hbar\right)}.
\end{align}

\section{$I$-functions of determinantal Calabi-Yau varieties}\label{sec:det_I_fn}

\begin{table}[t]
\begin{center}
\begin{tabular}{|c|c|c|c|}
\hline
Field & $U(k)$ & $U(\ell_p^{\vee})$ & $U(1)_R$ \\ \hline
$\Phi_a$ & $\mathbf{k}$ & $\mathbf{1}$ & $0$ \\
$X_i$ & $\mathbf{1}$ & $\overline{\boldsymbol\ell}^{\vee}_p$ & $0$ \\
$P$ & $\mathbf{\overline{R}}_p$ & ${\boldsymbol\ell}^{\vee}_p$ & $2$ \\
\hline
\end{tabular}
\quad$\overset{\textrm{dual}}{\longleftrightarrow}$\quad
\begin{tabular}{|c|c|c|c|}
\hline
Field & $U(k)$ & $U(\ell)$ & $U(1)_R$ \\ \hline
$\Phi_a$ & $\mathbf{k}$ & $\mathbf{1}$ & $0$ \\
$\widetilde{X}_i$ & $\mathbf{1}$ & ${\boldsymbol\ell}$ & $0$ \\
$\widetilde{Y}$ & $\mathbf{R}_p$ & $\overline{\boldsymbol\ell}$ & $0$ \\
$\widetilde{P}_i$ & $\mathbf{\overline{R}}_p$ & $\mathbf{1}$ & $2$ \\
\hline
\end{tabular}
\caption{The left table describes matter content of the PAX model for the desingularized determinantal Calabi-Yau variety in $G(k,n)$, 
where $a=1, \ldots, n$, $i=1, \ldots, p$, and $\mathbf{R}_p$ is 
a representation in the gauge group $U(k)$ which describes the vector bundle $\mathcal{F}_p$. 
$U(1)_R$ denotes an $R$-charge which is assigned to the matter content. 
The right table describes matter content of the PAXY model which is a dual GLSM of the PAX model.}
\label{PAX_det}
\end{center}
\end{table}

In this section, 
we describe how to utilize our formula \eqref{grass_gw_i} to compute genus-0 Gromov-Witten invariants of the determinantal Calabi-Yau varieties.  
Here we focus on the desingularized determinantal 
Calabi-Yau variety $X_A$ in \eqref{det_pax} 
which can be described by 
a $U(k) \times U(\ell_p^{\vee})$ 
GLSM with matter content in the left of Table \ref{PAX_det}.
This GLSM is called a PAX model and has a superpotential \cite{Jockers:2012zr}
\begin{align}
W_{PAX}=\sum_{i,j=1}^{p}\sum_{\alpha=1}^{\ell_p^{\vee}}
P_{\alpha i}A(\Phi)_{ij}X_{j\alpha}.
\label{pax_spot}
\end{align}

The PAX model has several distinct phases. 
Let $\xi_1$ and $\xi_2$ be the FI parameters associated with the central $U(1)$ factors of $U(k)$ and 
$U(\ell_p^{\vee})$, respectively. 
For example, 
a geometric phase called a ``$X_A$ phase'' with $\xi_1 > 0$ and $\xi_2 < 0$ of the PAX model in the IR describes 
the variety $X_A$ in \eqref{det_pax}, and 
another geometric phase ``$X_{A^{T}}$ phase'' with 
$k \xi_1 + \ell_p^{\vee} \xi_2> 0$ and $\xi_2 > 0$ 
corresponds to an incidence correspondence constructed from the transposed matrix $A(\phi)^T$.

\begin{remark}
Alternatively, one can consider Seiberg-like dual \cite{Hori:2006dk} 
of the PAX model as follows. 
The chiral matter multiplet $P$ in the fundamental representation ${\boldsymbol\ell}^{\vee}_p$ under the  $U(\ell^{\vee}_p)$ 
factor corresponds to the vector bundle $\mathcal{S}^*$ 
on $G(\ell_p^{\vee},p)$. 
By taking the Seiberg-like duality with respect to the gauge group $U(\ell_p^{\vee})$, 
$\mathcal{S}^*$ is mapped to $\mathcal{Q}$ on $G(\ell,p)$ and  
as indicated by the short exact sequence \eqref{ex_univ_grass}, 
the chiral matter multiplet $P$ is mapped to 
the dual chiral matter multiplets $\widetilde{Y}$ and $\widetilde{P}_i$ 
in the right of Table \ref{PAX_det}. 
This dualized GLSM is called a PAXY model with gauge group $U(k) \times U(\ell)$ 
and has a superpotential given by \cite{Jockers:2012zr}
\begin{align}
W_{PAXY}=\sum_{i,j=1}^{p}\widetilde{P}_{ij}
\Big(A(\Phi)_{ij}
-\sum_{\beta=1}^{\ell}\widetilde{Y}_{i \beta}\widetilde{X}_{\beta j}\Big).
\end{align}
\end{remark}

\subsection{$I$-functions and A-model correlators}\label{subsec:I_fn_det}

Let us consider the PAX model with massless matter multiplets shown in 
Table \ref{PAX_det}. 
From the building blocks 
\eqref{build_classic}, \eqref{build_vec}, \eqref{build_mat0} and \eqref{build_mat2}, 
the $I$-function of this model in the $X_A$ phase with FI parameters $\xi_1 > 0$ and 
$\xi_2 < 0$ is given by
\begin{align}
I_{X_A}(z,w;\mathbf{x},\mathbf{y};\hbar)=
\sum_{(\mathbf{q},\mathbf{r})\in ({\IZ}_{\ge 0})^k \times ({\IZ}_{\ge 0})^{\ell_p^{\vee}}}&
I_{\mathbf{q}}^{\mathrm{c}}(z,w;\mathbf{x},\mathbf{y};\hbar)\,
I_{\mathbf{q},\mathbf{r}}^{\mathrm{vec}}(\mathbf{x},\mathbf{y};\hbar)
\nonumber\\
&\times
\left(I_{\mathbf{q}}^{\Phi}(\mathbf{x};\hbar)\right)^n
\left(I_{\mathbf{r}}^{X}(\mathbf{y};\hbar)\right)^p
I_{\mathbf{q},\mathbf{r}}^{P}(\mathbf{x},\mathbf{y};\hbar),
\label{i_fun_det}
\end{align}
where
\begin{align}
\begin{split}
I_{\mathbf{q}}^{\mathrm{c}}(z,w;\mathbf{x},\mathbf{y};\hbar)&=
z^{\sum_{i=1}^k x_i/\hbar}
\left((-1)^{k-1}z\right)^{\sum_{i=1}^kq_i}
w^{\sum_{i=1}^{\ell_p^{\vee}} y_i/\hbar}
\left((-1)^{\ell_p^{\vee}-1}w\right)^{\sum_{i=1}^{\ell_p^{\vee}}r_i},
\\
I_{\mathbf{q},\mathbf{r}}^{\mathrm{vec}}(\mathbf{x},\mathbf{y};\hbar)&=
\bigg(\prod_{1\le i<j\le k}\frac{{x}_i-{x}_j+(q_i-q_j)\hbar}{{x}_i-{x}_j}\bigg)
\bigg(\prod_{1\le i<j\le \ell_p^{\vee}}\frac{{y}_i-{y}_j+(r_i-r_j)\hbar}{{y}_i-{y}_j}\bigg),
\\
I_{\mathbf{q}}^{\Phi}(\mathbf{x};\hbar)&=
\frac{1}
{\prod_{i=1}^k\prod_{d=1}^{q_i}({x}_i+d\hbar)},\qquad
I_{\mathbf{r}}^{X}(\mathbf{y};\hbar)=
\frac{1}
{\prod_{i=1}^{\ell_p^{\vee}}\prod_{d=1}^{r_i}({y}_i+d\hbar)},
\\
I_{\mathbf{q},\mathbf{r}}^{P}(\mathbf{x},\mathbf{y};\hbar)&=
\prod_{\rho \in \mathbf{R}_p}
\prod_{i=1}^{\ell_p^{\vee}}\prod_{d=1}^{\rho(\mathbf{q})+r_i}
\left(\rho(\mathbf{x})+y_i+d\hbar\right).
\end{split}
\end{align} 
Here $z=\mathrm{e}^{-2\pi \xi_1 + \sqrt{-1}\theta_1}$ and 
$w=\mathrm{e}^{2\pi \xi_2 - \sqrt{-1}\theta_2}$ are 
moduli parameters associated with the central 
$U(1)^2 \subset U(k) \times U(\ell_p^{\vee})$, 
and in particular $w$ parametrizes the blowing up in \eqref{det_pax_o}. 
$\mathbf{x}$ (\textit{resp.} $\mathbf{y}$) are identified with the degree 2 elements in the cohomology of 
$G(k,n)$ (\textit{resp.} $G(\ell_p^{\vee},p)$).

As performed in \eqref{gr_cy_I_expnd}, the $I$-function \eqref{i_fun_det} can be expanded around $\hbar=\infty$ in terms of Schur polynomials as
\begin{align}
I_{X_A}(z,w;\mathbf{x},\mathbf{y};\hbar)=
\sum_{|Q|,|R|=0}^{\infty} I_{Q;R}(z,w)\, 
\frac{s_Q(\mathbf{x})\, s_R(\mathbf{y})}{\hbar^{|Q|+|R|}}.
\label{det_cy_I_expnd}
\end{align}
The flat coordinates $q_z$ and $q_w$, 
which provide the exponentiated K\"ahler moduli parameters of $X_A$, 
are given by
\begin{align}
\log q_z = \frac{I_{1;0}(z,w)}{I_{0;0}(z,w)}=\log z + O(z,w),\qquad
\log q_w = \frac{I_{0;1}(z,w)}{I_{0;0}(z,w)}=\log w + O(z,w).
\end{align}
From our conjectural formula \eqref{grass_gw_i}, one can deduce a formula for 
the genus-0 1-point A-model correlator $\left<\mathcal{O}_{H}\right>_{{\IP}^1}$ 
in the $X_A$ phase as
\begin{align}
\begin{split}
\left<\mathcal{O}_{H}\right>_{{\IP}^1}&=
\int_{X_A} H \, \left(
\frac{I_{2;0}\left(z, w\right)}
{I_{0;0}\left(z, w\right)}\,\sigma_{2}+
\frac{I_{1,1;0}\left(z, w\right)}
{I_{0;0}\left(z, w\right)}\,\sigma_{1,1}
\right.
\\
&\qquad\qquad\quad\left.
+\frac{I_{1;1}\left(z, w\right)}
{I_{0;0}\left(z, w\right)}\,\sigma_{1} \, \tau_{1}
+\frac{I_{0;2}\left(z, w\right)}
{I_{0;0}\left(z, w\right)}\,\tau_{2}+
\frac{I_{0;1,1}\left(z, w\right)}
{I_{0;0}\left(z, w\right)}\,\tau_{1,1}
\right)
\\
&=\frac{\kappa_H^{\sigma}}{2} \left(\log q_z\right)^2 +
\kappa_H^{\sigma\tau}\, \log q_z\, \log q_w +
\frac{\kappa_H^{\tau}}{2} \left(\log q_w\right)^2 +
\sum_{d_1,d_2=1}^{\infty} n_{d_1,d_2}(H)\, \mathrm{Li}_2(q_z^{d_1}q_w^{d_2}).
\label{det_gr_gw_i}
\end{split}
\end{align}
Here
\begin{align}
\kappa_H^{\sigma}=
\int_{X_A} H \, \sigma_{1}^2,\quad\
\kappa_H^{\sigma\tau}=
\int_{X_A} H \, \sigma_{1} \, \tau_{1},\quad\
\kappa_H^{\tau}=
\int_{X_A} H \, \tau_{1}^2,
\label{det_classic_int}
\end{align}
are the classical intersection numbers associated with the Poincar\'e dual $H$ of a codimension 
$\dim X_A-2$ cycle in $X_A$, 
where $\sigma_P$ (\textit{resp.} $\tau_P$) is the Poincar\'e dual of 
a Schubert cycle of codimension $|P|$ in $G(k,n)$ 
(\textit{resp.} $G(\ell_p^{\vee},p)$). 
The genus-0 invariants $n_{d_1,d_2}(H)$ associated with $H$, 
which are related to Gromov-Witten invariants, are conjecturally integers.

\begin{remark}
When $\dim X_A=3$, 
analogous to the formula \eqref{0pt_ex1}, the above result \eqref{det_gr_gw_i} yields
\begin{align}
\begin{split}
\left< * \right>_{{\IP}^1}&=
\frac{\kappa_{\sigma_1}^{\sigma}}{3!} \left(\log q_z\right)^3 +
\frac{\kappa_{\sigma_1}^{\sigma\tau}}{2} \left(\log q_z\right)^2 \log q_w +
\frac{\kappa_{\tau_1}^{\sigma\tau}}{2}\, \log q_z \left(\log q_w\right)^2 +
\frac{\kappa_{\tau_1}^{\tau}}{3!} \left(\log q_w\right)^3 
\\
&\ \ \ +
\sum_{d_1,d_2=1}^{\infty} n_{d_1,d_2}\, \mathrm{Li}_3(q_z^{d_1}q_w^{d_2}),
\label{det_gr_gw_0}
\end{split}
\end{align} 
where $n_{d_1,d_2}=n_{d_1,d_2}(\sigma_1)/d_1$ (for $d_1\ne 0$) and 
$n_{d_1,d_2}=n_{d_1,d_2}(\tau_1)/d_2$ (for $d_2\ne 0$) each provide genus-0 integer invariants. 
\end{remark}

\subsection{An algorithm to compute genus-0 invariants}\label{subsec:I_fn_det_ex}

In a similar fashion to the computation \eqref{base_I_comp}, 
by taking the classes $\sigma_{1}$, $\sigma_{2}$, $\tau_{1}$ and $\tau_{2}$ 
for the special Schubert cycles, the 1-point correlator \eqref{det_gr_gw_i} can be evaluated with
\begin{align}
\begin{split}
&
\frac{I_{2;0}(z, w)}
{I_{0;0}(z, w)}\,\sigma_{2} +
\frac{I_{1,1;0}(z, w)}{I_{0;0}(z, w)}\,\sigma_{1,1} +
\frac{I_{1;1}(z, w)}{I_{0;0}(z, w)}\,\sigma_{1} \, \tau_{1} +
\frac{I_{0;2}(z, w)}{I_{0;0}(z, w)}\,\tau_{2} +
\frac{I_{0;1,1}(z, w)}{I_{0;0}(z, w)}\,\tau_{1,1}
\\
&=
\frac{-I_{X_A}[x_1^2] + I_{X_A}[x_1x_2]}{I_{X_A}[1]}\, 
\sigma_{1}^2 +
\frac{2I_{X_A}[x_1^2] - I_{X_A}[x_1x_2]}{I_{X_2}[1]}\, \sigma_{2} +
\frac{I_{X_A}[x_1y_1]}{I_{X_A}[1]}\, \sigma_{1} \, \tau_{1} 
\\
&\ \ +
\frac{-I_{X_A}[y_1^2] + I_{X_A}[y_1y_2]}{I_{X_A}[1]}\, 
\tau_{1}^2 +
\frac{2I_{X_A}[y_1^2] - I_{X_A}[y_1y_2]}{I_{X_2}[1]}\, \tau_{2},
\end{split}
\end{align}
where $I_{X_A}[t]$ denotes the coefficient of $t$ at $\hbar=1$ 
in the expansion \eqref{det_cy_I_expnd}. From 
the coefficients $I_{X_A}[t]$ and the classical intersection numbers
\begin{align}
\int_{X_A} H \, \sigma_{1}^2,\quad
\int_{X_A} H \, \sigma_{2},\quad
\int_{X_A} H \, \sigma_{1} \, \tau_{1},\quad
\int_{X_A} H \, \tau_{1}^2,\quad
\int_{X_A} H \, \tau_{2},
\label{classic_int_det}
\end{align}
one can compute the 1-point A-model correlator \eqref{det_gr_gw_i} 
and obtain the integer invariants. 
The classical intersection numbers can be computed by 
Martin's formula \eqref{martin_norm_bdl} as
\begin{align}
\int_{X_A}\prod_{|Q|+|R|=\dim X_A}\sigma_Q\, \tau_R &=
\frac{(-1)^{\frac12 k(k-1)+ \frac12 \ell_p^{\vee}(\ell_p^{\vee}-1)}}{k!\, \ell_p^{\vee}!}
\bigg(\prod_{i=1}^{k}\oint_{x_i=0}\frac{dx_i}{2\pi \sqrt{-1}}\bigg)
\bigg(\prod_{i=1}^{\ell_p^{\vee}}\oint_{y_i=0}\frac{dy_i}{2\pi \sqrt{-1}}\bigg)
\nonumber
\\
& \ \ \ \ \times
\frac{\prod_{1\le i<j\le k}(x_i-x_j)^2}{\prod_{i=1}^kx_i^n}\,
\frac{\prod_{1\le i<j\le \ell_p^{\vee}}(y_i-y_j)^2}{\prod_{i=1}^{\ell_p^{\vee}}y_i^p}\,
I^{P}(\mathbf{x}, \mathbf{y})
\nonumber
\\
& \ \ \ \ \times
\prod_{|Q|+|R|=\dim X_A}s_Q(\mathbf{x})\, s_R(\mathbf{y}),
\label{martin_det_cy}
\end{align}
where
$$
I^{P}(\mathbf{x}, \mathbf{y})=
\prod_{\rho \in \mathbf{R}_p}\prod_{i=1}^{\ell_p^{\vee}}
\left(\rho(\mathbf{x})+y_i\right).
$$

\subsection{Illustrative examples of the computations}

Here we will consider several examples of the desingularized determinantal Calabi-Yau 3-folds investigated in Section \ref{subsec:cy3} 
and compute their genus-0 invariants $n_{d_1,d_2}$ 
defined in \eqref{det_gr_gw_0}.\footnote{We only focus on the determinantal varieties described by 
$U(k) \times U(\ell_p^{\vee})$ PAX models with $k\leq 2$, $\ell_p^{\vee}\leq 2$. 
In Appendix \ref{app:gw_det} 
we summarize our computational results for several determinantal Calabi-Yau 4-folds.}

\subsubsection{Quintic family}\label{subsubapp:quintic}

The determinantal Calabi-Yau 3-folds in \eqref{class_cy3_1} are connected with 
the famous quintic Calabi-Yau 3-fold with $(h^{1,1}, h^{2,1})=(1, 101)$ which can be described 
as a ``trivial'' determinantal 
3-fold with $\mathcal{F}_p=\mathcal{O}_V(5)$ on $V={\IP}^4$. 
In terms of the parameters in Section \ref{subsub:cy_cp}, the quintic 3-fold is characterized as $X_1$ with $n=5$, $r=1$ and $d_1=5$. 
The classical intersection number \eqref{proj_cy_int} of 
$X_1$ is given by $\kappa=5$ 
and the genus-0 invariants $n_{d}$ in \eqref{0pt_ex1} are well-known to be \cite{Candelas:1990rm}
\begin{align}
n_{1}=2875,\ n_{2}=609250,\ n_{3}=317206375,\ 
n_{4}=242467530000,\ n_{5}=229305888887625,\ldots.
\label{quintic_inv}
\end{align}

The quintic family can be described by GLSMs with $U(1)\times U(1)$ gauge group. 
Following Section \ref{subsec:I_fn_det_ex} and Appendix \ref{app:hodge}, 
one can compute topological 
invariants of the quintic family as 
summarized in Table \ref{det_cy3_p4}, which is consistent with the previous works.
Here one can also check that $h^{1,0}=0$. 
By comparing \eqref{quintic_inv} with the entries 
$n_{d_1,d_2}$ in Table \ref{det_cy3_p4} of each 
determinantal 3-fold, we see that they exhibit a behavior of 
the extremal transition \cite{Li:1998hba} (see also \cite{Jockers:2012dk}):
\begin{align}
n_{d}=\sum_{d_2=0}^{N} n_{d,d_2},
\label{ext_trans}
\end{align}
where $N$ is a certain finite positive integer.

\begin{footnotesize}
\begin{longtable}[c]{|c|rrrrrrr|}
\caption{Genus-0 invariants of determinantal 
3-folds in \eqref{class_cy3_1} with $V={\IP}^4$}
\label{det_cy3_p4}
\endfirsthead
\hline
\multicolumn{8}{|c|}{$\mathcal{F}_p=\mathcal{O}_V(1)\oplus \mathcal{O}_V(4)$: $(h^{1,1}, h^{2,1})=(2, 86)$}
\\
\hline
\multicolumn{3}{|c|}{Intersection numbers}&
\multicolumn{5}{|c|}{$\sigma_1^3=5,\quad \sigma_1^2\tau_1=4,\quad \sigma_1\tau_1^2=0,\quad \tau_1^3=0$}
\\
\hline
$n_{d_1,d_2}$ & $d_1=0$ & 1 & 2 & 3 & 4 & 5 & 6 \\ \hline
$d_2=0$ &   & 640 & 10032 & 288384   & 10979984    & 495269504      
& 24945542832\\
     1  & 16& 2144& 231888& 23953120 & 2388434784  & 232460466048   
& 22229609118768\\
     2  & 0 & 120 & 356368& 144785584& 36512550816 & 7251261673320  
& 1242876017216016\\
     3  & 0 & -32 & 14608 & 144051072& 115675981232& 50833652046112 
& 16156774167471792\\
     4  & 0 &  3  & -4920 & 5273880  & 85456640608 & 106397389165188
& 69178537204963920\\
     5  & 0 &  0  & 1680  & -1505472 & 3018009984  & 62800738246496 
& 107220234702633360\\
     6  & 0 &  0  & -480  & 512136   & -748922304  & 2196615443648  
& 52910679981204144\\
\hhline{========}
\multicolumn{8}{|c|}{$\mathcal{F}_p=\mathcal{O}_V(2)\oplus \mathcal{O}_V(3)$: $(h^{1,1}, h^{2,1})=(2, 66)$}
\\
\hline
\multicolumn{3}{|c|}{Intersection numbers}&
\multicolumn{5}{|c|}{$\sigma_1^3=5,\quad \sigma_1^2\tau_1=6,\quad \sigma_1\tau_1^2=0,\quad \tau_1^3=0$}
\\
\hline
$n_{d_1,d_2}$ & $d_1=0$ & 1 & 2 & 3 & 4 & 5 & 6 \\ \hline
$d_2=0$ &   & 366 & 2670  & 35500    & 606264     & 12210702      
& 273649804 \\
     1  & 36& 1584& 73728 & 3286224  & 142523712  & 6060689280    
& 253954899504\\
     2  & 0 & 909 & 255960& 34736049 & 3387935304 & 273906849222  
& 19594379113848\\
     3  & 0 & 16  & 231336& 106245024& 23702767680& 3623779411776 
& 436922554063224\\
     4  & 0 &  0  & 45216 & 119474748& 66922830504& 19938817169442
& 4093759996324344\\
     5  & 0 &  0  & 360   & 48046176 & 85607985132& 53346064121712
& 19206910967576760\\
     6  & 0 &  0  & -20   & 5357838  & 49765200024& 74247746393898
& 49456242071288532\\
\hhline{========}
\multicolumn{8}{|c|}{$\mathcal{F}_p=\mathcal{O}_V(1)\oplus \mathcal{O}_V(2)^{\oplus 2}$: $(h^{1,1}, h^{2,1})=(2, 58)$}
\\
\hline
\multicolumn{3}{|c|}{Intersection numbers}&
\multicolumn{5}{|c|}{$\sigma_1^3=5,\quad \sigma_1^2\tau_1=8,\quad \sigma_1\tau_1^2=4,\quad \tau_1^3=0$}
\\
\hline
$n_{d_1,d_2}$ & $d_1=0$ & 1 & 2 & 3 & 4 & 5 & 6 \\ \hline
$d_2=0$ &   & 144 & 140   & 144      & 112        & 144      
& 140 \\
     1  & 44& 1120& 13520 & 107264   & 645048     & 3190528    
& 13669600\\
     2  & 0 & 1354& 113916& 3627224  & 68006448   & 901242596  
& 9287483360\\
     3  & 0 & 256 & 258840& 29390080 & 1463601384 & 44141205824 
& 937689927488\\
     4  & 0 &  1  & 183690& 89360780 & 11490671144& 741564140238
& 30303625673624\\
     5  & 0 &  0  & 37896 & 115185728& 41359928372& 5682155162688
& 434288936956304\\
     6  & 0 &  0  & 1248  & 64102328 & 74832601592& 22827028536708
& 3267218929443668\\
\hhline{========}
\multicolumn{8}{|c|}{$\mathcal{F}_p=\mathcal{O}_V(1)^{\oplus 2}\oplus \mathcal{O}_V(3)$: $(h^{1,1}, h^{2,1})=(2, 68)$}
\\
\hline
\multicolumn{3}{|c|}{Intersection numbers}&
\multicolumn{5}{|c|}{$\sigma_1^3=5,\quad \sigma_1^2\tau_1=7,\quad \sigma_1\tau_1^2=3,\quad \tau_1^3=0$}
\\
\hline
$n_{d_1,d_2}$ & $d_1=0$ & 1 & 2 & 3 & 4 & 5 & 6 \\ \hline
$d_2=0$ &   & 204 & 204   & 132      & 204        & 204      
& 132 \\
     1  & 34& 1348& 26843 & 338016   & 3050972    & 21359344    
& 123786248\\
     2  & 0 & 1290& 179490& 9621696  & 299056816  & 6401442680  
& 103385827082\\
     3  & 0 & 35  & 292557& 59496360 & 5101530190 & 260050051116 
& 9166825459347\\
     4  & 0 &  -2 & 108312& 127400436& 29874798664& 3367972159714
& 235659178171360\\
     5  & 0 &  0  & 1909  & 97863426 & 75032773743& 18958650256980
& 2557795024380895\\
     6  & 0 &  0  & -68   & 22115268 & 84738954674& 52879556793440
& 13984934136290076\\
\hhline{========}
\multicolumn{8}{|c|}{$\mathcal{F}_p=\mathcal{O}_V(1)^{\oplus 3}\oplus \mathcal{O}_V(2)$: $(h^{1,1}, h^{2,1})=(2, 56)$}
\\
\hline
\multicolumn{3}{|c|}{Intersection numbers}&
\multicolumn{5}{|c|}{$\sigma_1^3=5,\quad \sigma_1^2\tau_1=9,\quad \sigma_1\tau_1^2=7,\quad \tau_1^3=2$}
\\
\hline
$n_{d_1,d_2}$ & $d_1=0$ & 1 & 2 & 3 & 4 & 5 & 6 \\ \hline
$d_2=0$ &   & 84  & 10    & 0        & 0          & 0      
& 0 \\
     1  & 46& 865 & 4461  & 9380     & 9380       & 4461    
& 865\\
     2  & 0 & 1478& 60360 & 760580   & 4423324    & 14207450  
& 27724124\\
     3  & 0 & 438 & 211547& 10517154 & 200833886  & 1987023580 
& 11758507011\\
     4  & 0 &  10 & 238798& 51571964 & 2762153102 & 67275586298
& 926085646998\\
     5  & 0 &  0  & 86203 & 107216585& 16493768487& 916157767777
& 26171128616181\\
     6  & 0 &  0  & 7826  & 99623760 & 48905658096& 6224190580040
& 353098716104028\\
\hhline{========}
\multicolumn{8}{|c|}{$\mathcal{F}_p=\mathcal{O}_V(1)^{\oplus 5}$: $(h^{1,1}, h^{2,1})=(2, 52)$ \cite{Schoen,gross2001calabi,Hosono:2011np}}
\\
\hline
\multicolumn{3}{|c|}{Intersection numbers}&
\multicolumn{5}{|c|}{$\sigma_1^3=5,\quad \sigma_1^2\tau_1=10,\quad \sigma_1\tau_1^2=10,\quad \tau_1^3=5$}
\\
\hline
$n_{d_1,d_2}$ & $d_1=0$ & 1 & 2 & 3 & 4 & 5 & 6 \\ \hline
$d_2=0$ &   & 50  & 0     & 0        & 0          & 0      
& 0 \\
     1  & 50& 650 & 1475  & 650      & 50         & 0   
& 0\\
     2  & 0 & 1475& 29350 & 148525   & 250550     & 148525 
& 29350\\
     3  & 0 & 650 & 148525& 3270050  & 24162125   & 75885200 
& 110273275\\
     4  & 0 &  50 & 250550& 24162125 & 545403950  & 5048036025
& 22945154050\\
     5  & 0 &  0  & 148525& 75885200 & 5048036025 & 114678709000
& 1231494256550\\
     6  & 0 &  0  & 29350 & 110273275& 22945154050& 1231494256550
& 27995704239850\\
\hline
\end{longtable}
\end{footnotesize}

\subsubsection{Determinantal Calabi-Yau 3-folds in \eqref{class_cy3_2}}

Next, let us consider the determinantal Calabi-Yau 3-folds 
described in \eqref{class_cy3_2} with $p\ne 2$. These examples can be described by GLSMs with $U(1)\times U(2)$
gauge group. Using the methodology we established in Section \ref{subsec:I_fn_det_ex},
one can obtain the genus-0 invariants summarized in Table \ref{det_cy3_p7}. 
Here one can also check that $h^{1,0}=0$.

\begin{footnotesize}
\begin{longtable}[c]{|c|rrrrrrrr|}
\caption{Genus-0 invariants of determinantal 3-folds in \eqref{class_cy3_2} with $V={\IP}^7$}
\label{det_cy3_p7}
\endfirsthead
\hline
\multicolumn{9}{|c|}{$\mathcal{F}_p=\mathcal{O}_V(1)^{\oplus 2}\oplus \mathcal{O}_V(2)$: $(h^{1,1}, h^{2,1})=(2, 58)$ 
(Gulliksen-Neg{\aa}rd Calabi-Yau 3-fold \cite{Bertin0701,Jockers:2012zr})}
\\
\hline
\multicolumn{3}{|c|}{Intersection numbers}&
\multicolumn{6}{|c|}{$\sigma_1^3=17,\quad \sigma_1^2\tau_1=10,\quad \sigma_1\tau_1^2=4,\quad \sigma_1\tau_2=0,\quad \tau_1^3=0,\quad \tau_1\tau_2=0$}
\\
\hline
$n_{d_1,d_2}$ & $d_1=0$ & 1 & 2 & 3 & 4 & 5 & 6 & 7 \\ \hline
$d_2=0$ &   & 156& 116 & 156   & 112    & 156       & 116 
& 156 \\
     1  & 0 & 256& 6656& 63232 & 415232 & 2159360   & 9583104
& 37772288 \\
     2  & 0 & 1  & 1248& 193678& 8278144& 172114785 & 2326878112 
& 23641531470 \\
     3  & 0 & 0  & 0   & 10496 & 5211136& 592671744 & 28906081792
& 822717728768 \\
     4  & 0 & 0  & 0   & 0     & 111712 & 136564760 & 31768995672
& 2999009092032 \\
     5  & 0 & 0  & 0   & 0     & 0      & 1394944   & 3522539520
& 1444421355520 \\
     6  & 0 & 0  & 0   & 0     & 0      & 0         & 19318752
& 89779792749 \\
\hhline{=========}
\multicolumn{9}{|c|}{$\mathcal{F}_p=\mathcal{O}_V(1)^{\oplus 4}$: 
$(h^{1,1}, h^{2,1})=(2, 34)$ 
(Gulliksen-Neg{\aa}rd Calabi-Yau 3-fold \cite{gulliksen1971,Bertin0701,Kapustka0802,Jockers:2012dk})}
\\
\hline
\multicolumn{3}{|c|}{Intersection numbers}&
\multicolumn{6}{|c|}{$\sigma_1^3=20,\quad \sigma_1^2\tau_1=20,\quad \sigma_1\tau_1^2=16,\quad \sigma_1\tau_2=6,\quad \tau_1^3=8,\quad \tau_1\tau_2=4$}
\\
\hline
$n_{d_1,d_2}$ & $d_1=0$ & 1 & 2 & 3 & 4 & 5 & 6 &7 \\ \hline
$d_2=0$ &   & 56 & 0   & 0    & 0      & 0        & 0 
& 0 \\
     1  & 0 & 192& 896 & 192  & 0      & 0        & 0 
& 0 \\
     2  & 0 & 56 & 2544& 23016& 41056  & 23016    & 2544
& 56 \\
     3  & 0 & 0  & 896 & 52928& 813568 & 3814144  & 6292096
& 3814144 \\
     4  & 0 & 0  & 0   & 23016& 1680576& 35857016 & 284749056
& 933789504 \\
     5  & 0 & 0  & 0   & 192  & 813568 & 66781440 & 1784024064
& 20090433088 \\
     6  & 0 & 0  & 0   & 0    & 41056  & 35857016 & 3074369392
& 96591652016 \\
\hline
\end{longtable}
\end{footnotesize}

\subsubsection{Determinantal Calabi-Yau 3-folds in \eqref{det3_k2}}\label{subsubapp:det_gr3}

In a similar spirit to the quintic family discussed above, 
the determinantal Calabi-Yau 3-folds in \eqref{det3_k2} are connected with 
the complete intersection Calabi-Yau 3-fold with 
$(h^{1,1}, h^{2,1})=(1, 89)$ corresponding to 
the ``trivial'' determinantal Calabi-Yau 3-fold with 
$\mathcal{F}_p=\mathcal{O}_{V}(4)$ on $V=G(2,4)$, 
namely $X_2$ with $k=2$, $n=4$, $r=1$ and $d_1=4$ in the language of Section \ref{subsub:cy_gr}. 
This family is described by GLSMs with $U(2) \times U(1)$ gauge group. 

The classical intersection numbers \eqref{x2_intersection} of $X_2$ are given by
$\sigma_1^3=8$, $\sigma_1\sigma_2=4$,
and the genus-0 invariants $n_{d}=n_d(\sigma_1)/d$ in \eqref{grass_gw_i} 
are 
\begin{align}
n_{1}=1280,\ n_{2}=92288,\ n_{3}=15655168,\ 
n_{4}=3883902528,\ n_{5}=1190923282176,\ldots.
\end{align}
The genus-0 invariants for other determinantal Calabi-Yau 3-folds 
in \eqref{det3_k2} are summarized in Table \ref{det_cy3_g24}, where one can check that 
they exhibit the behavior \eqref{ext_trans} of the extremal transition and $h^{1,0}=0$. 

Note that, via the incidence correspondence \eqref{det_pax}, 
a geometric phase of the determinantal Calabi-Yau variety with 
$\mathcal{F}_p=\mathcal{O}_V(1)^{\oplus 4}$ 
on $V={\IP}^7$ in \eqref{class_cy3_2} 
can be identified with a geometric phase of 
the variety with $\mathcal{F}_p=\left(\mathcal{S}^*\right)^{\oplus 4}$ 
on $V=G(2,4)$ in \eqref{det3_k2} \cite{Jockers:2012zr}.
Indeed, by taking $d_1 \leftrightarrow d_2$, 
the genus-0 invariants $n_{d_1,d_2}$ of the former 
coincide with the genus-0 invariants of the latter \cite{Jockers:2012dk}.

\begin{footnotesize}
\begin{longtable}[c]{|c|rrrrrrrr|}
\caption{Genus-0 invariants of determinantal 3-folds in \eqref{det3_k2} with $V=G(2,4)$}
\label{det_cy3_g24}
\endfirsthead
\hline
\multicolumn{9}{|c|}{$\mathcal{F}_p=\mathcal{O}_V(1)\oplus \mathcal{O}_V(3)$: 
$(h^{1,1}, h^{2,1})=(2, 72)$}
\\
\hline
\multicolumn{3}{|c|}{Intersection numbers}&
\multicolumn{6}{|c|}{$\sigma_1^3=8,\quad \sigma_1\sigma_2=4,\quad \sigma_1^2\tau_1=6,\quad 
\sigma_1\tau_1^2=0,\quad \sigma_2\tau_1=3,\quad \tau_1^3=0$}
\\
\hline
$n_{d_1,d_2}$ & $d_1=0$ & 1 & 2 & 3 & 4 & 5 & 6 &7 \\ \hline
$d_2=0$ &   & 348& 2706 & 35416  & 606516    & 12209820     & 273653140  
& 6617946300 \\
     1  & 18& 900& 41778& 1871784& 81468792  & 3473471196   & 145835134092  
& 6050552127264 \\
     2  & 0 & 36 & 46548& 8009712& 864795636 & 74041264872  & 5497197606864  
& 370175324505012 \\
     3  & 0 & -4 & 1512 & 5604204& 1928672640& 363480492960 & 49681240379520  
& 5528217639011448 \\
     4  & 0 & 0  & -306 & 153936 & 985016556 & 530436671676 & 148552854522624  
& 28868137556536800 \\
     5  & 0 & 0  & 54   & -24768 & 25990110  & 214272257040 & 159209292083400  
& 60303976799146560 \\
     6  & 0 & 0  & -4   & 5940   & -3264792  & 5674351788   & 53439787982532  
& 50841527973755064 \\
\hhline{=========}
\multicolumn{9}{|c|}{$\mathcal{F}_p=\mathcal{O}_V(2)^{\oplus 2}$: 
$(h^{1,1}, h^{2,1})=(2, 58)$}
\\
\hline
\multicolumn{3}{|c|}{Intersection numbers}&
\multicolumn{6}{|c|}{$\sigma_1^3=8,\quad \sigma_1\sigma_2=4,\quad \sigma_1^2\tau_1=8,\quad 
\sigma_1\tau_1^2=0,\quad \sigma_2\tau_1=4,\quad \tau_1^3=0$}
\\
\hline
$n_{d_1,d_2}$ & $d_1=0$ & 1 & 2 & 3 & 4 & 5 & 6 & 7 \\ \hline
$d_2=0$ &   & 256& 1248 & 10496  & 111712    & 1394944      & 19318752  
& 288338176 \\
     1  & 32& 768& 21888& 591872 & 15653568  & 406723584    & 10427720448  
& 264554741760 \\
     2  & 0 & 256& 46016& 3851264& 229545472 & 11320801792  & 494003913216  
& 19776092919808 \\
     3  & 0 & 0  & 21888& 6747904& 952111808 & 90236788736  & 6690341483648  
& 419279237824512 \\
     4  & 0 & 0  & 1248 & 3851264& 1489057408& 286163875840 & 36930089276288  
& 3663867073538048 \\
     5  & 0 & 0  & 0    & 591872 & 952111808 & 414664112384 & 97746565623808  
& 15741994226581504 \\
     6  & 0 & 0  & 0    & 10496  & 229545472 & 286163875840 & 134131710670016  
& 36555466071304192 \\
\hhline{=========}
\multicolumn{9}{|c|}{$\mathcal{F}_p=\mathcal{O}_V(1)^{\oplus 2}\oplus \mathcal{O}_V(2)$: 
$(h^{1,1}, h^{2,1})=(2, 56)$}
\\
\hline
\multicolumn{3}{|c|}{Intersection numbers}&
\multicolumn{6}{|c|}{$\sigma_1^3=8,\quad \sigma_1\sigma_2=4,\quad \sigma_1^2\tau_1=10,\quad 
\sigma_1\tau_1^2=4,\quad \sigma_2\tau_1=5,\quad \tau_1^3=0$}
\\
\hline
$n_{d_1,d_2}$ & $d_1=0$ & 1 & 2 & 3 & 4 & 5 & 6 & 7 \\ \hline
$d_2=0$ &   & 140& 152  & 140    & 108       & 140         & 152   
& 140 \\
     1  & 34& 692& 8310 & 67644  & 424226    & 2179788     & 9628540 
& 37862432 \\
     2  & 0 & 436& 37266& 1201096& 23129444  & 318263924   & 3423444286 
& 30397041864 \\
     3  & 0 & 12 & 38424& 4809332& 251071058 & 7882006668  & 174584679336 
& 2978341361748 \\
     4  & 0 & 0  & 8072 & 6408160& 936362724 & 64838871368 & 2796104549608 
& 85836804179264 \\
     5  & 0 & 0  & 66   & 2838032& 1449869614& 230006825996 & 19186188980224 
& 1035385789366608 \\
     6  & 0 & 0  & -2   & 329036 & 956057192 & 393389988300 & 65626229819274 
& 6246121752675024 \\
\hhline{=========}
\multicolumn{9}{|c|}{$\mathcal{F}_p=\mathcal{S}^*\oplus\mathcal{O}_V(3)$: 
$(h^{1,1}, h^{2,1})=(2, 77)$}
\\
\hline
\multicolumn{3}{|c|}{Intersection numbers}&
\multicolumn{6}{|c|}{$\sigma_1^3=8,\quad \sigma_1\sigma_2=4,\quad \sigma_1^2\tau_1=7,\quad 
\sigma_1\tau_1^2=3,\quad \sigma_2\tau_1=3,\quad \tau_1^3=0$}
\\
\hline
$n_{d_1,d_2}$ & $d_1=0$ & 1 & 2 & 3 & 4 & 5 & 6 & 7 \\ \hline
$d_2=0$ &   & 195 & 195  & 150    & 195       & 195          & 150 
& 195 \\
     1  & 13& 1030& 24479& 330960 & 3035018   & 21301930     & 123660710 
& 622928364 \\
     2  & 0 & 78  & 65007& 5464206& 213740347 & 5220791429   & 91165319219 
& 1233231670475 \\
     3  & 0 & -26 & 3822 & 9502026& 1561721228& 114639515100 & 5115340545693 
& 159519319143362 \\
     4  & 0 & 3   & -1820& 503243 & 2028893885& 514721709028 & 58258127191937 
& 3983242948904679 \\
     5  & 0 & 0   & 858  & -215410& 103906805 & 535733030960 & 185997625577552 
& 29104035511228470 \\
     6  & 0 & 0   & -312 & 111267 & -38991863 & 27312140744  & 162043340071962 
& 71687188824610803 \\
\hhline{=========}
\multicolumn{9}{|c|}{$\mathcal{F}_p=\mathcal{S}^*(1)\oplus\mathcal{O}_V(1)$: 
$(h^{1,1}, h^{2,1})=(2, 49)$}
\\
\hline
\multicolumn{3}{|c|}{Intersection numbers}&
\multicolumn{6}{|c|}{$\sigma_1^3=8,\quad \sigma_1\sigma_2=4,\quad \sigma_1^2\tau_1=11,\quad 
\sigma_1\tau_1^2=5,\quad \sigma_2\tau_1=5,\quad \tau_1^3=0$}
\\
\hline
$n_{d_1,d_2}$ & $d_1=0$ & 1 & 2 & 3 & 4 & 5 & 6 & 7 \\ \hline
$d_2=0$ &   & 110& 113  & 113    & 110       & 94           & 110 
& 113 \\
     1  & 41& 632& 5449 & 32522  & 155463    & 628866       & 2256445 
& 7358644 \\
     2  & 0 & 486& 29680& 672004 & 9213931   & 91886539     & 730644383 
& 4890880851 \\
     3  & 0 & 52 & 40521& 3389134& 122021518 & 2682580356   & 42201281320 
& 518612135254 \\
     4  & 0 & 0  & 15206& 6089576& 584124117 & 27553828341  & 823110963896 
& 17728177368851 \\
     5  & 0 & 0  & 1318 & 4251622& 1230515498& 127127937012 & 7158680727853 
& 265381636196294 \\
     6  & 0 & 0  & 1    & 1125074& 1223539121& 297185353890 & 32148039886801 
& 2051420839803630 \\
\hhline{=========}
\multicolumn{9}{|c|}{$\mathcal{F}_p=\mathcal{O}_V(1)^{\oplus 4}$: 
$(h^{1,1}, h^{2,1})=(2, 50)$}
\\
\hline
\multicolumn{3}{|c|}{Intersection numbers}&
\multicolumn{6}{|c|}{$\sigma_1^3=8,\quad \sigma_1\sigma_2=4,\quad \sigma_1^2\tau_1=12,\quad 
\sigma_1\tau_1^2=8,\quad \sigma_2\tau_1=6,\quad \tau_1^3=2$}
\\
\hline
$n_{d_1,d_2}$ & $d_1=0$ & 1 & 2 & 3 & 4 & 5 & 6 & 7 \\ \hline
$d_2=0$ &   & 80 & 20   & 0      & 0         & 0            & 0 
& 0 \\
     1  & 40& 560& 2800 & 6800   & 9104      & 6800         & 2800 
& 560 \\
     2  & 0 & 560& 22220& 274784 & 1695200   & 6283360      & 15291620 
& 25650640 \\
     3  & 0 & 80 & 42208& 2102160& 40381840  & 417187840    & 2708790480 
& 12060977392 \\
     4  & 0 & 0  & 22220& 5443840& 299074880 & 7435705920   & 106637235608 
& 1000779043760 \\
     5  & 0 & 0  & 2800 & 5443840& 929117120 & 53663104048  & 1580847225600 
& 28485500761200 \\
     6  & 0 & 0  & 20   & 2102160& 1343346240& 187910411760 & 11181575861220 
& 371398929912800 \\
\hhline{=========}
\multicolumn{9}{|c|}{$\mathcal{F}_p=\mathcal{S}^*\oplus\mathcal{O}_V(1)\oplus \mathcal{O}_V(2)$: 
$(h^{1,1}, h^{2,1})=(2, 57)$}
\\
\hline
\multicolumn{3}{|c|}{Intersection numbers}&
\multicolumn{6}{|c|}{$\sigma_1^3=8,\quad \sigma_1\sigma_2=4,\quad \sigma_1^2\tau_1=11,\quad 
\sigma_1\tau_1^2=7,\quad \sigma_2\tau_1=5,\quad \tau_1^3=2$}
\\
\hline
$n_{d_1,d_2}$ & $d_1=0$ & 1 & 2 & 3 & 4 & 5 & 6 & 7 \\ \hline
$d_2=0$ &   & 88 & 5    & 0      & 0         & 0            & 0 
& 0 \\
     1  & 33& 634& 4048 & 10037  & 10037     & 4048         & 634 
& 33 \\
     2  & 0 & 548& 28997& 466086 & 3406527   & 13159772     & 28776716 
& 37157620 \\
     3  & 0 & 10 & 45347& 3181936& 80480431  & 1027716204   & 7647030133 
& 35755062323 \\
     4  & 0 & 0  & 13736& 6686966& 522689207 & 17306100970  & 313954566036 
& 3528240156238 \\
     5  & 0 & 0  & 165  & 4524366& 1312841562& 108179999795 & 4280740941876 
& 99081868036162 \\
     6  & 0 & 0  & -10  & 780282 & 1358341003& 305921060292 & 25988397030539 
& 1167267498525808 \\
\hhline{=========}
\multicolumn{9}{|c|}{$\mathcal{F}_p=\Sym^2 \mathcal{S}^*\oplus\mathcal{O}_V(1)$: 
$(h^{1,1}, h^{2,1})=(2, 32)$}
\\
\hline
\multicolumn{3}{|c|}{Intersection numbers}&
\multicolumn{6}{|c|}{$\sigma_1^3=8,\quad \sigma_1\sigma_2=4,\quad \sigma_1^2\tau_1=14,\quad 
\sigma_1\tau_1^2=12,\quad \sigma_2\tau_1=5,\quad \tau_1^3=4$}
\\
\hline
$n_{d_1,d_2}$ & $d_1=0$ & 1 & 2 & 3 & 4 & 5 & 6 & 7 \\ \hline
$d_2=0$ &   & 20 & 22   & 0      & 0        & 0           & 0 
& 0 \\
     1  & 58& 348& 870  & 1160   & 870      & 348         & 58 
& 0 \\
     2  & 0 & 844& 9460 & 42320  & 115744   & 200724      & 244280 
& 200724 \\
     3  & 0 & 68 & 35968& 541140 & 3646870  & 14883488    & 41436000 
& 83496920 \\
     4  & 0 & 0  & 34722& 2839040& 47787096 & 402821800   & 2153902504 
& 8105770980 \\
     5  & 0 & 0  & 11050& 5898656& 298453714& 5287652400  & 51848056504 
& 335849637824 \\
     6  & 0 & 0  & 196  & 4822716& 908058576& 37135584632 & 678692927028 
& 7409928380632 \\
\hhline{=========}
\multicolumn{9}{|c|}{$\mathcal{F}_p=\mathcal{S}^*\oplus\mathcal{O}_V(1)^{\oplus 3}$: 
$(h^{1,1}, h^{2,1})=(2, 49)$}
\\
\hline
\multicolumn{3}{|c|}{Intersection numbers}&
\multicolumn{6}{|c|}{$\sigma_1^3=8,\quad \sigma_1\sigma_2=4,\quad \sigma_1^2\tau_1=13,\quad 
\sigma_1\tau_1^2=11,\quad \sigma_2\tau_1=6,\quad \tau_1^3=5$}
\\
\hline
$n_{d_1,d_2}$ & $d_1=0$ & 1 & 2 & 3 & 4 & 5 & 6 & 7 \\ \hline
$d_2=0$ &   & 52 & 1    & 0      & 0         & 0           & 0 
& 0 \\
     1  & 41& 486& 1318 & 917    & 113       & 0           & 0 
& 0 \\
     2  & 0 & 632& 15206& 94206  & 216954    & 202196      & 72260 
& 7686 \\
     3  & 0 & 110& 40521& 1125074& 10519903  & 43910603    & 91555625 
& 99039844 \\
     4  & 0 & 0  & 29680& 4251622& 124486831 & 1484582184  & 8931510318 
& 29965206018 \\
     5  & 0 & 0  & 5449 & 6089576& 579108969 & 17340098333 & 242953144372 
& 1875605165389 \\
     6  & 0 & 0  & 113  & 3389134& 1223539121& 92586552714 & 2802737114627 
& 44031493485406 \\
\hhline{=========}
\multicolumn{9}{|c|}{$\mathcal{F}_p=\left(\mathcal{S}^*\right)^{\oplus 2}\oplus\mathcal{O}_V(2)$: 
$(h^{1,1}, h^{2,1})=(2, 56)$}
\\
\hline
\multicolumn{3}{|c|}{Intersection numbers}&
\multicolumn{6}{|c|}{$\sigma_1^3=8,\quad \sigma_1\sigma_2=4,\quad \sigma_1^2\tau_1=12,\quad 
\sigma_1\tau_1^2=10,\quad \sigma_2\tau_1=5,\quad \tau_1^3=5$}
\\
\hline
$n_{d_1,d_2}$ & $d_1=0$ & 1 & 2 & 3 & 4 & 5 & 6 & 7 \\ \hline
$d_2=0$ &   & 48 & -2   & 0      & 0         & 0            & 0 
& 0 \\
     1  & 34& 544& 1719 & 544    & 34        & 0            & 0 
& 0 \\
     2  & 0 & 688& 19704& 138352 & 291762    & 138352       & 19704 
& 688 \\
     3  & 0 & 0  & 48165& 1682784& 18006204  & 75544928     & 126642213 
& 75544928 \\
     4  & 0 & 0  & 22206& 5807280& 214145556 & 2945951712   & 18597811286 
& 57190487824 \\
     5  & 0 & 0  & 561  & 6279840& 910538594 & 34261029504  & 557526592367 
& 4630265286624 \\
     6  & 0 & 0  & -68  & 1729152& 1545311902& 167569246816 & 6356737689516 
& 116628229665712 \\
\hhline{=========}
\multicolumn{9}{|c|}{$\mathcal{F}_p=\left(\mathcal{S}^*\right)^{\oplus 2}\oplus\mathcal{O}_V(1)^{\oplus 2}$: 
$(h^{1,1}, h^{2,1})=(2, 46)$}
\\
\hline
\multicolumn{3}{|c|}{Intersection numbers}&
\multicolumn{6}{|c|}{$\sigma_1^3=8,\quad \sigma_1\sigma_2=4,\quad \sigma_1^2\tau_1=14,\quad 
\sigma_1\tau_1^2=14,\quad \sigma_2\tau_1=6,\quad \tau_1^3=9$}
\\
\hline
$n_{d_1,d_2}$ & $d_1=0$ & 1 & 2 & 3 & 4 & 5 & 6 & 7 \\ \hline
$d_2=0$ &   & 28 & 0    & 0      & 0        & 0           & 0 
& 0 \\
     1  & 44& 404& 579  & 28     & -2       & 0           & 0 
& 0 \\
     2  & 0 & 708& 9486 & 26276  & 15912    & 432         & -2 
& 0 \\
     3  & 0 & 140& 35891& 511640 & 2079058  & 2757236     & 1011037 
& 29956 \\
     4  & 0 & 0  & 36284& 2887060& 41253512 & 209384984   & 432768018 
& 355277816 \\
     5  & 0 & 0  & 9641 & 5964836& 295048376& 4172615020  & 24615473481 
& 67966106564 \\
     6  & 0 & 0  & 406  & 4749072& 923105328& 35005695588 & 489248692862 
& 3213917918364 \\
\hhline{=========}
\multicolumn{9}{|c|}{$\mathcal{F}_p=\left(\mathcal{S}^*\right)^{\oplus 3}\oplus\mathcal{O}_V(1)$: 
$(h^{1,1}, h^{2,1})=(2, 41)$}
\\
\hline
\multicolumn{3}{|c|}{Intersection numbers}&
\multicolumn{6}{|c|}{$\sigma_1^3=8,\quad \sigma_1\sigma_2=4,\quad \sigma_1^2\tau_1=15,\quad 
\sigma_1\tau_1^2=17,\quad \sigma_2\tau_1=6,\quad \tau_1^3=14$}
\\
\hline
$n_{d_1,d_2}$ & $d_1=0$ & 1 & 2 & 3 & 4 & 5 & 6 & 7 \\ \hline
$d_2=0$ &   & 10 & 0    & 0      & 0        & 0          & 0 
& 0 \\
     1  & 49& 308& 231  & 0      & 0        & 0          & 0 
& 0 \\
     2  & 0 & 794& 5349 & 5729   & 231      & 0          & 0 
& 0 \\
     3  & 0 & 168& 29491& 190382 & 287583   & 76182      & 78 
& 0 \\
     4  & 0 & 0  & 40547& 1681790& 10332969 & 18880381   & 9662787 
& 760431 \\
     5  & 0 & 0  & 15540& 5118106& 119727638& 699168640  & 1461234039 
& 1090271882 \\
     6  & 0 & 0  & 1120 & 5820116& 572514233& 9758035439 & 54759243098 
& 126157897721 \\
\hhline{=========}
\multicolumn{9}{|c|}{$\mathcal{F}_p=\left(\mathcal{S}^*\right)^{\oplus 4}$: 
$(h^{1,1}, h^{2,1})=(2, 34)$ \cite{Jockers:2012dk}}
\\
\hline
\multicolumn{3}{|c|}{Intersection numbers}&
\multicolumn{6}{|c|}{$\sigma_1^3=8,\quad \sigma_1\sigma_2=4,\quad \sigma_1^2\tau_1=16,\quad 
\sigma_1\tau_1^2=20,\quad \sigma_2\tau_1=6,\quad \tau_1^3=20$}
\\
\hline
$n_{d_1,d_2}$ & $d_1=0$ & 1 & 2 & 3 & 4 & 5 & 6 & 7 \\ \hline
$d_2=0$ &   & 0  & 0    & 0      & 0        & 0          & 0 
& 0 \\
     1  & 56& 192& 56   & 0      & 0        & 0          & 0 
& 0 \\
     2  & 0 & 896& 2544 & 896    & 0        & 0          & 0 
& 0 \\
     3  & 0 & 192& 23016& 52928  & 23016    & 192        & 0 
& 0 \\
     4  & 0 & 0  & 41056& 813568 & 1680576  & 813568     & 41056 
& 0 \\
     5  & 0 & 0  & 23016& 3814144& 35857016 & 66781440   & 35857016 
& 3814144 \\
     6  & 0 & 0  & 2544 & 6292096& 284749056& 1784024064 & 3074369392 
& 1784024064 \\
\hline
\end{longtable}
\end{footnotesize}

\hspace{1pc}

\section{Conclusions}\label{sec:conclusion}

In this paper we have examined a class of 
square determinantal Calabi-Yau varieties in Grassmannians 
satisfying appropriate conditions 
about dimension, a Calabi-Yau definition, duality $G(k,n)\cong G(n-k,n)$, 
and rank of the vector bundles. 
We found that infinite families of examples associated with
non-abelian quiver GLSMs might be possible. 
Furthermore, we explicitly demonstrated how to compute genus-0 integer invariants of the determinantal 
Calabi-Yau varieties via the Givental $I$-functions. By constructing the $I$-functions from the
supersymmetric localization formula for the GLSM on a supersymmetric 2-sphere, 
we provided a guideline for the evaluation of the genus-0
A-model correlators. 
We also found the handy formula for the 1-point correlators for Grassmannian Calabi-Yau varieties,
which turned out to be generalized into the cases with the determinantal varieties.
We hope that our results would give a clue to understand various properties of the less studied GLSMs with 
non-abelian gauge groups.

Finally, we comment on possible future research directions.
\begin{itemize}
\item
Since we have not imposed irreducibility 
as a requirement, 
to make our classification more rigorous, 
a comprehensive study of topological invariants such as Hodge numbers 
and Gromov-Witten invariants for the infinite families in (\ref{inf_det_3}), 
\eqref{inf_det_2}, and \eqref{inf_det_4} 
is indispensable to check whether they are appropriate irreducible Calabi-Yau varieties.
\item
We have classified the square determinantal varieties based on the requirement \eqref{det_grass}. 
It would be interesting to examine determinantal varieties with 
general vector bundles $\mathcal{E}_p$ 
such as $\mathcal{E}_p=L \otimes \mathcal{O}_V^{\oplus p}$, 
where $L$ is a line bundle, as studied in \cite{Bertin0701}.
\item
We conjectured the formula \eqref{grass_gw_i} for 
the genus-0 1-point A-model correlators 
for Grassmannian Calabi-Yau varieties, which generalizes 
the formula \eqref{1pt_ex1}. 
It would be interesting to find out the 3-point extension of our formula \eqref{grass_gw_i} 
as a natural generalization of the formula \eqref{3pt_ex1} 
studied in \cite{PopaZinger_co},
and give a proof of it.
\item
In \cite{Caldararu:2017usq}, GLSM realizations of 
the so-called Veronese embeddings and the Segre embeddings 
were proposed, and it opened up the possibility of more broad class of 
Calabi-Yau varieties. 
Various exotic Calabi-Yau examples including the constructions 
in \cite{Kanazawa:2012xya} have also been discussed, and 
it would be interesting to consider 
these examples and discuss their $I$-functions.
\end{itemize}



\subsection*{Acknowledgements}

The work of MM was supported by the ERC Starting Grant no. 335739 
``Quantum fields and knot homologies'' funded by the European Research Council under the European Union's Seventh Framework Programme, and by the Max-Planck-Institut f\"ur Mathematik in Bonn, and by the Australian Research Council.

\appendix

\section{Determinantal Calabi-Yau 2-folds and 4-folds}\label{app:det_cy24}

In Section \ref{subsec:cy3}, we have focused on the realization of 
a class of determinantal Calabi-Yau 3-folds of square type.
In a similar spirit, here 
we discuss the classification of determinantal Calabi-Yau 2-folds and 4-folds satisfying the requirements
(\ref{cy_dim_inc}) -- (\ref{cy_det_inc}).

\subsection{Determinantal Calabi-Yau 2-folds}\label{subapp:cy2}

When $\dim X_A = 2$, the dimensional condition (\ref{cy_dim_inc}) becomes
\begin{align}
\ell_p^{\vee} \left(k \mathfrak{c}_1(\mathcal{F}_p)-\ell_p^{\vee}\right)=k^2+2.
\label{cy_dim2_ns}
\end{align}
Note that, as mentioned in Remark \ref{rem:sing_det}, the generic determinantal Calabi-Yau 2-folds 
$X_A$ do not have the singular loci.
In the following, we clarify which type of choices for $\left(k,n;\ell_p^{\vee}, \mathfrak{c}_1(\mathcal{F}_p)\right)$
can be possible while changing the parameter $k$.

\subsubsection{$k=1$}

When $k=1$ we have $V=G(1,n)\cong {\IP}^{n-1}$, and the generic solution
(\ref{class_1}) with (\ref{class_1f}) provides a ``quartic family'' given by
\begin{align}
\boxed{
\left(k,n;\ell_p^{\vee}, \mathfrak{c}_1(\mathcal{F}_p)\right)=
\left(1,4;1,4 \right) \ \ \textrm{with} \ \ 
\mathcal{F}_p=\oplus_{i=1}^r\mathcal{O}_V(p_i),\ 
p_1\ge p_2 \ge \cdots \ge p_r > 0,\ \sum_{i=1}^rp_i=4.}
\end{align}
In addition, (\ref{class_2}) provides another class of solutions 
\begin{align}
\boxed{
\left(k,n;\ell_p^{\vee}, \mathfrak{c}_1(\mathcal{F}_p) \right)=
(1,12;3,4) \ \ \textrm{with} \ \ 
\mathcal{F}_p=
\mathcal{O}_V(1)^{\oplus 2}\oplus \mathcal{O}_V(2),\
\mathcal{O}_V(1)^{\oplus 4}.}
\label{cy2_class_2}
\end{align}
Here the Calabi-Yau 2-fold $X_A$ constructed from $\mathcal{F}_p=
\mathcal{O}_V(1)^{\oplus 2}\oplus \mathcal{O}_V(2)$ with $p=3$ (\textit{i.e.} $\ell=0$) 
can be described by the complete intersection Calabi-Yau 2-fold 
in ${\IP}^5$ with the vector bundle $\mathcal{O}_{{\IP}^5}(2)^{\oplus 3}$ as
$$
X_A\ \textrm{with}\ 
\mathcal{O}_V(1)^{\oplus 2}\oplus \mathcal{O}_V(2)\ \longleftrightarrow\
X_{2,2,2} \subset {\IP}^5.
$$

\subsubsection{$k\ge 2$}

When $k\ge 2$ we have $V=G(k,n)$. In this case, there exist two
``infinite families'' of solutions as determinantal 2-folds satisfying all the requirements
(\ref{cy_dim2_ns}), (\ref{cy_cd_inc}), (\ref{cy_dualc}), and (\ref{cy_det_inc}).
\begin{empheq}[box=\fbox]{equation}
\begin{split}
&
\left(k,n;\ell_p^{\vee}, \mathfrak{c}_1(\mathcal{F}_p)\right)=
(k_i, 4\ell_i ; \ell_i, 4),\ \ i\in{\IN}
\\
&\
\textrm{where}\ 
k_{1}=3,\ \ell_{1}=11,\
k_{i+1}=\ell_{i},\ \ell_{i+1}=-k_{i}+4\ell_{i}  \ \ \textrm{with} \ \
\mathcal{F}_p=\left(\mathcal{S}^*\right)^{\oplus 4},\
\mathcal{Q}^{\oplus 4}.
\label{inf_det_2}
\end{split}
\end{empheq}


\subsection{Determinantal Calabi-Yau 4-folds}\label{subapp:cy4}

When $\dim X_A = 4$, the dimensional condition (\ref{cy_dim_inc}) becomes
\begin{align}
\ell_p^{\vee} \left(k \mathfrak{c}_1(\mathcal{F}_p)-\ell_p^{\vee}\right)=k^2+4.
\label{cy_dim4_ns}
\end{align}

\subsubsection{$k=1$}

When $k=1$ we have $V=G(1,n)\cong {\IP}^{n-1}$, 
and the generic solution (\ref{class_1}) with (\ref{class_1f}) provides 
a ``sextic family'' given by
\begin{align}
\boxed{
\left(k,n;\ell_p^{\vee}, \mathfrak{c}_1(\mathcal{F}_p) \right)=
(1,6;1,6) \ \ \textrm{with} \ \ 
\mathcal{F}_p=\oplus_{i=1}^r\mathcal{O}_V(p_i),\ 
p_1\ge p_2 \ge \cdots \ge p_r > 0,\ \sum_{i=1}^rp_i=6.}
\label{class_cy4_1}
\end{align}
Via the incidence correspondence \eqref{det_pax}, 
the sextic family is connected each other through the extremal transitions (see Appendix \ref{app:sextic}). 

In addition, (\ref{class_2}) provides another class of solutions
\begin{align}
\boxed{
\left(k,n;\ell_p^{\vee}, \mathfrak{c}_1(\mathcal{F}_p) \right)=
(1,30;5,6) \ \ \textrm{with} \ \ 
\mathcal{F}_p=
\mathcal{O}_V(1)^{\oplus 4}\oplus \mathcal{O}_V(2),\
\mathcal{O}_V(1)^{\oplus 6}.}
\label{class_cy4_2}
\end{align}
Here the Calabi-Yau 4-fold $X_A$ constructed from 
$\mathcal{F}_p=\mathcal{O}_V(1)^{\oplus 4}\oplus \mathcal{O}_V(2)$ 
with $p=5$ (\textit{i.e.} $\ell=0$) can be described by 
the complete intersection Calabi-Yau 4-fold 
in ${\IP}^9$ with the vector bundle $\mathcal{O}_{{\IP}^5}(2)^{\oplus 5}$ as
$$
X_A\ \textrm{with}\ 
\mathcal{O}_V(1)^{\oplus 4}\oplus \mathcal{O}_V(2)\ \longleftrightarrow\
X_{2,2,2,2,2} \subset {\IP}^9.
$$

\subsubsection{$k=2$}

When $k=2$ we have $V=G(2,n)$. In this case there exists a class of determinantal 
4-folds which satisfy the conditions (\ref{cy_dim4_ns}), (\ref{cy_cd_inc}), (\ref{cy_dualc}), and (\ref{cy_det_inc})
given by\footnote{
In Appendix \ref{app:cy4_k2}, we see that the determinantal 4-fold associated with 
$\mathcal{F}_p=\Sym^2 \mathcal{S}^*$ is not a Calabi-Yau variety with 
$(h^{1,0},h^{2,0})=(0,0)$ but an irreducible holomorphic symplectic variety with 
$(h^{1,0},h^{2,0})=(0,1)$.}
\begin{empheq}[box=\fbox]{equation}
\begin{split}
&
\left(k,n;\ell_p^{\vee}, \mathfrak{c}_1(\mathcal{F}_p) \right)=
(2,6;2,3)  \ \ \textrm{with} \ \ \\
&\
\mathcal{F}_p=\mathcal{O}_V(1)\oplus \mathcal{O}_V(2),\
\mathcal{S}^*(1),\
\mathcal{O}_V(1)^{\oplus 3},\
\mathcal{S}^* \oplus \mathcal{O}_V(2),\
\Sym^2 \mathcal{S}^*,\
\mathcal{S}^* \oplus \mathcal{O}_V(1)^{\oplus 2},\\
&\qquad\ \
\left(\mathcal{S}^*\right)^{\oplus 2} \oplus \mathcal{O}_V(1),\
\left(\mathcal{S}^*\right)^{\oplus 3},\
\mathcal{Q} \oplus  \mathcal{O}_V(2),\
\mathcal{Q} \oplus \mathcal{O}_V(1)^{\oplus 2},\
\mathcal{Q}^{\oplus 2} \oplus \mathcal{O}_V(1),\
\mathcal{Q}^{\oplus 3},\\
&\qquad\ \
\wedge^2\mathcal{Q},\
\wedge^3\mathcal{Q},\
\mathcal{S}^* \oplus \mathcal{Q} \oplus  \mathcal{O}_V(1),\
\left(\mathcal{S}^*\right)^{\oplus 2} \oplus \mathcal{Q},\
\mathcal{S}^* \oplus \mathcal{Q}^{\oplus 2}.
\label{class_cy4_3}
\end{split}
\end{empheq}
Here the Calabi-Yau 4-fold $X_A$ constructed from 
$\mathcal{F}_p=\mathcal{O}_V(1)\oplus \mathcal{O}_V(2)$ 
(\textit{resp.} $\mathcal{S}^*(1)$) with $p=2$ (\textit{i.e.} $\ell=0$) can be described by 
the complete intersection Grassmannian Calabi-Yau 4-fold 
in $G(2,6)$ with the vector bundle 
$\mathcal{O}_{G(2,6)}(1)^{\oplus 2}\oplus \mathcal{O}_{G(2,6)}(2)^{\oplus 2}$ 
(\textit{resp.} $\mathcal{S}^*(1)^{\oplus 2}$ on $G(2,6)$).

We find that there exist another type of solutions 
satisfying all the requirements (\ref{cy_dim4_ns}), (\ref{cy_cd_inc}), (\ref{cy_dualc}), and (\ref{cy_det_inc}) given by
\begin{empheq}[box=\fbox]{equation}
\begin{split}
&
\left(k,n;\ell_p^{\vee}, \mathfrak{c}_1(\mathcal{F}_p) \right)=
(2,12;4,3) \ \ \textrm{with} \ \ \\
&\ 
\mathcal{F}_p=
\mathcal{S}^* \oplus \mathcal{O}_V(1)^{\oplus 2},\
\left(\mathcal{S}^*\right)^{\oplus 2} \oplus \mathcal{O}_V(1),\
\left(\mathcal{S}^*\right)^{\oplus 3},\
\mathcal{Q} \oplus \mathcal{O}_V(2),
\\
&\qquad\ \
\mathcal{Q} \oplus \mathcal{O}_V(1)^{\oplus 2},\
\mathcal{Q}^{\oplus 2} \oplus \mathcal{O}_V(1),\ 
\mathcal{Q}^{\oplus 3}.
\end{split}
\end{empheq}
Here the Calabi-Yau 4-fold $X_A$ constructed from 
$\mathcal{F}_p=\mathcal{S}^* \oplus \mathcal{O}_V(1)^{\oplus 2}$ 
with $p=4$ (\textit{i.e.} $\ell=0$) can be described by 
the complete intersection Grassmannian Calabi-Yau 4-fold 
in $G(2,8)$ with the vector bundle $\mathcal{O}_{G(2,8)}(1)^{\oplus 8}$.

\subsubsection{$k\ge 3$}

When $k\ge 3$ we have $V=G(k,n)$, and there exist six ``infinite families'' of determinantal 4-folds 
given by
\begin{empheq}[box=\fbox]{equation}
\begin{split}
&
\left(k,n;\ell_p^{\vee}, \mathfrak{c}_1(\mathcal{F}_p)\right)=
(k_i, 6\ell_i ; \ell_i, 6),\ \ i\in{\IN}
\\
&\
\textrm{with} \ \ 
k_{1}=5,\ \ell_{1}=29,\
k_{i+1}=\ell_{i},\ \ell_{i+1}=-k_{i}+6\ell_{i}: \ 
\mathcal{F}_p=\left(\mathcal{S}^*\right)^{\oplus 6},\
\mathcal{Q}^{\oplus 6},
\\
&
\left(k,n;\ell_p^{\vee}, \mathfrak{c}_1(\mathcal{F}_p)\right)=
(k_i, 3\ell_i ; \ell_i, 3),\ \ i\in{\IN}
\\
&\
\textrm{with} \ \ 
k_{1}=4,\ \ell_{1}=10,\
k_{i+1}=-3k_{i}+8\ell_{i},\ \ell_{i+1}=-8k_{i}+21\ell_{i}: \
\mathcal{F}_p=\left(\mathcal{S}^*\right)^{\oplus 3},\
\mathcal{Q}^{\oplus 3},
\\
&
\left(k,n;\ell_p^{\vee}, \mathfrak{c}_1(\mathcal{F}_p)\right)=
(k_i, 3\ell_i ; \ell_i, 3),\ \ i\in{\IN}
\\
&\
\textrm{with} \ \
k_{1}=10,\ \ell_{1}=26,\
k_{i+1}=-3k_{i}+8\ell_{i},\ \ell_{i+1}=-8k_{i}+21\ell_{i}: 
\mathcal{F}_p=\left(\mathcal{S}^*\right)^{\oplus 3},\
\mathcal{Q}^{\oplus 3},
\label{inf_det_4}
\end{split}
\end{empheq}
where the third (\textit{resp.} fourth) and the fifth (\textit{resp.} sixth) families of 4-folds associated
with $\mathcal{F}_p=\left(\mathcal{S}^*\right)^{\oplus 3}$ 
(\textit{resp.} $\mathcal{Q}^{\oplus 3}$) 
are given by the same recurrence relation with the different initial conditions. 
By using mathematical induction, one can 
check that the duality condition \eqref{cy_dualc}, 
the rank condition \eqref{cy_det_inc}, 
and in particular $\ell_p^{\vee} < \rank\,\mathcal{F}_p$, are maintained.


\section{Hodge number calculations via the Koszul complex}\label{app:hodge}

Following \cite{Hubsch:1992nu} (see also \textit{e.g.} 
\cite{Kuchle:1995,Anderson:2008ex,Blumenhagen:2010pv,Blumenhagen:2010ed}), 
 here we briefly review how to compute cohomologies and Hodge numbers of Calabi-Yau 
 varieties via the Koszul complex. We will demonstrate the explicit computations for several 
 examples.

\subsection{General algorithm}\label{subapp:hodge_gen}

Let $V$ be a complex manifold, 
$\mathcal{E}_p$ be a rank $p$ vector bundle over $V$ and consider the locus $X \subset V$ as a holomorphic 
section of $\mathcal{E}_p$. In this appendix we describe how to compute the cohomologies
\begin{align}
H^{\dim X-1,i}(X)=H^{i}(X,TX),\quad \textrm{or}\quad
H^{1,i}(X)=H^i(X,T^*X),\quad 
i=0,1,\ldots,\dim X,
\label{step2_goal}
\end{align}
via the Koszul complex.

\subsubsection{Step 1: Computation of \eqref{step1_goal}}

First we describe a method to compute bundle-valued cohomologies of $X$, 
\begin{align}
H^i(X,\mathcal{F}_V|_X),
\label{step1_goal}
\end{align}
by using the \textit{Koszul exact sequence}.

The Koszul exact sequence gives the resolution of $\mathcal{O}_X$ over $V$ as
\begin{align}
0 \longrightarrow \wedge^p \mathcal{E}_p^* \longrightarrow \cdots \longrightarrow
\wedge^2 \mathcal{E}_p^* \longrightarrow \mathcal{E}_p^* \longrightarrow
\mathcal{O}_V \longrightarrow \mathcal{O}_X \longrightarrow 0.
\label{koszul_seq}
\end{align}
For the Koszul exact sequence, the \textit{Koszul spectral sequence} 
(see \textit{e.g.} \cite{Griffiths:1978}),
$$
\left\{E_{r}^{i,q}\right\},\quad i=0,1,\ldots,\dim V,\ \
q=0,1,\ldots,p,\ \
r=0,1,2,\ldots,
$$
can be associated as follows. Starting from
\begin{align}
E_0^{i,q}=H^i(V,\wedge^q \mathcal{E}_p^*),
\end{align}
define $d_r$-cohomology recursively as
\begin{align}
E_{r+1}^{i,q}=\frac{\mathrm{ker}\big(d_r:\ E_{r}^{i,q}\ \longrightarrow \ E_{r}^{i-r,q-r-1}\big)}{\mathrm{im}\big(d_r:\ E_{r}^{i+r,q+r+1}\ \longrightarrow \ E_{r}^{i,q}\big)},
\end{align}
which is associated with \textit{differentials}
\begin{align}
d_r\ :\ \
E_{r}^{i,q}\ \longrightarrow\ 
E_{r}^{i-r,q-r-1},
\end{align}
with $d_r \circ d_r=0$. 
Here $E_{r}^{i,q}=0$ for $i, q <0$, $i>\dim V$, and $q>p$.
At finite $r=r_0$, $E_{r}^{i,q}$ converges to 
$E_{r_0}^{i,q}=E_{r_0+1}^{i,q}=\ldots=E_{\infty}^{i,q}$ and 
we obtain a cohomology of $X$ as
\begin{align}
\sum_{q=0}^p E_{\infty}^{i+q,q}\ \Longrightarrow\ H^i(X,\mathcal{O}_X)=H^{0,i}(X),
\label{k_sp_coh_0}
\end{align}
where the summation represents a formal sum.

\begin{remark}
By tensoring the Koszul exact sequence \eqref{koszul_seq} with 
a vector bundle $\mathcal{F}_V$ (\textit{e.g.} $\mathcal{E}_p$, $\mathcal{E}_p^*$, $TV$, $T^*V$, etc.) over $V$, one obtains the resolution of $\mathcal{F}_V|_X$ over $V$ as
\begin{align}
0 \longrightarrow \wedge^p \mathcal{E}_p^* \otimes \mathcal{F}_V \longrightarrow \cdots \longrightarrow
\wedge^2 \mathcal{E}_p^*  \otimes \mathcal{F}_V \longrightarrow 
\mathcal{E}_p^* \otimes \mathcal{F}_V \longrightarrow
\mathcal{F}_V \longrightarrow  \mathcal{F}_V|_X \longrightarrow 0.
\label{koszul_seq_t}
\end{align}
Then, by considering the Koszul spectral sequence associated with \eqref{koszul_seq_t}, 
one can obtain the bundle-valued cohomologies $H^i(X,\mathcal{F}_V|_X)$ in \eqref{step1_goal}.
\end{remark}

Therefore, by using the Koszul spectral sequence, the cohomologies $H^i(X,\mathcal{F}_V|_X)$ can be 
computed from the cohomologies $H^i(V,\wedge^q \mathcal{E}_p^*  \otimes \mathcal{F}_V)$. For computing 
these quantities, the Bott-Borel-Weil theorem \ref{thm:bott_borel_weil} is quite useful. To state the theorem, 
consider a flag manifold
\begin{align}
V=\frac{U(N)}{U(n_1)\times \cdots \times U(n_F)},\qquad
N=\sum_{i=1}^Fn_i.
\label{def_flag}
\end{align}
A holomorphic homogeneous vector bundle $\mathcal{F}_V$ over $V$ can be described by a representation of 
$U(n_1)\times \cdots \times U(n_F)$, where a representation of each $U(n)$ is described by a Young diagram 
which is given by a monotonically increasing sequence with length $n$ of integers as $(a_1,\ldots,a_n)$, $a_i \le a_{i+1}$. 
Then, a vector bundle $\mathcal{F}_V$ is described by a representation of $U(n_1)\times \cdots \times U(n_F)$ as 
\begin{align}
\mathcal{F}_V \sim (a_1,\ldots,a_{n_1}|b_1,\ldots,b_{n_2}|\cdots|r_1,\ldots,r_{n_F}).
\label{rep_vec_flag}
\end{align}


\begin{exam}\label{app_ex:pn}
For $V={\IP}^{n-1}=U(n)/[U(1)\times U(n-1)]$, in terms of the representations of $U(1)\times U(n-1)$, one can describe \textit{e.g.},
\begin{align}
\begin{split}
&
\mathcal{S} \sim (1|0,\ldots,0),\qquad
\mathcal{S}^* \sim (-1|0,\ldots,0),
\\
&
\mathcal{O}_V(p)=(\mathcal{S}^*)^{\otimes p} \sim  (-p|0,\ldots,0),\qquad
\mathcal{O}_V(p)^*=\mathcal{S}^{\otimes p} \sim  (p|0,\ldots,0),
\\
&
TV \sim (-1|0,\ldots,0,1),\qquad
T^*V \sim (1|-1,0,\ldots,0),
\label{rep_proj_v}
\end{split}
\end{align}
where $\mathcal{S}$ is the universal subbundle on $V={\IP}^{n-1}$. The representations of tensor product and wedge product are obtained as \textit{e.g.},
$$
\mathcal{O}_V(p)^* \otimes T^*V \sim (p+1|-1,0,\ldots,0),\qquad
\wedge^2 TV \sim (-2|0,\ldots,0,1,1).
$$
\end{exam}

\begin{exam}\label{app_ex:gkn}
For $V=G(k,n)=U(n)/[U(k)\times U(n-k)]$, in terms of the representations of $U(k)\times U(n-k)$, one can describe \textit{e.g.},
\begin{align}
\begin{split}
&
\mathcal{S} \sim (0,\ldots,0,1|0,\ldots,0),\qquad
\mathcal{S}^* \sim (-1,0,\ldots,0|0,\ldots,0),
\\
&
\mathcal{O}_V(p)=(\det \mathcal{S}^*)^{\otimes p} \sim  (-p,\ldots,-p|0,\ldots,0),\\
&
\mathcal{O}_V(p)^*=(\det \mathcal{S})^{\otimes p} \sim  (p,\ldots,p|0,\ldots,0),
\\
&
TV \sim (-1,0,\ldots,0|0,\ldots,0,1),\qquad
T^*V \sim (0,\ldots,0,1|-1,0,\ldots,0),
\label{rep_proj_v2}
\end{split}
\end{align}
where $\mathcal{S}$ is the rank $k$ universal subbundle on $V=G(k,n)$. The representations of tensor product and wedge product are obtained \textit{e.g.} for $V=G(2,5)$ as
\begin{align}
\begin{split}
&
\mathcal{S}(-1)=\mathcal{S} \otimes \mathcal{O}_V(1)^* \sim (1,2|0,\ldots,0),\qquad
\wedge^2 \mathcal{S}(-1) \sim (3,3|0,\ldots,0,0),
\\
&
\mathcal{S}(-1) \wedge \mathcal{O}_V(2)^* \sim (3,4|0,\ldots,0,0),\qquad
\left(\wedge^2 \mathcal{S}(-1)\right) \wedge \mathcal{O}_V(2)^* \sim (5,5|0,\ldots,0,0).
\nonumber
\end{split}
\end{align}
\end{exam}

\begin{exam}\label{app_ex:p2g24}
For product manifold $V={\IP}^2 \times G(2,4)$, one obtains \textit{e.g.},
\begin{align}
\begin{split}
&
\mathcal{S}(0,-1)=\mathcal{S} \otimes \mathcal{O}_V(0,1)^* \sim 
\left(
\begin{array}{cc|cc}
 & 1 & 0 & 0 \\
1 & 2 & 0 & 0
\end{array}
\right),\qquad
\wedge^2 \mathcal{S}(0,-1) \sim 
\left(
\begin{array}{cc|cc}
 & 2 & 0 & 0 \\
3 & 3 & 0 & 0
\end{array}
\right),
\\
&
\mathcal{S}(0,-1) \wedge \mathcal{O}_V(1,1)^* \sim 
\left(
\begin{array}{cc|cc}
 & 2 & 0 & 0 \\
3 & 2 & 0 & 0
\end{array}
\right),\qquad
\left(\wedge^2 \mathcal{S}(0,-1)\right) \wedge \mathcal{O}_V(1,1)^* \sim 
\left(
\begin{array}{cc|cc}
 & 3 & 0 & 0 \\
4 & 4 & 0 & 0
\end{array}
\right),
\nonumber
\end{split}
\end{align}
where $\mathcal{S}$ is the rank $(1,2)$ universal subbundle on $V$.
\end{exam}

Using the above representations \eqref{rep_vec_flag} for vector bundles, the Bott-Borel-Weil theorem is stated as follows.

\begin{thm}[Bott-Borel-Weil]\label{thm:bott_borel_weil}
Let $\mathcal{F}_V$ be a holomorphic homogeneous vector bundle, represented as \eqref{rep_vec_flag}, over a flag manifold $V$ in \eqref{def_flag}. Then at most only one of the cohomologies $H^i(V, \mathcal{F}_V)$ is non-trivial ($\cong {\IC}^D$), and $D$ is given by the dimension of an irreducible representation $(y_1,\ldots,y_N)$ of $U(N)$ determined as follows:
\begin{itemize}
\item[{\bf 1.}]
For the sequence \eqref{rep_vec_flag}, add the sequence $(1,2,\ldots,N)$ as
$$
(a_1+1,a_2+2,\ldots,a_{n_1}+n_1,b_1+n_1+1,\ldots,b_{n_2}+n_1+n_2,\ldots,r_{n_F}+N).
$$

\item[{\bf 2.}]
If the above sequence contains any same number, the cohomologies $H^i(V, \mathcal{F}_V)$ are trivial, if not;

\item[{\bf 3.}]
Minimally swap the above sequence, with the minimal swapping number $I$, 
so that the resulting sequence gives a strictly increasing sequence $(y_1',\ldots,y_N')$, $y_i'<y_{i+1}'$.

\item[{\bf 4.}]
For the above swapped sequence, subtracting the sequence $(1,2,\ldots,N)$ as
$$
(y_1,y_2,\ldots,y_N)=(y_1'-1,y_2'-2,\ldots,y_N'-N),
$$
one obtains a representation $(y_1,\ldots,y_N)$ of $U(N)$ which gives $H^I(V, \mathcal{F}_V)$.
\end{itemize}
\end{thm}

\begin{remark}
The dimension $D$ of a representation $(y_1,\ldots,y_N)$ of $U(N)$, which is given by a Young diagram $Y$ with length $y_i$ for the $i$-th row, is computed by
\begin{align}
D=\prod_{s \in Y}\frac{F_Y(s)}{H_Y(s)},
\end{align}
where $H_Y(s)$ is the hook length of $s$ in $Y$, and $F_Y(s)=N-i+j$ for $s=(i,j)$ (the box of $i$-th row and $j$-th column).
\end{remark}

\begin{remark}
When $V$ is a product manifold $V=V_1\times V_2$ of two flag manifolds $V_1$ and $V_2$, for computing the cohomologies of $V$ one can use the \textit{K\"unneth formula}
\begin{align}
H^i(V, \mathcal{F}_V)=\bigoplus_{i_1+i_2=i}
H^{i_1}(V_1, \mathcal{F}_V|_{V_1}) \otimes H^{i_2}(V_2, \mathcal{F}_V|_{V_2}),
\end{align}
where $\mathcal{F}_V$ is a vector bundle over $V$ and each $\mathcal{F}_V|_{V_i}$ is the restricted vector bundle over $V_i$.
\end{remark}

\subsubsection{Step 2: Computation of \eqref{step2_goal}}

By Step 1 one can compute, in particular, $H^{0,i}(X)=H^{i}(X,\mathcal{O}_X)$, $H^i(X,TV|_X)$, $H^i(X,\mathcal{E}_p|_X)$, $H^i(X,T^*V|_X)$, and $H^i(X,\mathcal{E}_p^*|_X)$. Now, using these results one can compute the cohomologies \eqref{step2_goal} via the short exact sequence 
(adjunction formula)
\begin{align}
0 \longrightarrow TX \longrightarrow
TV |_X \longrightarrow \mathcal{E}_p|_X \longrightarrow 0,
\end{align}
or its dual
\begin{align}
0 \longrightarrow \mathcal{E}_p^*|_X \longrightarrow
T^*V |_X \longrightarrow  T^*X \longrightarrow 0.
\label{ex_seq_vx}
\end{align}
The former and the latter in \eqref{step2_goal} are related by the Hodge dual, and then in the following we only describe the computation of the latter by using the exact sequence \eqref{ex_seq_vx}. The exact sequence \eqref{ex_seq_vx} induces the following long exact sequence:
\begin{align}
0 & \longrightarrow & H^0(X,\mathcal{E}_p^*|_X) & \longrightarrow &
H^0(X,T^*V|_X) & \longrightarrow & H^0(X,T^*X) & \longrightarrow
\nonumber
\\
& \longrightarrow & H^1(X,\mathcal{E}_p^*|_X) & \longrightarrow &
H^1(X,T^*V|_X) & \longrightarrow & H^1(X,T^*X) & \longrightarrow \dots
\label{ex_seq_vx_coh}
\\
\dots & \longrightarrow & H^{\dim X}(X,\mathcal{E}_p^*|_X) & \longrightarrow &
H^{\dim X}(X,T^*V|_X) & \longrightarrow & H^{\dim X}(X,T^*X) & \longrightarrow 0.
\nonumber
\end{align}
Then, using the above exact sequence, from the cohomologies $H^i(X,T^*V|_X)$ and $H^i(X,\mathcal{E}_p^*|_X)$ obtained in Step 1, one can compute the cohomologies $H^i(X,T^*X)$.

\begin{remark}\label{rem:dualities}
To find other cohomologies, one can use the following well-known relations
\begin{align}
&
\textrm{\textit{Complex conjugate}:}\qquad
H^{i,j}(X) \cong H^{j,i}(X),
\nonumber\\
&
\textrm{\textit{Hodge duality}:}\qquad
H^{i,j}(X) \cong H^{\dim X-i,\dim X-j}(X).
\nonumber
\end{align}
Furthermore, if $X$ admits the unique holomorphic 3-form in 
$H^{\dim X,0}(X)={\IC}$, \textit{Serre duality} gives $H^{0,i}(X) \cong H^{\dim X,i}(X)$.
\end{remark}

\begin{remark}\label{rem:HRR}
To compute the Hodge numbers of $X$, one can also use a formula for 
the \textit{Hirzebruch $\chi_y$-genus} (see \cite{Klemm:1996ts} for 
the explicit formulae of $\dim X=2,3,4$ written in terms of the Chern 
classes of $X$):
\begin{align}
\chi_y=
\sum_{i,j=0}^{\dim X}\left(-1\right)^j\dim H^j(X,\wedge^iT^*X)\,y^i=
\int_X \prod_{k=1}^{p}\frac{\left(1+y \mathrm{e}^{-x_k}\right)x_k}{1-\mathrm{e}^{-x_k}},
\label{index_formula}
\end{align}
which is derived from \textit{Hirzebruch-Riemann-Roch index theorem}. Here $x_k$, $k=1,\ldots,p$, are the Chern roots of $X$ in \eqref{chern_rt}.
\end{remark}

\subsection{Examples}\label{subapp:hodge_ex}

We demonstrate the explicit computations of cohomologies for some examples based on the strategy in Appendix \ref{subapp:hodge_gen}.

\subsubsection{Quintic Calabi-Yau 3-fold: $\mathcal{E}_1=\mathcal{O}_V(5)$ on $V={\IP}^4$}

As a famous example, consider the quintic Calabi-Yau 3-fold $X$ defined by the zero locus of a holomorphic section of $\mathcal{E}_1=\mathcal{O}_V(5)$ on $V={\IP}^4$ \cite{Candelas:1990rm}. Using the representations in Example \ref{app_ex:pn}, the Koszul exact sequence \eqref{koszul_seq} is given by
\begin{align}
0 \longrightarrow (5|0,0,0,0) \longrightarrow
(0|0,0,0,0) \longrightarrow \mathcal{O}_X \longrightarrow 0.
\label{koszul_seq_quintic}
\end{align}
By \eqref{k_sp_coh_0} and the Bott-Borel-Weil theorem \ref{thm:bott_borel_weil}, one finds that
\begin{align}
H^0(V,\mathcal{O}_V)={\IC}\ \Longrightarrow\ H^{0,0}(X)={\IC},\quad
H^4(V,\mathcal{O}_V(5)^*)={\IC}\ \Longrightarrow\ H^{0,3}(X)={\IC},
\end{align}
and $H^{0,1}(X)$ and $H^{0,2}(X)$ are trivial, \textit{i.e.} $H^{0,1}(X)=H^{0,2}(X)=0$.

For $\mathcal{F}_V=\mathcal{O}_V(5)^*$ and $\mathcal{F}_V=T^*V$ the exact sequence \eqref{koszul_seq_t} yields
\begin{align}
&
0 \longrightarrow (10|0,0,0,0) \longrightarrow
(5|0,0,0,0) \longrightarrow \mathcal{O}_V(5)^*|_X \longrightarrow 0,
\label{koszul_seq_quintic_t1}
\\
&
0 \longrightarrow (6|-1,0,0,0) \longrightarrow
(1|-1,0,0,0) \longrightarrow T^*V|_X \longrightarrow 0.
\label{koszul_seq_quintic_t2}
\end{align}
From \eqref{koszul_seq_quintic_t1} one finds that
\begin{align}
&
\mathrm{ker}\big(d_0:\ 
H^4(V,\mathcal{O}_V(5)^* \otimes \mathcal{O}_V(5)^*)={\IC}^{126}\
\longrightarrow\ 
H^4(V,\mathcal{O}_V(5)^*)={\IC}\big)
\nonumber
\\
&
\Longrightarrow\
H^3(X,\mathcal{O}_V(5)^*|_X)={\IC}^{125},
\end{align}
and $H^i(X,\mathcal{O}_V(5)^*|_X)$, $i=0,1,2$, are trivial.
From \eqref{koszul_seq_quintic_t2} one finds that
\begin{align}
\begin{split}
&
H^1(V,T^*V)={\IC}\ \Longrightarrow\ H^1(X,T^*V|_X)={\IC},\\
&
H^4(V,\mathcal{O}_V(5)^* \otimes T^*V)={\IC}^{24}\ \Longrightarrow\ 
H^3(X,T^*V|_X)={\IC}^{24},
\end{split}
\end{align}
and $H^0(X,T^*V|_X)$ and $H^2(X,T^*V|_X)$ are trivial.
Using these results, the exact sequence \eqref{ex_seq_vx_coh} yields 
the following exact sequences:
\begin{align}
\begin{split}
&
0 \longrightarrow H^0(X,T^*X) \longrightarrow 0,
\\
&
0 \longrightarrow {\IC} \longrightarrow H^1(X,T^*X) \longrightarrow 0,
\\
&
0 \longrightarrow H^2(X,T^*X) \longrightarrow {\IC}^{125} \mathop{\longrightarrow}^{f} 
{\IC}^{24} \longrightarrow H^3(X,T^*X) \longrightarrow 0.
\end{split}
\end{align}
Then one obtains
\begin{align}
\begin{split}
&
H^{1,0}(X)=H^0(X,T^*X)=0,\quad
H^{1,1}(X)=H^1(X,T^*X)={\IC},\\
&
H^{1,2}(X)=H^2(X,T^*X)=\mathrm{ker}(f)={\IC}^{101},\quad
H^{1,3}(X)=H^3(X,T^*X)=\mathrm{coker}(f)=0.
\end{split}
\end{align}

As a result, the Hodge diamond is obtained as
\begin{center}
\begin{small}
\begin{tabular}{ccccccc}
{}&{}&{}&$h^{0,0}$&{}&{}&{}\\ {}&{}&$h^{1,0}$&{}&$h^{0,1}$&{}&{}\\
{}&$h^{2,0}$&{}&$h^{1,1}$&{}&$h^{0,2}$&{}\\ $h^{3,0}$&{}&$h^{2,1}$&{}&$h^{1,2}$ &{}&$h^{0,3}$\\ {}&$h^{3,1}$&{}&$h^{2,2}$&{}&$h^{1,3}$&{}\\ 
{}&{}&$h^{3,2}$&{}&$h^{2,3}$&{}&{}\\ {}&{}&{}&$h^{3,3}$&{}&{}&{}
\end{tabular}
$=$
\begin{tabular}{ccccccc}
{}&{}&{}&1&{}&{}&{}\\ {}&{}&0&{}&0&{}&{}\\
{}&0&{}&$1$&{}&0&{}\\ 1&{}&$101$&{}&$101$ &{}&1\\ {}&0&{}&$1$&{}&0&{}\\ 
{}&{}&0&{}&0&{}&{}\\ {}&{}&{}&1&{}&{}&{}
\end{tabular}.
\end{small}
\end{center}

\subsubsection{Grassmannian Calabi-Yau 3-fold: $\mathcal{E}_3=\mathcal{S}^*(1)\oplus \mathcal{O}_V(2)$ on $V=G(2,5)$}

As a non-abelian example, consider a Grassmannian Calabi-Yau 3-fold $X$ defined by the zero locus of a holomorphic section of $\mathcal{E}_3=\mathcal{S}^*(1)\oplus \mathcal{O}_V(2)$ on $V=G(2,5)$ \cite{Inoue16_1,Benedetti:2016}. 
Using the representations in Example \ref{app_ex:gkn}, the Koszul exact sequence \eqref{koszul_seq} is given by
\begin{align}
0 \to 
(5,5|0,0,0) \to 
\begin{array}{c}
(3, 3| 0, 0, 0) \\
\oplus \\
(3, 4| 0, 0, 0)
\end{array}
\to 
\begin{array}{c}
(1, 2| 0, 0, 0) \\
\oplus \\
(2, 2| 0, 0, 0)
\end{array}
\to
(0,0|0,0,0) \to \mathcal{O}_X \to 0.
\label{koszul_seq_gr_cy3_1}
\end{align}
By \eqref{k_sp_coh_0} and the Bott-Borel-Weil theorem \ref{thm:bott_borel_weil}, one finds that
\begin{align}
H^0(V,\mathcal{O}_V)={\IC}\ \Longrightarrow\ H^{0,0}(X)={\IC},\quad
H^6(V,\wedge^3\mathcal{E}_3^*)={\IC}\ \Longrightarrow\ H^{0,3}(X)={\IC},
\end{align}
and $H^{0,1}(X)$ and $H^{0,2}(X)$ are trivial, \textit{i.e.} $H^{0,1}(X)=H^{0,2}(X)=0$.

For $\mathcal{F}_V=\mathcal{S}(1)$, $\mathcal{O}_V(2)^*$, and $T^*V$ the exact sequence \eqref{koszul_seq_t} yields respectively,
\begin{align}
&
0 \to 
(6,7|0,0,0) \to 
\begin{array}{c}
(4, 5| 0, 0, 0) \\
\oplus \\
(5, 5| 0, 0, 0) \\
\oplus \\
(4, 6| 0, 0, 0)
\end{array}
\to 
\begin{array}{c}
(2, 4| 0, 0, 0) \\
\oplus \\
(3, 3| 0, 0, 0) \\
\oplus \\
(3, 4| 0, 0, 0) 
\end{array}
\to
(1,2|0,0,0) \to \mathcal{S}(1)|_X \to 0,
\label{koszul_seq_gr_cy3_1_t1}
\\
&
0 \to 
(7,7|0,0,0) \to 
\begin{array}{c}
(5, 5| 0, 0, 0) \\
\oplus \\
(5, 6| 0, 0, 0)
\end{array}
\to 
\begin{array}{c}
(3, 4| 0, 0, 0) \\
\oplus \\
(4, 4| 0, 0, 0)
\end{array}
\to
(2,2|0,0,0) \to \mathcal{O}(2)^*|_X \to 0,
\label{koszul_seq_gr_cy3_1_t2}
\end{align}
and
\begin{align}
0 \to 
(5,6|-1,0,0) \to 
\begin{array}{c}
(3, 4| -1, 0, 0) \\
\oplus \\
(3, 5| -1, 0, 0) \\
\oplus \\
(4, 4| -1, 0, 0)
\end{array}
\to 
\begin{array}{c}
(1, 3| -1, 0, 0) \\
\oplus \\
(2, 2| -1, 0, 0) \\
\oplus \\
(2, 3| -1, 0, 0)
\end{array}
\to
(0,1|-1,0,0) \to T^*V|_X \to 0.
\label{koszul_seq_gr_cy3_1_t3}
\end{align}
From \eqref{koszul_seq_gr_cy3_1_t1} one finds that
\begin{align}
&
\mathrm{ker}\big(d_0:\ 
H^6(V,\wedge^3\mathcal{E}_3^* \otimes \mathcal{S}(1))={\IC}^{40}\
\longrightarrow\ 
H^6(V,\wedge^2\mathcal{E}_3^* \otimes \mathcal{S}(1))={\IC}\big)
\nonumber
\\
&
\Longrightarrow\
H^3(X,\mathcal{S}(1)|_X)={\IC}^{39},
\label{gr1_s1}
\end{align}
and $H^i(X,\mathcal{S}(1)|_X)$, $i=0,1,2$, are trivial. From \eqref{koszul_seq_gr_cy3_1_t2} one finds that
\begin{align}
&
\mathrm{ker}\big(d_0:\ 
H^6(V,\wedge^3\mathcal{E}_3^* \otimes \mathcal{O}_V(2)^*)={\IC}^{50}\
\longrightarrow\ 
H^6(V,\wedge^2\mathcal{E}_3^* \otimes \mathcal{O}_V(2)^*)={\IC}^{6}\big)
\nonumber
\\
&
\Longrightarrow\
H^3(X,\mathcal{O}_V(2)^*|_X)={\IC}^{44},
\label{gr1_o2}
\end{align}
and $H^i(X,\mathcal{O}_V(2)^*|_X)$, $i=0,1,2$, are trivial. 
Then one gets $H^3(X,\mathcal{E}_3^*|_X)={\IC}^{83}$ by \eqref{gr1_s1} and \eqref{gr1_o2}. 
From \eqref{koszul_seq_gr_cy3_1_t3} one finds that
\begin{align}
\begin{split}
&
H^1(V,T^*V)={\IC}\ \Longrightarrow\ H^1(X,T^*V|_X)={\IC},\\
&
H^6(V,\wedge^3\mathcal{E}_3^* \otimes T^*V)={\IC}^{24}\ \Longrightarrow\ 
H^3(X,T^*V|_X)={\IC}^{24},
\end{split}
\end{align}
and $H^0(X,T^*V|_X)$ and $H^2(X,T^*V|_X)$ are trivial.
Using these results, the exact sequence \eqref{ex_seq_vx_coh} yields 
the following exact sequences:
\begin{align}
\begin{split}
&
0 \longrightarrow H^0(X,T^*X) \longrightarrow 0,
\\
&
0 \longrightarrow {\IC} \longrightarrow H^1(X,T^*X) \longrightarrow 0,
\\
&
0 \longrightarrow H^2(X,T^*X) \longrightarrow {\IC}^{83} \mathop{\longrightarrow}^{f} 
{\IC}^{24} \longrightarrow H^3(X,T^*X) \longrightarrow 0.
\end{split}
\end{align}
Then one obtains
\begin{align}
\begin{split}
&
H^{1,0}(X)=H^0(X,T^*X)=0,\quad
H^{1,1}(X)=H^1(X,T^*X)={\IC},\\
&
H^{1,2}(X)=H^2(X,T^*X)=\mathrm{ker}(f)={\IC}^{59},\quad
H^{1,3}(X)=H^3(X,T^*X)=\mathrm{coker}(f)=0.
\end{split}
\end{align}

As a result, the Hodge diamond is obtained as
\begin{center}
\begin{small}
\begin{tabular}{ccccccc}
{}&{}&{}&$h^{0,0}$&{}&{}&{}\\ {}&{}&$h^{1,0}$&{}&$h^{0,1}$&{}&{}\\
{}&$h^{2,0}$&{}&$h^{1,1}$&{}&$h^{0,2}$&{}\\ $h^{3,0}$&{}&$h^{2,1}$&{}&$h^{1,2}$ &{}&$h^{0,3}$\\ {}&$h^{3,1}$&{}&$h^{2,2}$&{}&$h^{1,3}$&{}\\ 
{}&{}&$h^{3,2}$&{}&$h^{2,3}$&{}&{}\\ {}&{}&{}&$h^{3,3}$&{}&{}&{}
\end{tabular}
$=$
\begin{tabular}{ccccccc}
{}&{}&{}&1&{}&{}&{}\\ {}&{}&0&{}&0&{}&{}\\
{}&0&{}&$1$&{}&0&{}\\ 1&{}&$59$&{}&$59$ &{}&1\\ {}&0&{}&$1$&{}&0&{}\\ 
{}&{}&0&{}&0&{}&{}\\ {}&{}&{}&1&{}&{}&{}
\end{tabular}.
\end{small}
\end{center}

\subsubsection{Determinantal Calabi-Yau 3-fold in \eqref{det3_k2} 
with $\mathcal{F}_3=\mathcal{S}^*(1)\oplus \mathcal{O}_V(1)$}

Consider a determinantal Calabi-Yau 3-fold in \eqref{det3_k2} with $\mathcal{F}_3=\mathcal{S}^*(1)\oplus \mathcal{O}_V(1)$. We especially consider a geometric phase, and then the desingularized determinantal Calabi-Yau 3-fold $X$ is defined by the locus of a holomorphic section of 
$\mathcal{E}_3=\mathcal{E}_3^{(2)}\oplus \mathcal{E}_3^{(1)}$ on $V'=G(2,4)\times {\IP}^2$, where 
$\mathcal{E}_3^{(2)}=\mathcal{S}^* \otimes \mathcal{O}_{V'}(1,0)$ and 
$\mathcal{E}_3^{(1)}=\mathcal{O}_{V'}(1,1)$. 
Using the representations in Example \ref{app_ex:p2g24}, the Koszul exact sequence \eqref{koszul_seq} is given by
\begin{align}
\scalebox{0.85}{$\displaystyle
0 \to 
\left(
\begin{array}{cc|cc}
4 & 4 & 0 & 0 \\
  & 3 & 0 & 0
\end{array}
\right)
\to 
\begin{array}{c}
\left(
\begin{array}{cc|cc}
3 & 3 & 0 & 0 \\
  & 2 & 0 & 0
\end{array}
\right)
\\
\oplus \\
\left(
\begin{array}{cc|cc}
2 & 3 & 0 & 0 \\
  & 2 & 0 & 0
\end{array}
\right)
\end{array}
\to 
\begin{array}{c}
\left(
\begin{array}{cc|cc}
1 & 2 & 0 & 0 \\
  & 1 & 0 & 0
\end{array}
\right)
\\
\oplus \\
\left(
\begin{array}{cc|cc}
1 & 1 & 0 & 0 \\
  & 1 & 0 & 0
\end{array}
\right)
\end{array}
\to
\left(
\begin{array}{cc|cc}
0 & 0 & 0 & 0 \\
  & 0 & 0 & 0
\end{array}
\right)
\to \mathcal{O}_X \to 0.$}
\label{koszul_seq_det_cy3_1}
\end{align}
By \eqref{k_sp_coh_0} and the Bott-Borel-Weil theorem \ref{thm:bott_borel_weil}, one finds that
\begin{align}
H^0(V',\mathcal{O}_{V'})={\IC}\ \Longrightarrow\ H^{0,0}(X)={\IC},\quad
H^6(V',\wedge^3\mathcal{E}_3^*)={\IC}\ \Longrightarrow\ H^{0,3}(X)={\IC},
\end{align}
and $H^{0,1}(X)$ and $H^{0,2}(X)$ are trivial, \textit{i.e.} $H^{0,1}(X)=H^{0,2}(X)=0$.

The exact sequence \eqref{koszul_seq_t} gives, for 
$\mathcal{F}_{V'}=\mathcal{E}_3^{(2)\,*}$,
\begin{align}
\scalebox{0.85}{$\displaystyle
0 \to 
\left(
\begin{array}{cc|cc}
5 & 6 & 0 & 0 \\
  & 4 & 0 & 0
\end{array}
\right)
\to 
\begin{array}{c}
\left(
\begin{array}{cc|cc}
4 & 5 & 0 & 0 \\
  & 3 & 0 & 0
\end{array}
\right)
\\
\oplus \\
\left(
\begin{array}{cc|cc}
3 & 5 & 0 & 0 \\
  & 3 & 0 & 0
\end{array}
\right)
\\
\oplus \\
\left(
\begin{array}{cc|cc}
4 & 4 & 0 & 0 \\
  & 3 & 0 & 0
\end{array}
\right)
\end{array}
\to 
\begin{array}{c}
\left(
\begin{array}{cc|cc}
2 & 4 & 0 & 0 \\
  & 2 & 0 & 0
\end{array}
\right)
\\
\oplus \\
\left(
\begin{array}{cc|cc}
3 & 3 & 0 & 0 \\
  & 2 & 0 & 0
\end{array}
\right)
\\
\oplus \\
\left(
\begin{array}{cc|cc}
2 & 3 & 0 & 0 \\
  & 2 & 0 & 0
\end{array}
\right)
\end{array}
\to
\left(
\begin{array}{cc|cc}
1 & 2 & 0 & 0 \\
  & 1 & 0 & 0
\end{array}
\right)
\to \mathcal{E}_3^{(2)\,*}|_X \to 0,$}
\label{koszul_seq_det_cy3_t1}
\end{align}
for $\mathcal{F}_{V'}=\mathcal{E}_3^{(1)\,*}$,
\begin{align}
\scalebox{0.85}{$\displaystyle
0 \to 
\left(
\begin{array}{cc|cc}
5 & 5 & 0 & 0 \\
  & 4 & 0 & 0
\end{array}
\right)
\to 
\begin{array}{c}
\left(
\begin{array}{cc|cc}
4 & 4 & 0 & 0 \\
  & 3 & 0 & 0
\end{array}
\right)
\\
\oplus \\
\left(
\begin{array}{cc|cc}
3 & 4 & 0 & 0 \\
  & 3 & 0 & 0
\end{array}
\right)
\end{array}
\to 
\begin{array}{c}
\left(
\begin{array}{cc|cc}
2 & 3 & 0 & 0 \\
  & 2 & 0 & 0
\end{array}
\right)
\\
\oplus \\
\left(
\begin{array}{cc|cc}
2 & 2 & 0 & 0 \\
  & 2 & 0 & 0
\end{array}
\right)
\end{array}
\to
\left(
\begin{array}{cc|cc}
1 & 1 & 0 & 0 \\
  & 1 & 0 & 0
\end{array}
\right)
\to \mathcal{E}_3^{(1)\,*}|_X \to 0,$}
\label{koszul_seq_det_cy3_t2}
\end{align}
for $\mathcal{F}_{V'}=T^*G(2,4)$,
\begin{align}
\scalebox{0.83}{$\displaystyle
0 \to 
\left(
\begin{array}{cc|cc}
4 & 5 & -1 & 0 \\
  & 3 & 0 & 0
\end{array}
\right)
\to 
\begin{array}{c}
\left(
\begin{array}{cc|cc}
3 & 4 & -1 & 0 \\
  & 2 & 0 & 0
\end{array}
\right)
\\
\oplus \\
\left(
\begin{array}{cc|cc}
2 & 4 & -1 & 0 \\
  & 2 & 0 & 0
\end{array}
\right)
\\
\oplus \\
\left(
\begin{array}{cc|cc}
3 & 3 & -1 & 0 \\
  & 2 & 0 & 0
\end{array}
\right)
\end{array}
\to 
\begin{array}{c}
\left(
\begin{array}{cc|cc}
1 & 3 & -1 & 0 \\
  & 1 & 0 & 0
\end{array}
\right)
\\
\oplus \\
\left(
\begin{array}{cc|cc}
2 & 2 & -1 & 0 \\
  & 1 & 0 & 0
\end{array}
\right)
\\
\oplus \\
\left(
\begin{array}{cc|cc}
1 & 2 & -1 & 0 \\
  & 1 & 0 & 0
\end{array}
\right)
\end{array}
\to
\left(
\begin{array}{cc|cc}
0 & 1 & -1 & 0 \\
  & 0 & 0 & 0
\end{array}
\right)
\to T^*G(2,4)|_X \to 0,$}
\label{koszul_seq_det_cy3_t3}
\end{align}
and for $\mathcal{F}_{V'}=T^*{\IP}^2$,
\begin{align}
\scalebox{0.85}{$\displaystyle
0 \to 
\left(
\begin{array}{cc|cc}
4 & 4 & 0 & 0 \\
  & 4 & 0 & 0
\end{array}
\right)
\to 
\begin{array}{c}
\left(
\begin{array}{cc|cc}
3 & 3 & 0 & 0 \\
  & 3 & -1 & 0
\end{array}
\right)
\\
\oplus \\
\left(
\begin{array}{cc|cc}
2 & 3 & 0 & 0 \\
  & 3 & -1 & 0
\end{array}
\right)
\end{array}
\to 
\begin{array}{c}
\left(
\begin{array}{cc|cc}
1 & 2 & 0 & 0 \\
  & 2 & -1 & 0
\end{array}
\right)
\\
\oplus \\
\left(
\begin{array}{cc|cc}
1 & 1 & 0 & 0 \\
  & 2 & -1 & 0
\end{array}
\right)
\end{array}
\to
\left(
\begin{array}{cc|cc}
0 & 0 & 0 & 0 \\
  & 1 & -1 & 0
\end{array}
\right)
\to T^*{\IP}^2|_X \to 0.$}
\label{koszul_seq_det_cy3_t4}
\end{align}
From \eqref{koszul_seq_det_cy3_t1} one finds that
\begin{align}
&
\mathrm{ker}\big(d_0:\ 
H^6(V',\wedge^3\mathcal{E}_3^* \otimes \mathcal{E}_3^{(2)\,*})={\IC}^{60}\
\longrightarrow\ 
H^6(V',\wedge^2\mathcal{E}_3^* \otimes \mathcal{E}_3^{(2)\,*})={\IC}^{5}\big)
\nonumber
\\
&
\Longrightarrow\
H^3(X,\mathcal{E}_3^{(2)\,*}|_X)={\IC}^{55},
\label{det1_s1}
\end{align}
and $H^i(X,\mathcal{E}_3^{(2)\,*}|_X)$, $i=0,1,2$, are trivial. 
From \eqref{koszul_seq_det_cy3_t2} one finds that
\begin{align}
&
\mathrm{ker}\big(d_0:\ 
H^6(V',\wedge^3\mathcal{E}_3^* \otimes \mathcal{E}_3^{(1)\,*})={\IC}^{18}\
\longrightarrow\ 
H^6(V',\wedge^2\mathcal{E}_3^* \otimes \mathcal{E}_3^{(1)\,*})={\IC}\big)
\nonumber
\\
&
\Longrightarrow\
H^3(X,\mathcal{E}_3^{(1)\,*}|_X)={\IC}^{17},
\label{det1_s2}
\end{align}
and $H^i(X,\mathcal{E}_3^{(1)\,*}|_X)$, $i=0,1,2$, are trivial.
Then one gets $H^3(X,\mathcal{E}_3^*|_X)={\IC}^{72}$ by \eqref{det1_s1} and \eqref{det1_s2}. 
From \eqref{koszul_seq_det_cy3_t3} one finds that
\begin{align}
\begin{split}
&
H^1(V',T^*G(2,4))={\IC}\ \Longrightarrow\ H^1(X,T^*G(2,4)|_X)={\IC},\\
&
H^6(V',\wedge^3\mathcal{E}_3^* \otimes T^*G(2,4))={\IC}^{15}\ \Longrightarrow\ 
H^3(X,T^*G(2,4)|_X)={\IC}^{15},
\label{det1_t1}
\end{split}
\end{align}
and $H^0(X,T^*G(2,4)|_X)$ and $H^2(X,T^*G(2,4)|_X)$ are trivial. 
From \eqref{koszul_seq_det_cy3_t4} one finds that
\begin{align}
\begin{split}
&
H^1(V',T^*{\IP}^2)={\IC}\ \Longrightarrow\ H^1(X,T^*{\IP}^2|_X)={\IC},\\
&
H^6(V',\wedge^3\mathcal{E}_3^* \otimes T^*{\IP}^2)={\IC}^{8}\ \Longrightarrow\ 
H^3(X,T^*{\IP}^2|_X)={\IC}^{8},
\label{det1_t2}
\end{split}
\end{align}
and $H^0(X,T^*{\IP}^2|_X)$ and $H^2(X,T^*{\IP}^2|_X)$ are trivial. 
Then one gets $H^1(X,T^*V'|_X)={\IC}^{2}$ and $H^3(X,T^*V'|_X)={\IC}^{23}$ by \eqref{det1_t1} and \eqref{det1_t2}. 
Using these results, the exact sequence \eqref{ex_seq_vx_coh} yields 
the following exact sequences:
\begin{align}
\begin{split}
&
0 \longrightarrow H^0(X,T^*X) \longrightarrow 0,
\\
&
0 \longrightarrow {\IC}^2 \longrightarrow H^1(X,T^*X) \longrightarrow 0,
\\
&
0 \longrightarrow H^2(X,T^*X) \longrightarrow {\IC}^{72} \mathop{\longrightarrow}^{f} 
{\IC}^{23} \longrightarrow H^3(X,T^*X) \longrightarrow 0.
\end{split}
\end{align}
Then one obtains
\begin{align}
\begin{split}
&
H^{1,0}(X)=H^0(X,T^*X)=0,\quad
H^{1,1}(X)=H^1(X,T^*X)={\IC}^2,\\
&
H^{1,2}(X)=H^2(X,T^*X)=\mathrm{ker}(f)={\IC}^{49},\quad
H^{1,3}(X)=H^3(X,T^*X)=\mathrm{coker}(f)=0.
\end{split}
\end{align}

As a result, the Hodge diamond is obtained as
\begin{center}
\begin{small}
\begin{tabular}{ccccccc}
{}&{}&{}&$h^{0,0}$&{}&{}&{}\\ {}&{}&$h^{1,0}$&{}&$h^{0,1}$&{}&{}\\
{}&$h^{2,0}$&{}&$h^{1,1}$&{}&$h^{0,2}$&{}\\ $h^{3,0}$&{}&$h^{2,1}$&{}&$h^{1,2}$ &{}&$h^{0,3}$\\ {}&$h^{3,1}$&{}&$h^{2,2}$&{}&$h^{1,3}$&{}\\ 
{}&{}&$h^{3,2}$&{}&$h^{2,3}$&{}&{}\\ {}&{}&{}&$h^{3,3}$&{}&{}&{}
\end{tabular}
$=$
\begin{tabular}{ccccccc}
{}&{}&{}&1&{}&{}&{}\\ {}&{}&0&{}&0&{}&{}\\
{}&0&{}&$2$&{}&0&{}\\ 1&{}&$49$&{}&$49$ &{}&1\\ {}&0&{}&$2$&{}&0&{}\\ 
{}&{}&0&{}&0&{}&{}\\ {}&{}&{}&1&{}&{}&{}
\end{tabular}.
\end{small}
\end{center}

\section{Genus-0 invariants of determinantal Calabi-Yau 4-folds}\label{app:gw_det}

Following the techniques introduced in Section \ref{subsec:I_fn_det_ex}, here we consider 
genus-0 invariants of the desingularized determinantal Calabi-Yau 
4-folds clarified in Appendix \ref{subapp:cy4} 
which are described by 
$U(k) \times U(\ell_p^{\vee})$ PAX models with $k, \ell_p^{\vee}=1,2$, 
while ignoring the examples with the universal quotient bundle $\mathcal{Q}$ in $\mathcal{F}_p$. 
For the computation, we need to evaluate the classical intersection numbers 
\eqref{classic_int_det} simply denoted by
\begin{align}
H \sigma_{1}^2,\quad
H \sigma_{2},\quad
H \sigma_{1} \tau_{1},\quad
H \tau_{1}^2,\quad
H \tau_{2}.
\label{classic_int_det_c}
\end{align}
We also compute their Hodge numbers by analysing 
the Koszul complex 
introduced in Appendix \ref{app:hodge}. 
Note that here the genus-0 invariants $n_{d_1,d_2}(H)$ in \eqref{det_gr_gw_i} 
are denoted as
\begin{align}
\begin{split}
&
n_{d_1,d_2,11}=n_{d_1,d_2}(\sigma_1^2),\quad
n_{d_1,d_2,\sigma}=n_{d_1,d_2}(\sigma_2),\quad
n_{d_1,d_2,12}=n_{d_1,d_2}(\sigma_1\tau_1),
\\
&
n_{d_1,d_2,22}=n_{d_1,d_2}(\tau_1^2),\quad
n_{d_1,d_2,\tau}=n_{d_1,d_2}(\tau_2).
\label{gw_cy4_sp}
\end{split}
\end{align}

\subsection{Sextic family}\label{app:sextic}

As a higher dimensional analogue of the quintic family discussed
in Section \ref{subsubapp:quintic}, we consider the determinantal Calabi-Yau 4-folds in \eqref{class_cy4_1} which 
can be described by $U(1)\times U(1)$ GLSMs. 
The ``trivial'' determinantal Calabi-Yau 4-fold with $\mathcal{F}_p=\mathcal{O}_V(6)$ on $V={\IP}^5$ corresponds 
to the sextic Calabi-Yau 4-fold $X_1$ with
$n=6$, $r=1$, $d_1=6$ in Section \ref{subsub:cy_cp}. 
The Hodge numbers are $(h^{1,1}, h^{2,1}, h^{2,2}, h^{3,1})=(1, 0, 1752, 426)$. 
The classical intersection number \eqref{proj_cy_int} 
and the genus-0 invariants \eqref{1pt_ex1} of the sextic $X_1$ are given by 
$\kappa=6$ and
\begin{align}
n_{1,11}=60480,\ n_{2,11}=440884080,\ n_{3,11}=6255156277440,\ 
n_{4,11}=117715791990353760,\ldots,
\label{sextic_inv}
\end{align}
respectively \cite{Greene:1993vm}. 

Several genus-0 invariants of the sextic family \eqref{class_cy4_1} are 
summarized in Table \ref{det_cy4_p5}, 
where one can check that $(h^{1,0},h^{2,0})=(0,0)$ 
and there is a relation originated from the extremal transition:
\begin{align}
n_{d,11}=\sum_{d_2=0}^{N} n_{d,d_2,11},
\label{ext_trans_cy4}
\end{align}
where $N$ is a certain finite positive integer.

\begin{footnotesize}
\begin{longtable}[c]{|c|rrrrrr|}
\caption{Genus-0 invariants of determinantal 4-folds in \eqref{class_cy4_1} with $V={\IP}^5$}
\label{det_cy4_p5}
\endfirsthead
\hline
\multicolumn{7}{|c|}{$\mathcal{F}_p=\mathcal{O}_V(1)\oplus \mathcal{O}_V(5)$: 
$(h^{1,1}, h^{2,1}, h^{2,2}, h^{3,1})=(2, 0, 1452, 350)$, ($n_{d_1,d_2,22}=0$)}
\\
\hline
\multicolumn{3}{|c|}{Intersection numbers}&
\multicolumn{4}{|c|}{$\sigma_1^4=6,\quad \sigma_1^3\tau_1=5,\quad \sigma_1^2\tau_1^2=0,\quad \sigma_1\tau_1^3=0,\quad \tau_1^4=0$}
\\
\hline
$n_{d_1,d_2,11}$ & $d_1=0$ & 1 & 2 & 3 & 4 & 5 \\ \hline
$d_2=0$ &   & 11100 & 5974850   & 5337637100    & 5961261947000
& 7549696778037500  \\
     1  & 0 & 47800 & 139595300 & 341903160900  & 781526104500800
& 1722498037214056500  \\
     2  & 0 & 2300  & 288301400 & 2474705048600 & 12772788325116200
& 51691531760557694400  \\
     3  & 0 & -900  & 10709800  & 3363595465000 & 51229393390313200
& 425107528698920155100  \\
     4  & 0 & 200   & -5618400  & 103567454100  & 51958819718170400
& 1158355364337024993600  \\
     5  & 0 & -20   & 2835300   & -51911590000  & 1403818415592500
& 938149531037521616000  \\
\hline
$n_{d_1,d_2,12}$ & $d_1=0$ & 1 & 2 & 3 & 4 & 5 \\ \hline
$d_2=0$ &    & 2875  & 1218500   & 951619125     & 969870120000
& 1146529444438125  \\
     1  & 25 & 43025 & 80799950  & 156102470525  & 304442819735350
& 596487343049391900  \\
     2  & 0  & 7075  & 268094350 & 1716513933050 & 7342810580729600
& 25898280425210696100  \\
     3  & 0  & -3325 & 27921700  & 3182702667725 & 38694830186103150
& 274863504753902753625  \\
     4  & 0  & 850   & -16827350 & 244126695475  & 49743335407652800
& 920672555667202043750  \\
     5  & 0  & -100  & 9475875   & -139333207500 & 3049558752331250
& 905924121779310315625  \\
\hhline{=======}
\multicolumn{7}{|c|}{$\mathcal{F}_p=\mathcal{O}_V(2)\oplus \mathcal{O}_V(4)$: 
$(h^{1,1}, h^{2,1}, h^{2,2}, h^{3,1})=(2, 0, 984, 233)$, ($n_{d_1,d_2,22}=0$)}
\\
\hline
\multicolumn{3}{|c|}{Intersection numbers}&
\multicolumn{4}{|c|}{$\sigma_1^4=6,\quad \sigma_1^3\tau_1=8,\quad \sigma_1^2\tau_1^2=0,\quad \sigma_1\tau_1^3=0,\quad \tau_1^4=0$}
\\
\hline
$n_{d_1,d_2,11}$ & $d_1=0$ & 1 & 2 & 3 & 4 & 5 \\ \hline
$d_2=0$ &   & 5152  & 933968    & 274818272     & 100238592192
& 41343866067168  \\
     1  & 0 & 30464 & 30631168  & 24983703040   & 18823860029184
& 13607365845297920  \\
     2  & 0 & 24384 & 148136832 & 344078200064  & 540875866571264
& 689210122091722112  \\
     3  & 0 & 512   & 196648704 & 1484112439552 & 5090782137162496
& 11687877699117056512  \\
     4  & 0 & -32   & 63767008  & 2480102598912 & 20463045985953792
& 88125209117491672192  \\
     5  & 0 & 0     & 833024    & 1597722995712 & 39020950213890816
& 338009904921145655552  \\
\hline
$n_{d_1,d_2,12}$ & $d_1=0$ & 1 & 2 & 3 & 4 & 5 \\ \hline
$d_2=0$ &    & 1280  & 184576    & 46965504      & 15535610112
& 5954616410880  \\
     1  & 64 & 32000 & 19441024  & 12147379712   & 7666954166848
& 4863492485707008  \\
     2  & 0  & 45696 & 155375232 & 264748238336  & 338696685934592
& 370700658349715200  \\
     3  & 0  & 1792  & 288667776 & 1555786807552 & 4265241324428224
& 8297783461923275008  \\
     4  & 0  & -128  & 121767232 & 3295974933504 & 21444381835426304
& 77477798899226917376  \\
     5  & 0  & 0     & 2638208   & 2582769371136 & 49128078374461248
& 354147270418259729152  \\
\hhline{=======}
\multicolumn{7}{|c|}{$\mathcal{F}_p=\mathcal{O}_V(3)^{\oplus 2}$: 
$(h^{1,1}, h^{2,1}, h^{2,2}, h^{3,1})=(2, 0, 780, 182)$, ($n_{d_1,d_2,22}=0$)}
\\
\hline
\multicolumn{3}{|c|}{Intersection numbers}&
\multicolumn{4}{|c|}{$\sigma_1^4=6,\quad \sigma_1^3\tau_1=9,\quad \sigma_1^2\tau_1^2=0,\quad \sigma_1\tau_1^3=0,\quad \tau_1^4=0$}
\\
\hline
$n_{d_1,d_2,11}$ & $d_1=0$ & 1 & 2 & 3 & 4 & 5 \\ \hline
$d_2=0$ &   & 3996  & 528012    & 111620808     & 29176888824
& 8616413173572  \\
     1  & 0 & 26244 & 18834444  & 10994448492   & 5928552658692
& 3066843382569540  \\
     2  & 0 & 26244 & 107617896 & 174298692024  & 193364991313056
& 174792376622296872  \\
     3  & 0 & 3996  & 186923376 & 916489110132  & 2154915386009316
& 3451110870456005940  \\
     4  & 0 & 0     & 107617896 & 2025684267264 & 10761703969222224
& 31389410244093246936  \\
     5  & 0 & 0     & 18834444  & 2025684267264 & 27258432537609648
& 151496100969322358520  \\
\hline
$n_{d_1,d_2,12}$ & $d_1=0$ & 1 & 2 & 3 & 4 & 5 \\ \hline
$d_2=0$ &    & 1053  & 105624    & 19272978      & 4557793536
& 1248939462915  \\
     1  & 81 & 27945 & 12158991  & 5429580417    & 2449467003132
& 1110716561847951  \\
     2  & 0  & 50787 & 114975450 & 136362364218  & 122998311431838
& 95426230272412266  \\
     3  & 0  & 10935 & 280385064 & 978583930209  & 1836173061024606
& 2489535822420553203  \\
     4  & 0  & 0     & 207878238 & 2745855802998 & 11484979618590612
& 28070822227226686614  \\
     5  & 0  & 0     & 44344341  & 3331196998794 & 34978559566645782
& 161618974169879654460  \\
\hhline{=======}
\multicolumn{7}{|c|}{$\mathcal{F}_p=\mathcal{O}_V(2)^{\oplus 3}$: 
$(h^{1,1}, h^{2,1}, h^{2,2}, h^{3,1})=(2, 0, 600, 137)$}
\\
\hline
\multicolumn{3}{|c|}{Intersection numbers}&
\multicolumn{4}{|c|}{$\sigma_1^4=6,\quad \sigma_1^3\tau_1=12,\quad \sigma_1^2\tau_1^2=8,\quad \sigma_1\tau_1^3=0,\quad \tau_1^4=0$}
\\
\hline
$n_{d_1,d_2,11}$ & $d_1=0$ & 1 & 2 & 3 & 4 & 5 \\ \hline
$d_2=0$ &   & 1104  & 13464     & 196848        & 3102144
& 52343184  \\
     1  & 0 & 14016 & 1708224   & 120184704     & 6485946432
& 298466405568  \\
     2  & 0 & 30240 & 24792672  & 6198463872    & 882834675456
& 89849127844224  \\
     3  & 0 & 14016 & 107708352 & 89554908096   & 31167666411840
& 6550759785428544  \\
     4  & 0 & 1104  & 172438656 & 529075766016  & 444446786009856
& 187337558146915488  \\
     5  & 0 & 0     & 107708352 & 1473092769024 & 3125024856532800
& 2640802670799084864  \\
\hline
$n_{d_1,d_2,12}$ & $d_1=0$ & 1 & 2 & 3 & 4 & 5 \\ \hline
$d_2=0$ &    & 384   & 3744      & 47232         & 670272
& 10462080  \\
     1  & 96 & 16128 & 1240896   & 68822016      & 3180759840
& 130766711040  \\
     2  & 0  & 60480 & 28308096  & 5320032000    & 629362595328
& 55970237060352  \\
     3  & 0  & 39936 & 168950976 & 102040729344  & 28860224852448
& 5218075942288896  \\
     4  & 0  & 4032  & 344877312 & 753418660608  & 505820047097088
& 181160699096012928  \\
     5  & 0  & 0     & 261882432 & 2521023722496 & 4224146727978144
& 3003266065628143872  \\
\hline
$n_{d_1,d_2,22}$ & $d_1=0$ & 1 & 2 & 3 & 4 & 5 \\ \hline
$d_2=0$ &    & 0     & 0         & 0             & 0
& 0  \\
     1  & 96 & 7296  & 358080    & 14528256      & 528892320
& 17932977792  \\
     2  & 0  & 38592 & 13180992  & 1930140672    & 186374277120
& 13980674976768  \\
     3  & 0  & 31104 & 97081920  & 47789472384   & 11376521292384
& 1772136040869504  \\
     4  & 0  & 3648  & 226720032 & 411451649280  & 237591482788608
& 74578392862419072  \\
     5  & 0  & 0     & 190013376 & 1533694993920 & 2239140736880928
& 1413152060075199360  \\
\hhline{=======}
\multicolumn{7}{|c|}{$\mathcal{F}_p=\mathcal{O}_V(1)\oplus \mathcal{O}_V(2)\oplus \mathcal{O}_V(3)$: 
$(h^{1,1}, h^{2,1}, h^{2,2}, h^{3,1})=(2, 0, 732, 170)$}
\\
\hline
\multicolumn{3}{|c|}{Intersection numbers}&
\multicolumn{4}{|c|}{$\sigma_1^4=6,\quad \sigma_1^3\tau_1=11,\quad \sigma_1^2\tau_1^2=6,\quad \sigma_1\tau_1^3=0,\quad \tau_1^4=0$}
\\
\hline
$n_{d_1,d_2,11}$ & $d_1=0$ & 1 & 2 & 3 & 4 & 5 \\ \hline
$d_2=0$ &   & 1547  & 29197     & 664966        & 16655276
& 449773471  \\
     1  & 0 & 17415 & 3387141   & 390771675     & 34733397231
& 2631133285191  \\
     2  & 0 & 31234 & 40960891  & 16736425757   & 3941526901738
& 666261117547152  \\
     3  & 0 & 10138 & 143247860 & 196780670876  & 113318694822063
& 39594013933183682  \\
     4  & 0 & 147   & 172936880 & 924949591952  & 1299468913703502
& 913025985733713523  \\
     5  & 0 & -1    & 72200686  & 1974502966996 & 7224038700165558
& 10275054838460143332  \\
\hline
$n_{d_1,d_2,12}$ & $d_1=0$ & 1 & 2 & 3 & 4 & 5 \\ \hline
$d_2=0$ &    & 546   & 8022      & 159708        & 3637752
& 91579530  \\
     1  & 85 & 19550 & 2414660   & 220345310     & 16823027671
& 1141797701894  \\
     2  & 0  & 61679 & 45928018  & 14111157767   & 2763493715428
& 408754388583466  \\
     3  & 0  & 28503 & 221344804 & 220417729691  & 103132562736885
& 31012895322615811  \\
     4  & 0  & 607   & 341620521 & 1297053558298 & 1454510648416542
& 868094813403547075  \\
     5  & 0  & -5    & 173699427 & 3333597269546 & 9614835008030966
& 11495407244589530600  \\
\hline
$n_{d_1,d_2,22}$ & $d_1=0$ & 1 & 2 & 3 & 4 & 5 \\ \hline
$d_2=0$ &    & 0     & 0         & 0             & 0
& 0  \\
     1  & 66 & 7944  & 665916    & 45550248      & 2770233906
& 155976830808  \\
     2  & 0  & 33774 & 19165296  & 4739973306    & 774617782512
& 98109047877348  \\
     3  & 0  & 18270 & 110465976 & 92373759582   & 37217080568574
& 9812878163478678  \\
     4  & 0  & 498   & 190515834 & 619593755556  & 610866205882020
& 325210132883359866  \\
     5  & 0  & -6    & 104930694 & 1742691953364 & 4476709752274068
& 4833888549187725312  \\
\hhline{=======}
\multicolumn{7}{|c|}{$\mathcal{F}_p=\mathcal{O}_V(1)^{\oplus 2}\oplus \mathcal{O}_V(4)$: 
$(h^{1,1}, h^{2,1}, h^{2,2}, h^{3,1})=(2, 0, 1068, 254)$}
\\
\hline
\multicolumn{3}{|c|}{Intersection numbers}&
\multicolumn{4}{|c|}{$\sigma_1^4=6,\quad \sigma_1^3\tau_1=9,\quad \sigma_1^2\tau_1^2=4,\quad \sigma_1\tau_1^3=0,\quad \tau_1^4=0$}
\\
\hline
$n_{d_1,d_2,11}$ & $d_1=0$ & 1 & 2 & 3 & 4 & 5 \\ \hline
$d_2=0$ &   & 2796  & 111420    & 5415876       & 297906744
& 17836490652  \\
     1  & 0 & 25368 & 11074512  & 2923919928    & 597405875232
& 103952142270864  \\
     2  & 0 & 31452 & 96756426  & 92450088216   & 50890408972608
& 20094224706004404  \\
     3  & 0 & 1020  & 215383416 & 746257722660  & 1032988712010672
& 857190260480341476  \\
     4  & 0 & -168  & 115603488 & 2184801219792 & 7902484863731640
& 13590303106004642712  \\
     5  & 0 & 12    & 2269284   & 2399750411904 & 27079947685057788
& 100137122291641121868  \\
\hline
$n_{d_1,d_2,12}$ & $d_1=0$ & 1 & 2 & 3 & 4 & 5 \\ \hline
$d_2=0$ &    & 960   & 30096     & 1297728       & 65879904
& 3714521280  \\
     1  & 57 & 27438 & 7644177   & 1605192702    & 282936273240
& 44282085901044  \\
     2  & 0  & 3657  & 104754642 & 75519972306   & 34661964976188
& 12004979836835943  \\
     3  & 0  & -681  & 320554332 & 807765491547  & 910677818722380
& 651630952247594127  \\
     4  & 0  & 57    & 222052140 & 2956519017480 & 8551880283800160
& 12515007949041304332  \\
     5  & 0  & 0     & 7374009   & 3919508517492 & 34811597561726766
& 108347819585126752047  \\
\hline
$n_{d_1,d_2,22}$ & $d_1=0$ & 1 & 2 & 3 & 4 & 5 \\ \hline
$d_2=0$ &    & 0     & 0         & 0             & 0
& 0  \\
     1  & 36 & 10104 & 2000580   & 323373624     & 46020698976
& 6018914291280  \\
     2  & 0  & 28260 & 39402408  & 23483405064   & 9172938062064
& 2759901496567260  \\
     3  & 0  & 2340  & 140674032 & 304708034988  & 301404502100016
& 191986430652018684  \\
     4  & 0  & -420  & 108023472 & 1249191338400 & 3231021483376128
& 4278942916449123696  \\
     5  & 0  & 36    & 4447332   & 1793863512336 & 14384155887545592
& 40973480285079165564  \\
\hhline{=======}
\multicolumn{7}{|c|}{$\mathcal{F}_p=\mathcal{O}_V(1)^{\oplus 2}\oplus \mathcal{O}_V(2)^{\oplus 2}$: 
$(h^{1,1}, h^{2,1}, h^{2,2}, h^{3,1})=(2, 0, 588, 134)$}
\\
\hline
\multicolumn{3}{|c|}{Intersection numbers}&
\multicolumn{4}{|c|}{$\sigma_1^4=6,\quad \sigma_1^3\tau_1=13,\quad \sigma_1^2\tau_1^2=12,\quad \sigma_1\tau_1^3=4,\quad \tau_1^4=0$}
\\
\hline
$n_{d_1,d_2,11}$ & $d_1=0$ & 1 & 2 & 3 & 4 & 5 \\ \hline
$d_2=0$ &   & 624   & 1512      & 2448         & 2352
& 3696  \\
     1  & 0 & 10552 & 560168    & 12419664     & 167237208
& 1611465704  \\
     2  & 0 & 29420 & 12216604  & 1270882648   & 63182183032
& 1933845747256  \\
     3  & 0 & 17892 & 74261296  & 28590861496  & 4065328815272
& 310191664249228  \\
     4  & 0 & 1988  & 162756004 & 245615224064 & 91643735242360
& 15636111812478864  \\
     5  & 0 & 4     & 139994444 & 965140012064 & 959192743343064
& 351280339787531780  \\
\hline
$n_{d_1,d_2,12}$ & $d_1=0$ & 1 & 2 & 3 & 4 & 5 \\ \hline
$d_2=0$ &    & 256   & 504       & 768           & 784
& 1280  \\
     1  & 97 & 12412 & 431558    & 7712632       & 90771993
& 798269252  \\
     2  & 0  & 59597 & 14304660  & 1139364456    & 47795706040
& 1296194947220  \\
     3  & 0  & 51377 & 118235744 & 33425512482   & 3910275895911
& 259643765174029  \\
     4  & 0  & 7379  & 329186320 & 355897081724  & 107040862884600
& 15662234678584878  \\
     5  & 0  & 19    & 343523755 & 1673793068984 & 1321513374907908
& 410052719920052585  \\
\hline
$n_{d_1,d_2,22}$ & $d_1=0$ & 1 & 2 & 3 & 4 & 5 \\ \hline
$d_2=0$ &     & 0     & 0         & 0             & 0
& 0  \\
     1  & 132 & 7696  & 167672    & 2165664       & 19882116
& 143012720  \\
     2  & 0   & 51428 & 8977216   & 553197376     & 18798577216
& 427186381936  \\
     3  & 0   & 53076 & 91577920  & 21040910824   & 2061819081980
& 117330402886948  \\
     4  & 0   & 8732  & 290650960 & 261518705648  & 67487950758592
& 8623613017337400  \\
     5  & 0   & 28    & 332988380 & 1368920496992 & 941519514353136
& 258784559873434676  \\
\hhline{=======}
\multicolumn{7}{|c|}{$\mathcal{F}_p=\mathcal{O}_V(1)^{\oplus 3}\oplus \mathcal{O}_V(3)$: 
$(h^{1,1}, h^{2,1}, h^{2,2}, h^{3,1})=(2, 0, 744, 173)$}
\\
\hline
\multicolumn{3}{|c|}{Intersection numbers}&
\multicolumn{4}{|c|}{$\sigma_1^4=6,\quad \sigma_1^3\tau_1=12,\quad \sigma_1^2\tau_1^2=10,\quad \sigma_1\tau_1^3=3,\quad \tau_1^4=0$}
\\
\hline
$n_{d_1,d_2,11}$ & $d_1=0$ & 1 & 2 & 3 & 4 & 5 \\ \hline
$d_2=0$ &   & 876   & 2754      & 2340          & 4506
& 6384  \\
     1  & 0 & 13224 & 1111680   & 40000284      & 854694672
& 12548849748  \\
     2  & 0 & 31692 & 20814618  & 3502913724    & 285343689342
& 14296324624164  \\
     3  & 0 & 14388 & 105558792 & 66167398224   & 15392391723972
& 1933287656228904  \\
     4  & 0 & 312   & 182631774 & 469032117948  & 288248187932658
& 80990413510330656  \\
     5  & 0 & -12   & 111190416 & 1474197621264 & 2474334577233804
& 1501801964420459808  \\
\hline
$n_{d_1,d_2,12}$ & $d_1=0$ & 1 & 2 & 3 & 4 & 5 \\ \hline
$d_2=0$ &    & 363   & 726       & 702           & 1452
& 1815  \\
     1  & 84 & 15267 & 846654    & 24454455      & 451560702
& 5971621263  \\
     2  & 0  & 63390 & 24022410  & 3100366203    & 213030667314
& 9432564106950  \\
     3  & 0  & 40614 & 165803226 & 76310133777   & 14614556335185
& 1597429012014378  \\
     4  & 0  & 1383  & 364788126 & 670587309870  & 332222425569714
& 80079498824062290  \\
     5  & 0  & -57   & 269640918 & 2524188530034 & 3364141372570932
& 1730133151583505285  \\
\hline
$n_{d_1,d_2,22}$ & $d_1=0$ & 1 & 2 & 3 & 4 & 5 \\ \hline
$d_2=0$ &     & 0     & 0         & 0             & 0
& 0  \\
     1  & 102 & 8976  & 325302    & 6898368       & 99988848
& 1084024620  \\
     2  & 0   & 51372 & 14396598  & 1462589520    & 82497883614
& 3089416732524  \\
     3  & 0   & 38808 & 121109886 & 45893323332   & 7447857934212
& 704572082465418  \\
     4  & 0   & 1716  & 301039440 & 466225271184  & 200138738462880
& 42499979609169498  \\
     5  & 0   & -72   & 242665410 & 1939341503028 & 2272289621718792
& 1043316736192230672  \\
\hhline{=======}
\multicolumn{7}{|c|}{$\mathcal{F}_p=\mathcal{O}_V(1)^{\oplus 4}\oplus \mathcal{O}_V(2)$: 
$(h^{1,1}, h^{2,1}, h^{2,2}, h^{3,1})=(2, 0, 552, 125)$}
\\
\hline
\multicolumn{3}{|c|}{Intersection numbers}&
\multicolumn{4}{|c|}{$\sigma_1^4=6,\quad \sigma_1^3\tau_1=14,\quad \sigma_1^2\tau_1^2=16,\quad \sigma_1\tau_1^3=9,\quad \tau_1^4=2$}
\\
\hline
$n_{d_1,d_2,11}$ & $d_1=0$ & 1 & 2 & 3 & 4 & 5 \\ \hline
$d_2=0$ &   & 360   & 110       & 0            & 0
& 0  \\
     1  & 0 & 7780  & 181660    & 1182060      & 3226100
& 4267500  \\
     2  & 0 & 27260 & 5703230   & 247915040    & 4199645570
& 35730530600  \\
     3  & 0 & 21540 & 47022520  & 8368536780   & 487011519380
& 13321298891500  \\
     4  & 0 & 3500  & 137972490 & 101219111200 & 16761773920870
& 1158268337627600  \\
     5  & 0 & 40    & 161154360 & 546681379380 & 252762644451260
& 40097079604396400  \\
\hline
$n_{d_1,d_2,12}$ & $d_1=0$ & 1 & 2 & 3 & 4 & 5 \\ \hline
$d_2=0$ &     & 165   & 40        & 0            & 0
& 0  \\
     1  & 100 & 9345  & 147740    & 785690       & 1894775
& 2320375  \\
     2  & 0   & 55990 & 6844020   & 231492010    & 3350412730
& 25503215825  \\
     3  & 0   & 62310 & 76006130  & 10030946845  & 485539930370
& 11668962937750  \\
     4  & 0   & 13125 & 282277260 & 149233093550 & 20076931277380
& 1199570758766725  \\
     5  & 0   & 185   & 399094290 & 960929261945 & 354879437832140
& 48004138643512850  \\
\hline
$n_{d_1,d_2,22}$ & $d_1=0$ & 1 & 2 & 3 & 4 & 5 \\ \hline
$d_2=0$ &     & 0     & 0         & 0            & 0
& 0  \\
     1  & 170 & 6860  & 65700     & 250120       & 468350
& 468350  \\
     2  & 0   & 58020 & 5042710   & 129717540    & 1502417400
& 9512031410  \\
     3  & 0   & 77400 & 70019970  & 7403781260   & 296605169360
& 6050035785700  \\
     4  & 0   & 18700 & 298541880 & 129860240700 & 14833108148590
& 766859594261640  \\
     5  & 0   & 300   & 465232050 & 936876897400 & 298655994540650
& 35488204982996150  \\
\hhline{=======}
\multicolumn{7}{|c|}{$\mathcal{F}_p=\mathcal{O}_V(1)^{\oplus 6}$: 
$(h^{1,1}, h^{2,1}, h^{2,2}, h^{3,1})=(2, 0, 492, 110)$, ($n_{d_1,d_2,22}=n_{d_2,d_1,11}$) \cite{Honma:2013hma}}
\\
\hline
\multicolumn{3}{|c|}{Intersection numbers}&
\multicolumn{4}{|c|}{$\sigma_1^4=6,\quad \sigma_1^3\tau_1=15,\quad \sigma_1^2\tau_1^2=20,\quad \sigma_1\tau_1^3=15,\quad \tau_1^4=6$}
\\
\hline
$n_{d_1,d_2,11}$ & $d_1=0$ & 1 & 2 & 3 & 4 & 5 \\ \hline
$d_2=0$ &   & 210   & 0         & 0            & 0
& 0  \\
     1  & 0 & 5670  & 59430     & 100170       & 34650
& 1680  \\
     2  & 0 & 24360 & 2579640   & 47382930     & 264433680
& 546221760  \\
     3  & 0 & 24360 & 28015260  & 2324403900   & 55841697870
& 539959428960  \\
     4  & 0 & 5670  & 107096220 & 38404166850  & 2848564316640
& 80315543697900  \\
     5  & 0 & 210   & 165382980 & 277070715810 & 60035324018880
& 4163431890254700  \\
\hline
$n_{d_1,d_2,12}$ & $d_1=0$ & 1 & 2 & 3 & 4 & 5 \\ \hline
$d_2=0$ &     & 105   & 0         & 0            & 0
& 0  \\
     1  & 105 & 6930  & 50715     & 71085        & 21420
& 945  \\
     2  & 0   & 50715 & 3166800   & 45928155     & 221593050
& 413457450  \\
     3  & 0   & 71085 & 45928155  & 2851172100   & 57546197940
& 493317415605  \\
     4  & 0   & 21420 & 221593050 & 57546197940  & 3492450469200
& 85788539294850  \\
     5  & 0   & 945   & 413457450 & 493317415605 & 85788539294850
& 5102793274479600  \\
\hline
\end{longtable}
\end{footnotesize}

\subsection{Determinantal Calabi-Yau 4-folds in \eqref{class_cy4_3}}\label{app:cy4_k2}

Finally, we consider the determinantal Calabi-Yau 4-folds with $p\ne 2$ in 
\eqref{class_cy4_3} 
which are described by $U(2)\times U(2)$ GLSMs,
while ignoring the examples with the universal quotient bundle $\mathcal{Q}$ 
in $\mathcal{F}_p$. 
We summarized the genus-0 invariants of \eqref{gw_cy4_sp} in Table \ref{det_cy4_g26}, 
where one can check that $(h^{1,0},h^{2,0})=(0,0)$. 
For the Calabi-Yau 4-folds with $\mathcal{F}_p=\mathcal{S}^* \oplus \mathcal{O}_V(1)^{\oplus 2}$, $\left(\mathcal{S}^*\right)^{\oplus 2} \oplus \mathcal{O}_V(1)$, $\left(\mathcal{S}^*\right)^{\oplus 3}$, 
just due to a technical complexity, 
instead of the Hodge numbers we give the $\chi_y$-genera 
$\chi_i=\sum_{j=0}^4(-1)^j h^{i,j}$, $i=0,1,2$ obtained 
by the formula \eqref{index_formula}. 

Note that the determinantal 4-fold with 
$\mathcal{F}_p=\Sym^2 \mathcal{S}^*$ in \eqref{class_cy4_3} 
is an irreducible holomorphic symplectic variety with 
$(h^{1,0},h^{2,0})=(0,1)$, and all the genus-0 invariants 
vanished. This property is known to be a general phenomenon for 
irreducible holomorphic symplectic varieties 
(see \textit{e.g.} \cite{Maulik:2010jw}).

\begin{footnotesize}
\begin{longtable}[c]{|c|rrrrrrr|}
\caption{Genus-0 invariants of determinantal 4-folds in \eqref{class_cy4_3} with $V=G(2,6)$}
\label{det_cy4_g26}
\endfirsthead
\hline
\multicolumn{8}{|c|}{$\mathcal{F}_p=\mathcal{O}_V(1)^{\oplus 3}$: 
$(h^{1,1}, h^{2,1}, h^{2,2}, h^{3,1})=(2, 0, 384, 83)$, ($n_{d_1,d_2,\tau}=0$)}
\\
\hline
\multicolumn{3}{|c|}{Intersection numbers}&
\multicolumn{5}{|c|}{$\sigma_1^4=84,\quad \sigma_1^2\sigma_2=54,
\quad \sigma_1^3\tau_1=42,\quad \sigma_1^2\tau_1^2=14,\quad 
\sigma_1^2\tau_2=0,\quad \sigma_2^2=36,$}
\\
\hhline{---}
\multicolumn{8}{|c|}{$\sigma_1\sigma_2\tau_1=27,\quad \sigma_2\tau_1^2=9,
\quad \sigma_2\tau_2=0,\quad \sigma_1\tau_1^3=0,\quad 
\sigma_1\tau_1\tau_2=0,\quad \tau_1^4=0,\quad \tau_1^2\tau_2=0,
\quad \tau_2^2=0$}
\\
\hline
$n_{d_1,d_2,11}$ & $d_1=0$ & 1 & 2 & 3 & 4 & 5 & 6 \\ \hline
$d_2=0$ &   & 966 & 6258  & 40194   & 313992    & 2465694 
& 20471724 \\
     1  & 0 & 966 & 79464 & 2850624 & 73342920  & 1577557254 
& 30264388560 \\
     2  & 0 & 0   & 6258  & 2850624 & 353216472 & 23351152860 
& 1075419836442 \\
     3  & 0 & 0   & 0     & 40194   & 73342920  & 23351152860 
& 3280923722160 \\
     4  & 0 & 0   & 0     & 0       & 313992    & 1577557254 
& 1075419836442 \\
     5  & 0 & 0   & 0     & 0       & 0         & 2465694 
& 30264388560 \\
\hline
$n_{d_1,d_2,\sigma}$ & $d_1=0$ & 1 & 2 & 3 & 4 & 5 & 6 \\ \hline
$d_2=0$ &   & 639 & 3987  & 25857   & 201888    & 1584999 
& 13160502 \\
     1  & 0 & 639 & 51804 & 1846260 & 47378196  & 1017817191 
& 19510365672 \\
     2  & 0 & 0   & 3987  & 1846260 & 229064418 & 15125263182 
& 695710713879 \\
     3  & 0 & 0   & 0     & 25857   & 47378196  & 15125263182 
& 2125753214616 \\
     4  & 0 & 0   & 0     & 0       & 201888    & 1017817191 
& 695710713879 \\
     5  & 0 & 0   & 0     & 0       & 0         & 1584999 
& 19510365672 \\
\hline
$n_{d_1,d_2,12}$ & $d_1=0$ & 1 & 2 & 3 & 4 & 5 & 6 \\ \hline
$d_2=0$ &   & 210 & 1218  & 7182    & 50400     & 369810 
& 2894220 \\
     1  & 0 & 756 & 39732 & 1136940 & 25336080  & 491193528 
& 8693374452 \\
     2  & 0 & 0   & 5040  & 1713684 & 176608236 & 10236028824 
& 426294634584 \\
     3  & 0 & 0   & 0     & 33012   & 48006840  & 13115124036 
& 1640461861080 \\
     4  & 0 & 0   & 0     & 0       & 263592    & 1086363726 
& 649125201858 \\
     5  & 0 & 0   & 0     & 0       & 0         & 2095884 
& 21571014108 \\
\hline
$n_{d_1,d_2,22}$ & $d_1=0$ & 1 & 2 & 3 & 4 & 5 & 6 \\ \hline
$d_2=0$ &   & 0   & 0     & 0      & 0        & 0 
& 0 \\
     1  & 0 & 546 & 13692 & 260484 & 4432344  & 69555990 
& 1036360500 \\
     2  & 0 & 0   & 3822  & 837228 & 64074948 & 2961523404 
& 102768656970 \\
     3  & 0 & 0   & 0     & 25830  & 27103104 & 5840618616 
& 604275131880 \\
     4  & 0 & 0   & 0     & 0      & 213192   & 664726188 
& 325599224244 \\
     5  & 0 & 0   & 0     & 0      & 0        & 1726074 
& 13914000156 \\
\hhline{========}
\multicolumn{8}{|c|}{$\mathcal{F}_p=\mathcal{S}^* \oplus \mathcal{O}_V(2)$: 
$(h^{1,1}, h^{2,1}, h^{2,2}, h^{3,1})=(2, 0, 636, 146)$, ($n_{d_1,d_2,\tau}=0$)}
\\
\hline
\multicolumn{3}{|c|}{Intersection numbers}&
\multicolumn{5}{|c|}{$\sigma_1^4=48,\quad \sigma_1^2\sigma_2=31,
\quad \sigma_1^3\tau_1=24,\quad \sigma_1^2\tau_1^2=8, \quad 
\sigma_1^2\tau_2=0,\quad \sigma_2^2=21,$}
\\
\hhline{---}
\multicolumn{8}{|c|}{$\sigma_1\sigma_2\tau_1=14,\quad \sigma_2\tau_1^2=4,
\quad \sigma_2\tau_2=0,\quad \sigma_1\tau_1^3=0,\quad 
\sigma_1\tau_1\tau_2=0,\quad \tau_1^4=0,\quad \tau_1^2\tau_2=0,
\quad \tau_2^2=0$}
\\
\hline
$n_{d_1,d_2,11}$ & $d_1=0$ & 1 & 2 & 3 & 4 & 5 & 6 \\ \hline
$d_2=0$ &   & 1536 & 17280  & 244224   & 3772608    & 62805504 
& 1099018368 \\
     1  & 0 & 1536 & 288768 & 23427072 & 1353129984 & 64724846592 
& 2742129192960 \\
     2  & 0 & 0    & 17280  & 23427072 & 7040804352 & 1086348288000 
& 114929521132032 \\
     3  & 0 & 0    & 0      & 244224   & 1353129984 & 1086348288000 
& 366458865408000 \\
     4  & 0 & 0    & 0      & 0        & 3772608    & 64724846592 
& 114929521132032 \\
     5  & 0 & 0    & 0      & 0        & 0          & 62805504 
& 2742129192960 \\
\hline
$n_{d_1,d_2,\sigma}$ & $d_1=0$ & 1 & 2 & 3 & 4 & 5 & 6 \\ \hline
$d_2=0$ &   & 864  & 9576   & 133920   & 2053872    & 34018272 
& 592976376 \\
     1  & 0 & 1152 & 186624 & 14379264 & 808399872  & 38013725952 
& 1591301187072 \\
     2  & 0 & 0    & 13248  & 16001280 & 4557463488 & 680963318784 
& 70502641947072 \\
     3  & 0 & 0    & 0      & 190080   & 952611840  & 726901318656 
& 237300376059648 \\
     4  & 0 & 0    & 0      & 0        & 2965896    & 46471290624 
& 78693300038592 \\
     5  & 0 & 0    & 0      & 0        & 0          & 49723776 
& 1996675868160 \\
\hline
$n_{d_1,d_2,12}$ & $d_1=0$ & 1 & 2 & 3 & 4 & 5 & 6 \\ \hline
$d_2=0$ &   & 384  & 3744   & 47232    & 670272     & 10462080 
& 173868768 \\
     1  & 0 & 1152 & 144384 & 9550848  & 484282368  & 21085670400 
& 830565021696 \\
     2  & 0 & 0    & 13536  & 13876224 & 3520402176 & 481923477504 
& 46543883110656 \\
     3  & 0 & 0    & 0      & 196992   & 868847616  & 604424810496 
& 183229432704000 \\
     4  & 0 & 0    & 0      & 0        & 3102336    & 43639176192 
& 68385638021376 \\
     5  & 0 & 0    & 0      & 0        & 0          & 52343424 
& 1911564171264 \\
\hline
$n_{d_1,d_2,22}$ & $d_1=0$ & 1 & 2 & 3 & 4 & 5 & 6 \\ \hline
$d_2=0$ &   & 0   & 0     & 0       & 0          & 0 
& 0 \\
     1  & 0 & 768 & 49152 & 2224128 & 86114304   & 3039080448 
& 100762681344 \\
     2  & 0 & 0   & 9792  & 6549504 & 1255686144 & 138466443264 
& 11215284827136 \\
     3  & 0 & 0   & 0     & 149760  & 470679552  & 260967776256 
& 66108681437184 \\
     4  & 0 & 0   & 0     & 0       & 2432064    & 25592586240 
& 33057039737856 \\
     5  & 0 & 0   & 0     & 0       & 0          & 41881344 
& 1181761830912 \\
\hhline{========}
\multicolumn{8}{|c|}{$\mathcal{F}_p=\mathcal{S}^* \oplus \mathcal{O}_V(1)^{\oplus 2}$: 
$(\chi_0, \chi_1, \chi_2)=(2, -80, 364)$}
\\
\hline
\multicolumn{3}{|c|}{Intersection numbers}&
\multicolumn{5}{|c|}{$\sigma_1^4=86,\quad \sigma_1^2\sigma_2=55,
\quad \sigma_1^3\tau_1=66,\quad \sigma_1^2\tau_1^2=42, \quad 
\sigma_1^2\tau_2=17,\quad \sigma_2^2=37,$}
\\
\hhline{---}
\multicolumn{8}{|c|}{$\sigma_1\sigma_2\tau_1=41,\quad \sigma_2\tau_1^2=25,
\quad \sigma_2\tau_2=10,\quad \sigma_1\tau_1^3=18,\quad 
\sigma_1\tau_1\tau_2=9,\quad \tau_1^4=4,\quad \tau_1^2\tau_2=2,
\quad \tau_2^2=2$}
\\
\hline
$n_{d_1,d_2,11}$ & $d_1=0$ & 1 & 2 & 3 & 4 & 5 & 6 \\ \hline
$d_2=0$ &   & 496  & 244   & 0       & 0         & 0 
& 0 \\
     1  & 0 & 1312 & 38880 & 238740  & 605136    & 785832 
& 546816 \\
     2  & 0 & 44   & 41500 & 3158040 & 65572008  & 586267944 
& 2862291270 \\
     3  & 0 & 0    & 760   & 1641368 & 253291560 & 11296776072 
& 221197911448 \\
     4  & 0 & 0    & 0     & 23284   & 88370324  & 21272865380 
& 1596400086708 \\
     5  & 0 & 0    & 0     & -592    & 863776    & 5440233652 
& 1834451480648 \\
\hline
$n_{d_1,d_2,\sigma}$ & $d_1=0$ & 1 & 2 & 3 & 4 & 5 & 6 \\ \hline
$d_2=0$ &   & 292 & 140   & 0       & 0         & 0 
& 0 \\
     1  & 0 & 880 & 24204 & 144470  & 361496    & 466204 
& 323180 \\
     2  & 0 & 34  & 27450 & 2002612 & 40636882  & 358061288 
& 1731003173 \\
     3  & 0 & 0   & 556   & 1087240 & 162344808 & 7095866952 
& 137011548860 \\
     4  & 0 & 0   & 0     & 16830   & 58584756  & 13723707354 
& 1011627956132 \\
     5  & 0 & 0   & 0     & -440    & 622184    & 3609161614 
& 1188798184500 \\
\hline
$n_{d_1,d_2,12}$ & $d_1=0$ & 1 & 2 & 3 & 4 & 5 & 6 \\ \hline
$d_2=0$ &   & 168  & 72    & 0       & 0         & 0 
& 0 \\
     1  & 0 & 1176 & 23688 & 119628  & 271752    & 328080 
& 216216 \\
     2  & 0 & 72   & 37524 & 2229240 & 39561432  & 317016480 
& 1428444174 \\
     3  & 0 & 0    & 1056  & 1492992 & 191488464 & 7501134312 
& 133180541856 \\
     4  & 0 & 0    & 0     & 31128   & 80807544  & 16742371056 
& 1125010922424 \\
     5  & 0 & 0    & 0     & -864    & 1140240   & 4994099532 
& 1482866868960 \\
\hline
$n_{d_1,d_2,22}$ & $d_1=0$ & 1 & 2 & 3 & 4 & 5 & 6 \\ \hline
$d_2=0$ &   & 0    & 0     & 0       & 0         & 0 
& 0 \\
     1  & 0 & 1040 & 10516 & 37316   & 66648     & 66648 
& 37316 \\
     2  & 0 & 112  & 31924 & 1296048 & 17835110  & 117630928 
& 452438518 \\
     3  & 0 & 0    & 1356  & 1271728 & 125327752 & 4009228688 
& 60469656020 \\
     4  & 0 & 0    & 0     & 37768   & 69214312  & 11674274648 
& 663812607488 \\
     5  & 0 & 0    & 0     & -1184   & 1367852   & 4298534956 
& 1076178498476 \\
\hline
$n_{d_1,d_2,\tau}$ & $d_1=0$ & 1 & 2 & 3 & 4 & 5 & 6 \\ \hline
$d_2=0$ &   & 0   & 0     & 0      & 0        & 0 
& 0 \\
     1  & 0 & 320 & 4618  & 18418  & 34404    & 34404 
& 18418 \\
     2  & 0 & 28  & 10858 & 522792 & 7879519  & 55031576 
& 219730439 \\
     3  & 0 & 0   & 382   & 432616 & 47989060 & 1663887680 
& 26568226946 \\
     4  & 0 & 0   & 0     & 11040  & 23512884 & 4349537480 
& 265076269952 \\
     5  & 0 & 0   & 0     & -320   & 402214   & 1457274582 
& 393989709022 \\
\hhline{========}
\multicolumn{8}{|c|}{$\mathcal{F}_p=\left(\mathcal{S}^*\right)^{\oplus 2} \oplus \mathcal{O}_V(1)$: 
$(\chi_0, \chi_1, \chi_2)=(2, -62, 292)$}
\\
\hline
\multicolumn{3}{|c|}{Intersection numbers}&
\multicolumn{5}{|c|}{$\sigma_1^4=92,\quad \sigma_1^2\sigma_2=58,
\quad \sigma_1^3\tau_1=92,\quad \sigma_1^2\tau_1^2=80, \quad 
\sigma_1^2\tau_2=19,\quad \sigma_2^2=40,$}
\\
\hhline{---}
\multicolumn{8}{|c|}{$\sigma_1\sigma_2\tau_1=56,\quad \sigma_2\tau_1^2=46,
\quad \sigma_2\tau_2=22,\quad \sigma_1\tau_1^3=56,\quad 
\sigma_1\tau_1\tau_2=32,\quad \tau_1^4=32,\quad \tau_1^2\tau_2=19,
\quad \tau_2^2=13$}
\\
\hline
$n_{d_1,d_2,11}$ & $d_1=0$ & 1 & 2 & 3 & 4 & 5 & 6 \\ \hline
$d_2=0$ &   & 180  & 0     & 0       & 0         & 0 
& 0 \\
     1  & 0 & 1280 & 11520 & 0       & 0         & 0 
& 0 \\
     2  & 0 & 180  & 39420 & 725760  & 1285920   & 110180 
& -36660 \\
     3  & 0 & 0    & 11520 & 1981440 & 54604800  & 288737280 
& 294952960 \\
     4  & 0 & 0    & 0     & 725760  & 127668480 & 4632572700 
& 44638440480 \\
     5  & 0 & 0    & 0     & 0       & 54604800  & 9651020800 
& 425483704320 \\
\hline
$n_{d_1,d_2,\sigma}$ & $d_1=0$ & 1 & 2 & 3 & 4 & 5 & 6 \\ \hline
$d_2=0$ &   & 90  & 0     & 0       & 0        & 0 
& 0 \\
     1  & 0 & 800 & 6560  & 0       & 0        & 0 
& 0 \\
     2  & 0 & 130 & 24730 & 428000  & 729020   & 59850 
& -19710 \\
     3  & 0 & 0   & 7840  & 1244480 & 32729280 & 167506560 
& 166957440 \\
     4  & 0 & 0   & 0     & 482400  & 80204960 & 2803401590 
& 26273744220 \\
     5  &0  & 0   & 0     & 0       & 35821760 & 6063510400 
& 259122088640 \\
\hline
$n_{d_1,d_2,12}$ & $d_1=0$ & 1 & 2 & 3 & 4 & 5 & 6 \\ \hline
$d_2=0$ &   & 60   & 0     & 0       & 0         & 0 
& 0 \\
     1  & 0 & 1280 & 7680  & 0       & 0         & 0 
& 0 \\
     2  & 0 & 300  & 39420 & 562560  & 846120    & 58940 
& -18660 \\
     3  & 0 & 0    & 15360 & 1981440 & 45327360  & 209679360 
& 192616960 \\
     4  & 0 & 0    & 0     & 888960  & 127668480 & 4000335300 
& 34440454440 \\
     5  & 0 & 0    & 0     & 0       & 63882240  & 9651020800 
& 376969835520 \\
\hline
$n_{d_1,d_2,22}$ & $d_1=0$ & 1 & 2 & 3 & 4 & 5 & 6 \\ \hline
$d_2=0$ &   & 0    & 0     & 0       & 0         & 0 
& 0 \\
     1  & 0 & 1280 & 3840  & 0       & 0         & 0 
& 0 \\
     2  & 0 & 480  & 37200 & 360960  & 439440    & 19040 
& -5520 \\
     3  & 0 & 0    & 19200 & 1835520 & 32509440  & 125291520 
& 98252800 \\
     4  & 0 & 0    & 0     & 1013760 & 117776640 & 3046100640 
& 22457270160 \\
     5  & 0 & 0    & 0     & 0       & 69619200  & 8892083200 
& 297725928960 \\
\hline
$n_{d_1,d_2,\tau}$ & $d_1=0$ & 1 & 2 & 3 & 4 & 5 & 6 \\ \hline
$d_2=0$ &   & 0   & 0     & 0      & 0        & 0 
& 0 \\
     1  & 0 & 560 & 2480  & 0      & 0        & 0 
& 0 \\
     2  & 0 & 160 & 17200 & 207920 & 270080   & 14280 
& -4140 \\
     3  & 0 & 0   & 7600  & 863840 & 17630880 & 73506240 
& 60984000 \\
     4  & 0 & 0   & 0     & 425520 & 55648880 & 1598550680 
& 12693468720 \\
     5  & 0 & 0   & 0     & 0      & 30000800 & 4206510400 
& 153166332320 \\
\hhline{========}
\multicolumn{8}{|c|}{$\mathcal{F}_p=\left(\mathcal{S}^*\right)^{\oplus 3}$: 
$(\chi_0, \chi_1, \chi_2)=(2, -22, 132)$, 
($n_{d_1,d_2,22}=n_{d_2,d_1,11}$, $n_{d_1,d_2,\tau}=n_{d_2,d_1,\sigma}$)}
\\
\hline
\multicolumn{3}{|c|}{Intersection numbers}&
\multicolumn{5}{|c|}{$\sigma_1^4=102,\quad \sigma_1^2\sigma_2=63,
\quad \sigma_1^3\tau_1=120,\quad \sigma_1^2\tau_1^2=128, \quad 
\sigma_1^2\tau_2=72,\quad \sigma_2^2=45,$}
\\
\hhline{---}
\multicolumn{8}{|c|}{$\sigma_1\sigma_2\tau_1=72,\quad \sigma_2\tau_1^2=72,
\quad \sigma_2\tau_2=36,\quad \sigma_1\tau_1^3=120,\quad 
\sigma_1\tau_1\tau_2=72,\quad \tau_1^4=102,\quad \tau_1^2\tau_2=63,
\quad \tau_2^2=45$}
\\
\hline
$n_{d_1,d_2,11}$ & $d_1=0$ & 1 & 2 & 3 & 4 & 5 & 6 \\ \hline
$d_2=0$ &   & 0   & 0     & 0       & 0        & 0 
& 0 \\
     1  & 0 & 960 & 1200  & 0       & 0        & 0 
& 0 \\
     2  & 0 & 420 & 20160 & 42300   & 600      & 0 
& 0 \\
     3  & 0 & 0   & 22800 & 668160  & 1867200  & 197520 
& 0 \\
     4  & 0 & 0   & 210   & 1206540 & 28032600 & 94238940 
& 29975670 \\
     5  & 0 & 0   & 0     & 91440   & 67122240 & 1368583200 
& 5269901040 \\
\hline
$n_{d_1,d_2,\sigma}$ & $d_1=0$ & 1 & 2 & 3 & 4 & 5 & 6 \\ \hline
$d_2=0$ &   & 0   & 0     & 0      & 0        & 0 
& 0 \\
     1  & 0 & 540 & 540   & 0      & 0        & 0 
& 0 \\
     2  & 0 & 270 & 11340 & 21330  & 270      & 0 
& 0 \\
     3  & 0 & 0   & 14580 & 382320 & 984960   & 95580 
& 0 \\
     4  & 0 & 0   & 135   & 756270 & 16113600 & 50752710 
& 15091785 \\
     5  & 0 & 0   & 0     & 58860  & 41366160 & 787977720 
& 2873723940 \\
\hline
$n_{d_1,d_2,12}$ & $d_1=0$ & 1 & 2 & 3 & 4 & 5 & 6 \\ \hline
$d_2=0$ &   & 0   & 0     & 0       & 0        & 0 
& 0 \\
     1  & 0 & 960 & 720   & 0       & 0        & 0 
& 0 \\
     2  & 0 & 720 & 20160 & 31920   & 360      & 0 
& 0 \\
     3  & 0 & 0   & 31920 & 679680  & 1547760  & 137280 
& 0 \\
     4  & 0 & 0   & 360   & 1547760 & 28646400 & 81883440 
& 22591320 \\
     5  & 0 & 0   & 0     & 137280  & 81883440 & 1400849280 
& 4715787120 \\
\hline
\end{longtable}
\end{footnotesize}

\hspace{1pc}



\begin{thebibliography}{99}
\bibitem{Martin:1999ng} 
  S.~Martin,
  ``Symplectic quotients by a nonabelian group and by its maximal torus,''
  [math/0001002 [math-sg]].
\end{thebibliography}
\end{document}